\documentclass[fleqn,usenatbib]{mnras}

\usepackage{newtxtext,newtxmath}

\usepackage[T1]{fontenc}
\usepackage[super]{nth}

\DeclareRobustCommand{\VAN}[3]{#2}
\let\VANthebibliography\thebibliography
\def\thebibliography{\DeclareRobustCommand{\VAN}[3]{##3}\VANthebibliography}

\usepackage{graphicx}	
\usepackage{amsmath}	

\usepackage{subcaption}
\usepackage{caption}

\title[SE3D RT emulator]{SE3D: Building a radiative transfer emulator to fit panchromatic resolved galaxy observations with 3D models of dust and stars}

\author[Steven Ramnichal et al.]{
Steven Ramnichal,$^{1}$\thanks{E-mail: sdr43@bath.ac.uk}
Junkai Zhang,$^{2}$
Stijn Wuyts,$^{1}$
Cheng Li$^{2}$
\\
$^{1}$Department of Physics, University of Bath, Claverton Down, Bath BA2 7AY, UK\\
$^{2}$Department of Astronomy, Tsinghua University, Beijing 1000084, China\\
}

\date{Accepted 2026 March 13. Received 2026 March 11; in original form 2025 November 24}

\pubyear{2026}

\begin{document}
\label{firstpage}
\pagerange{\pageref{firstpage}--\pageref{lastpage}}
\maketitle

\begin{abstract}
We present a framework for analysing panchromatic and spatially resolved galaxy observations, dubbed {\tt SE3D}.  {\tt SE3D} simultaneously and self-consistently models a galaxy's spectral energy distribution and its spectral distributions of global structural parameters: the wavelength-dependent galaxy size, light profile and projected axis ratio.  To this end, it employs a machine learning emulator trained on a large library of toy model galaxies processed with 3D dust radiative transfer and mock-observed under a range of viewing angles.  The toy models vary in their stellar and dust geometries, and include radial stellar population gradients.  The computationally efficient machine learning emulator uses a Bayesian neural network architecture, and reproduces the spectral distributions at an accuracy of $\sim0.05$ dex or less across the dynamic range of input parameters, and across the rest-frame {\it UVJ} colour space spanned by observed galaxies.  We carry out a sensitivity analysis demonstrating that the emulator has successfully learned the intricate mappings between galaxy physical properties and direct observables (fluxes, colours, sizes, size ratios between different wavebands, ...).  We further discuss the physical conditions giving rise to a range of total-to-selective attenuation ratios, $R_V$, with among them most prominently the projected dust surface mass density. 
\end{abstract}




\section{Introduction}
\label{Intro}

The past few decades have seen a flourishing of efforts to model the spectral energy distributions (SEDs) of galaxies with the aim of translating direct observables to intrinsic physical properties \citep[e.g.,][]{Conroy2013}.  In essence, the technique of modelling a galaxy's integrated emission as a composite of stellar spectra traces back to the pioneering work by \citet{Tinsley1968}.  Commonly used libraries with spectra of the basic building blocks, the so-called single (or mono-age) stellar populations, have been composed by, e.g., \citet{Bruzual2003}, \citet{Maraston2005} and \citet{Conroy2009}.  These can be integrated according to a particular star formation history, and subject to dust attenuation.  When paired with an energy balance approach (i.e., energy absorbed equals energy re-emitted) the predicted SEDs can further be extended into the far-infrared(FIR)/sub-mm regime \citep{daCunha2008,Boquien2019}.

For the most part, applications of the above technique have focussed on galaxy-integrated SEDs, and necessarily imposed major simplifications compared to the true and complex nature of galaxy systems.  These notably include, but are not limited to: (1) a simplified star formation history (SFH; see, e.g., \citealt{Carnall2019, Leja2019} for a discussion on parametrization), (2) the assumption of a universal attenuation law \citep[e.g.,][]{Calzetti2000}, and (3) a uniform foreground screen as dust configuration.  Here, the attenuation law encodes the wavelength dependence of the combined effects of scattering and absorption, and incorporates the net effect of the typical star-dust geometry for the sample of galaxies on which it was calibrated.  Its application via an assumed foreground screen of dust implies that dust reddening and attenuation are scaling linearly in the modelling.  Where energy balance arguments are applied, these inherently assume isotropic emission across all wavelengths, or at least their basic ansatz is understood to only hold in an angle-averaged sense.

That the above assumptions break down when considering galaxies in more detail should come as no surprise.  Indeed, evidence for non-universality of the attenuation law shape has been presented by, e.g., \citet{Salim2020} and \citet{Reddy2023}, although its inference is not free of modelling degeneracies \citep{Qin2022}.  Among sources of variation in attenuation law shape are intrinsic galaxy properties, such as their specific star formation rate \citep{Kriek2013, Reddy2018, Reddy2023}, but also observer-specific conditions such as the viewing angle.  More inclined galaxies have been shown to feature greyer attenuation laws (i.e., with shallower wavelength dependence) compared to face-on counterparts, both on the basis of statistical samples of observed galaxies \citep{Wild2011, Salim2018, Barisic2020} and as demonstrated via radiative transfer calculations \citep{Trayford2020, Zhang2023}.  As for the energy balance approach, isotropic emission is a reasonable expectation for all but the most extreme systems in the FIR regime \citep{Lovell2022}, but does not generally hold at shorter wavelengths \citep{Zhang2023}.  Integrating over $4\pi$ steradian, energy conservation dictates that the amount of dust-obscured starlight must match the emerging dust reprocessed radiation.  However, this argument does not need to hold for any individual viewing angle.

Spatially resolved studies of galaxy structure across cosmic time have also progressed dramatically over the past two decades, thanks to space telescopes ({\it HST}, {\it JWST}) and ground-based interferometers (ALMA, NOEMA).  Often, such studies address galaxy structural properties monochromatically and in 2D.  Translation to 3D intrinsic shapes is feasible in a statistical sense for ensembles of galaxies that can be treated as similar aside from their orientation \citep{van2014b, Zhang2019, Zhang2023, Pandya2024}.  With the advance from {\it HST}/ACS to {\it HST}/WFC3 and ultimately {\it JWST}/NIRCam, the rest-wavelengths probed for galaxies at cosmic noon ($1 \lesssim z \lesssim 3$) progressed from rest-UV to rest-optical and now rest-NIR emission.  Increasingly, such multi-wavelength structural measurements are complemented by longer wavelength resolved probes, in the mid-infrared (MIR) regime probing Polycyclic Aromatic Hydrocarbons (PAHs), a tracer of dust-obscured star formation \citep[e.g.,][]{Magnelli2023}, or on the Rayleigh-Jeans tail of the dust continuum \citep[e.g.,][]{Tadaki2020, Tan2024}.

One approach to leverage this multi-wavelength resolved information is to perform a pixel-by-pixel (or bin-by-bin) SED modelling \citep[e.g.,][]{Zibetti2009, Wuyts2012, Suess2019}.  However, the ability to do so panchromatically (i.e., across UV-to-submm wavelengths) remains largely restricted to well-resolved, nearby galaxies \citep[see][]{Abdurrouf2022}.  Even for nearby galaxies, such modelling is normally carried out in 2D, with few exceptions.\footnote{For examples of modelling in 3D, see the high-resolution dust radiative transfer applications by \citet{DeLooze2014, Verstocken2020, Nersesian2020a, Nersesian2020b, Viaene2020, Pricopi2025}.} The reason for the paucity of resolved panchromatic SED modelling studies at high-z owes to the heterogeneous resolution of the aforementioned instruments, the often marginally resolved nature of MIR and submm observations, and the frequently modest wavelength sampling of resolved observations for high-z galaxies.  Consequently, rather than having robust UV-to-submm SEDs for every galaxy pixel, the more common situation high-z observers face is one in which a well-sampled galaxy-integrated SED is available, and sizes (as well as S\'{e}rsic indices under good signal-to-noise ratio conditions) are measured across a subset of the wavebands.  Projected axis ratios can be extracted from high-resolution imaging too.  Given an assumed intrinsic 3D shape, or one inferred via the aforementioned ensemble modelling approach, they can provide constraints on the observer's viewing angle with respect to the galaxy under consideration.  

In this paper, we lay out a framework to exploit the full wealth of the accumulating multi-wavelength (semi-)resolved observations of distant galaxies.  What is it that one may wish to learn from a joint analysis of the galaxy-integrated SED ($F_{\lambda}(\lambda)$), wavelength-dependent size ($R_e(\lambda)$), S\'{e}rsic index ($n(\lambda)$) and axis ratio ($q(\lambda)$)?  

First, there are resolved properties, unattainable from the integrated SED alone, which can shed crucial light on the internal build-up of galaxies over cosmic time.  What are the stellar age gradients within galaxies, if any?  How is the dust distributed with respect to the stars?  How clumpy is the dust distribution, and how extended or compact is it with respect to the stellar distribution?  Both spatial variations in the SFH and spatial variations in the attenuation can give rise to colour gradients \citep{Guo2011, Guo2012, Wuyts2012, Liu2016, Liu2017, Suess2019, Miller2023, vanderWel2024, Martorano2025, Martorano2026}, and breaking these degeneracies is key to understanding the inside-out (or other) growth scenarios for galaxies across cosmic time.  As for star-dust geometries, \citet{Zhang2023} modelled the dust attenuation, dust mass and structural constraints on observed galaxy populations out to cosmic noon using toy model galaxies treated with dust radiative transfer.  They found evidence for clumpier dust geometries towards higher redshift, and highlighted how enhanced central dust columns (even for dust and stellar distributions of equal scalelength, and in the absence of stellar population gradients) can yield significant wavelength-dependent size differences.  Along a similar vein, but using a more empirical approach, \citet{Miller2022} and \citet{Martorano2026} argued that dust is the dominant driver of colour gradients in distant galaxies.  

Secondly, a more self-consistent treatment of the SED and structural constraints also has the potential of improving the accuracy of recovered galaxy-integrated properties.  After all, the approach we will outline should naturally account for non-uniform attenuation across the galaxy, and even systematic changes in attenuation law shape with viewing angle and star-dust geometry.

The objective of this first paper of the {\tt SE3D} project is to introduce a tool to simultaneously model observations of $F_{\lambda}(\lambda)$, $R_e(\lambda)$, $n(\lambda)$ and $q(\lambda)$, hereafter jointly referred to as spectral distributions (SDs), with an eye on constraining galaxies' stellar and dust content as well as distributions (i.e., age gradients and star-dust geometry).  At the heart of this tool is a Machine Learning (ML) emulator designed to speed up the computation of 3D dust radiative transfer on toy model galaxies.  The merit of ML-based emulators to reproduce the outputs of computationally expensive calculations of the actual physics has been illustrated across a wide range of fields (see \citealt{Mathews2023} and \citealt{Lovell2025} for applications in a similar context as this work, and \citealt{Kasim2021} for a range of multidisciplinary examples).  Specifically, the computational speed-up allows calling the emulator thousands of times within the model function of a Monte Carlo Markov Chain (MCMC) algorithm, enabling Bayesian inference.  Here, we describe how the emulator is constructed, assess its performance and carry out a sensitivity analysis illustrating how input parameters used to describe the physical make-up of toy model galaxies connect to accessible observables (fluxes, colours, and structural properties in different wavebands).  In a companion paper \citep[][hereafter Z25]{Zhang2025}, we further carry out tests of how well intrinsic physical properties are recovered from mock observations with realistic waveband coverage and measurement errors, when applied to toy model and simulated galaxies.

Our paper is structured as follows.  In Section\ \ref{sec:methodology}, we introduce the {\tt SE3D} methodology, from definition of the parametrized toy model galaxies to creation of a dust radiative transfer library, construction of the ML emulator and ultimately integration into a fitting algorithm. Section\ \ref{sec:results} then documents the emulator's performance, and the mapping between input physical parameters and output observables that the emulator has learned.  In Section\ \ref{sec:discussion}, we discuss the science that can be conducted with {\tt SE3D}, the connection between dust reddening and attenuation, and potential alternative ML approaches to facilitate connecting physical properties and observables. Finally, we summarize our results in Section\ \ref{sec:summary}.

Throughout the paper, we adopt a \citet{Chabrier2003} stellar initial mass function (IMF) and a flat $\Lambda$CDM cosmology with $\Omega_{\Lambda} = 0.7$, $\Omega_m = 0.3$ and $H_0 = 70\ {\rm km}\ {\rm s}^{-1}\ {\rm Mpc}^{-1}$.

\begin{figure*}
    \centering
    \includegraphics[width=\linewidth]{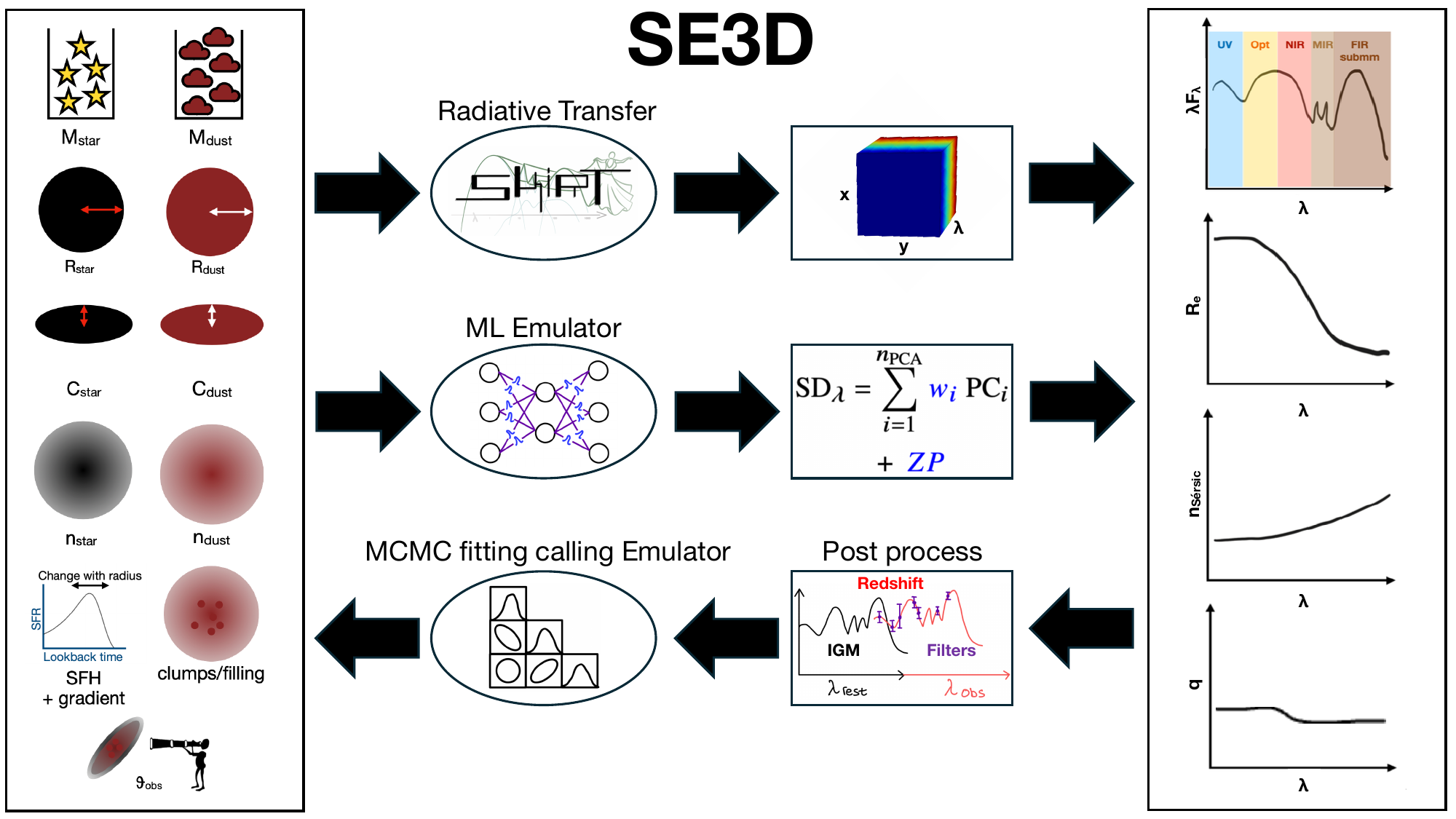}
    \caption{A visual schematic displaying the workflow of {\tt SE3D} modelling. \textit{Top:} radiative transfer is applied to toy model galaxies to produce 3D mock data cubes, which are distilled into 4 spectral distributions (flux, size, S\'{e}rsic index and axis ratio as a function of wavelength). \textit{Middle:} a Bayesian Neural Network (BNN) emulator is trained to improve computational efficiency. \textit{Bottom:} the flexible and efficient emulator, along with a post-processing step, is placed within a fitting framework to extract physical properties from observed galaxies. The post-processing step accounts for redshift, IGM absorption and filter band convolution.
    }
    \label{fig:schematic}
\end{figure*}

\section{Methodology}
\label{sec:methodology}

\subsection{SE3D framework}

We start by outlining the general approach to modelling panchromatic and spatially resolved galaxy observations followed by {\tt SE3D}.  The major steps of this framework are pictorially displayed in Figure\ \ref{fig:schematic}.  In first instance, a toy model galaxy is defined by a finite set of input parameters describing the stellar and dust content, their 3D spatial distribution, the star formation history and any radial variations it may feature, and an observer-specific viewing angle.  The specific parametrization adopted in this paper is detailed further in Section\ \ref{sec:toymodel}.

We subsequently run 3D dust radiative transfer on the toy model galaxy, and extract four spectral distributions (SDs) from its output datacube: the spectral energy distribution (SED), spectral size distribution (SRD), S\'{e}rsic index as a function of wavelength (SND) and projected axis ratio vs wavelength (SQD).  These steps are documented to greater depth in Section\ \ref{sec:SDs}.

Once a large library of toy model galaxies has been processed to observed SDs, the latter are decomposed into Principal Components.  A Bayesian neural network (BNN) is then trained to predict the weights of each component as well as a zero point, allowing us to reproduce the respective SDs given the values of input parameters describing the toy model (Section\ \ref{sec:emulator}).

With the trained emulator in hand, observed data points populating the SD diagrams can then be interpreted by searching for the best-fitting model description.  Specifically, in each model realization the continuous SDs predicted by the emulator are redshifted and convolution with the relevant filter throughput curves is applied.  The likelihood of the model describing the observations is computed and after an exploration of parameter space, posterior distributions for the various toy model parameters are obtained (Section\ \ref{sec:fitting}).

We note that in this framework there is no requirement for the different SDs to be sampled by the same set of wavebands.  For example, if galaxy-integrated {\it Herschel} photometry is available this can be included in the fit, despite there being no associated structural measurements at the same wavelengths, on account of {\it Herschel}'s lower resolution.

\subsection{Toy model galaxy definition}
\label{sec:toymodel}

\begin{table*}
\centering 
\resizebox{\linewidth}{!}
{
\begin{tabular}{l|l|l|c|c|c|c|c} 
\hline \hline
\# & Parameter & Definition & Min & Max & Type & Peak & $\sigma$ \\
\hline 
\multicolumn{8}{|c|}{Stellar properties - Mass and Structure} \\
\hline
1. & $\log(M_{\rm star})$ & Total stellar mass [$\log(M_\odot)$] & 9.5 & 11.5 & U & - & - \\
2. & $\log(R_{\rm star})$ & Size (half-mass radius) of stellar disk [$\log({\rm kpc})$] & -0.5 & 1.0 & U & - & - \\
3. & $C_{\rm{star}}/R_{\rm{star}}$ & Stellar disk thickness & 0.1 & 1.0 & N & 0.2 & 0.3 \\
4. & $\log(n_{\rm{star}})$ & S\'{e}rsic index of projected stellar distribution & -0.7 & 1.0 & N & 0.15 & 0.6 \\
\hline
\multicolumn{8}{|c|}{Stellar properties - Stellar population} \\
\hline
5. & $\log({\rm Age})$ & Time since the onset of star formation [$\log({\rm Gyr})]$ & 0.0 & 1.1 & U & - & - \\
6. & $t_{\rm peak,R_{\rm{star}}}$ & SFH peak time, evaluated at $R_{\rm star}$ [Gyr] & $0.1\times{\rm Age}$ & $1.2\times{\rm Age}$ & U & - & - \\
7. & ${\rm{fwhm}_{R_{\rm{star}}}}$ & Full-width-at-half-maximum of the log-normal SFH, evaluated at $R_{\rm star}$ [Gyr]  & $0.1\times{\rm Age}$ & $1.2\times{\rm Age}$ & U & - & - \\
8. & $k_{\rm peak}$ & Radial gradient of $t_{\rm peak}$ [dex] & -0.3 & 0.3 & N & 0.0 & 0.2 \\
9. & $k_{\rm fwhm}$ & Radial gradient of ${\rm{fwhm}}$ [dex] & -0.3 & 0.3 & N & 0.0 & 0.2 \\
10. & $Z_{\rm star}$ & Stellar metallicity [$Z_{\odot}$] & 0.2 & 2.5 & N & 1.0 & 0.5 \\
\hline
\multicolumn{8}{|c|}{ISM dust properties - Mass and Structure} \\
\hline
11. & $\log(M_{\rm dust}/M_{\rm star})$ & Ratio of dust over stellar mass & -4.0 & -2.0 & U & - & - \\
12. & $R_{\rm dust}/R_{\rm star}$ & Size ratio of dust and stellar disk & 0.1 & 4.0 & N & 1.0 & 1.0 \\
13. & $C_{\rm dust}/R_{\rm dust}$ & Dust disk thickness & 0.1 & 1.0 & N & 0.2 & 0.3 \\
14. & $\log(n_{\rm{dust}})$ & S\'{e}rsic index of projected dust distribution & -0.7 & 1.0 & N & 0.0 & 0.6\\
\hline
\multicolumn{8}{|c|}{Birth cloud properties} \\
\hline
15. & $f_{\rm{cov}}$ & Fraction of young stars (<$t_{\rm{BC}}$) in birth clouds & 0 & 1 & U & - & - \\ 
16. & $t_{\rm{BC}}$ & Lifetime of birth clouds [Myr] & 10 & 10 & U & - & - \\ 
17. & $\rm{GTS_{BC}}$ & Gas-to-star ratio in birth clouds & 10 & 10 & U & - & - \\ 
18. & $\rm{DTG_{BC}}$ & Dust-to-gas ratio in birth clouds & 0.01 & 0.01 & U & - & - \\ 
19. & $\Sigma_{\rm{gas, BC}}$ & Gas surface-density in birth clouds [$M_\odot/{\rm{pc}^2}$]& 35 & 35 & U & - & - \\ 
\hline
\multicolumn{8}{|c|}{Observer's perspective} \\
\hline
20. & $\cos(\theta)$ & Inclination ($\theta = 0^{\circ}$ is face-on) & 0 & 1 & U & - & - \\ 
\hline  
\end{tabular}
}
\caption{Parametrization of toy model galaxies.  In constructing the library of SKIRT runs, we draw parameter values within hard bounds (Min/Max) from a uniform (Type = U) or normal (Type = N) distribution.  In the normal case, values of peak location and width ($\sigma$) are also indicated.  These values were chosen to ensure a library that is well sampled where we desire the ML emulator to be trained for the most accurate performance.  Note that different priors can be set when using {\tt SE3D} to fit observed spectral distributions.
}
\label{tab:param_library}
\end{table*}

Real galaxies feature a range in structural properties, star-dust geometries and star formation histories (SFHs), including, e.g., episodic star formation and its variation across sub-structures such as spiral arms and bulges.  While the available panchromatic observations may not necessarily allow reconstructing this full complexity, we do aim to introduce sufficient flexibility in our toy model galaxy definition to capture (a) the fact that stars and dust are spatially mixed (leading to non-uniform attenuation as opposed to the uniform foreground screen of dust implemented in conventional SED modelling), (b) the fact that not all of the stars and dust will make up smooth, volume-filling components, and (c) the fact that star formation histories may vary radially.  We believe that adding this complexity with respect to traditional SED modelling -albeit captured in a simplified form by a limited set of parameters- is justified (see also Z25) because the observational constraints on high-z galaxies have become increasingly rich, including global structural parameters quantified across UV-to-mm wavelengths.

In this Section, we cover the stellar (Section\ \ref{sec:toy_stars}) and dust (Section\ \ref{sec:toy_dust}) properties of our toy models, which are rendered as tables containing the coordinates and physical properties (x, y, z, mass, age, metallicity) of stellar and (x, y, z, mass) of dust particles, respectively, for easy ingestion into the SKIRT radiative transfer code by\ \citet{Camps2020}.  We summarize all parameters in Table \ref{tab:param_library}.

\subsubsection{Stellar properties}
\label{sec:toy_stars}

The overall geometry adopted for the stellar distribution is an axisymmetric, oblate ellipsoid, characterized by a half-mass radius ($R_{\rm star}$) and disk thickness ($C_{\rm star}/R_{\rm star}$).  The 3D density profile is constant over ellipsoidal surfaces and projects to a 2D S\'{e}rsic profile of index $n_{\rm star}$.

At any radius, the stellar age distribution is characterized by a log-normal star formation history, as described in and motivated by \citet{Gladders2013}:
\begin{equation}
    \begin{aligned}
    {\rm{SFR}}(t) \propto \frac{1}{t\sqrt{2\pi\tau^2}}\ \exp \left[ -\frac{[\ln(t)-\ln(t_0)]^2}{2\tau^2} \right] .
    \end{aligned}
\label{eq:lognormalSFH}
\end{equation}
The time interval between the onset of star formation ($t = 0$) and the time of observation is captured by the parameter Age (\#5 in Table\ \ref{tab:param_library}).  Following its implementation in the Bagpipes stellar population modelling code \citep{Carnall2018}, we convert the above parametrization in terms of $t_0$ and $\tau$ into the more intuitive parameters $t_{\rm peak}$ and fwhm.  Here, $t_{\rm peak}$ corresponds to the time at which star formation peaked as measured from the onset of star formation, and fwhm is the full width at half maximum of the SFH curve.  The parameter conversion is given by:
\begin{align}
    t_{\rm{peak}} &= \exp \left[ \ln(t_0)-\tau^2 \right],\\
    {\rm{fwhm}} &= t_{\rm{peak}}\left( \exp\left[0.5\sqrt{8\ln(2)\tau^2}\right] 
   - \exp\left[-0.5\sqrt{8\ln(2)\tau^2}\right] \right). 
\label{eq:parameter_swap_lognormal}
\end{align}

We allow the star formation history to vary from the galaxy centre to the outskirts, but in a regularized way.  Specifically, we express $\log(t_{\rm{peak}})$ and $\log(\rm{fwhm})$ as linear functions of radial position $R$:
\begin{align}
        \log(t_{\rm{peak}}) &= k_{\rm{peak}} (R/R_{\rm{star}}-1) + \log(t_{{\rm{peak}}, {{\rm R_{star}}}}), \\
        \log({\rm{fwhm}}) &= k_{\rm{fwhm}} (R/R_{\rm{star}}-1) + \log({\rm{fwhm}}_{{\rm R_{star}}}),
\label{eq:SFHgradient}
\end{align}
where $t_{\rm{peak, R_{\rm{star}}}}$ and ${\rm{fwhm}}_{{\rm R_{star}}}$ are the star formation peak time and fwhm at the stellar half-mass radius, while $k_{\rm{peak}}$ and $k_{\rm{fwhm}}$ encode the SFH gradient.  Positive values of $k_{\rm{peak}}$ refer to galaxy outskirts being younger than galaxy centres, whereas positive $k_{\rm{fwhm}}$ encode a larger spread of ages in the outskirts compared to the centre.

In practice, the (radially varying) SFH is sampled by a large number of discrete stellar particles, each of which represents a mono-age Single Stellar Population (SSP).  Once SSP ages have been assigned to the stellar particles, we account for the age-dependent stellar mass loss of the SSPs to compute the current mass locked up in stars for every particle.  Their sum constitutes the galaxy's current stellar mass, which is what the emulator will be trained on, as opposed to the integral over the SFH (the latter we refer to as $M_{\rm formed}$).  We likewise recompute the structural properties (\#2 - 4 in Table\ \ref{tab:param_library}) on the basis of this current stellar mass distribution.  In the absence of stellar population gradients the two ways of defining the stellar structural parameters (based on $M_{\rm formed}$ or $M_{\rm current}$) yield identical results.  For the range in SFH gradients defined in Table\ \ref{tab:param_library}, they typically differ by less than 0.01 dex.

Finally, to reduce the complexity of the galaxy model, we assume all stars in a given toy model to have an identical metallicity, $Z_{\rm star}$.  Throughout the paper, we use the \citet{Bruzual2003} library of SSPs.

\subsubsection{Dust properties}
\label{sec:toy_dust}

The global dust geometry follows a similar parametrization as that of the stars, but their spatial distribution is not tied.  In other words, the S\'{e}rsic index ($n_{\rm dust}$) and thickness of the dust disk ($C_{\rm dust}/R_{\rm dust}$) are set independently of the equivalent parameters for the stars.  The radial extent may likewise vary, which is captured by the size ratio $R_{\rm dust}/R_{\rm star}$.  Finally, for convenience of drawing realistic parameter values for our library of toy model galaxies, we parametrize the dust content in units normalized to the amount of stellar mass, $M_{\rm dust}/M_{\rm star}$.

Having defined the overall geometry of stellar and dust components, we now introduce clumpiness via a rudimentary two-component dust model. A fraction $f_{\rm cov}$ of young stellar particles with ages $<t_{\rm BC}$ are embedded in dust-rich birth clouds.  By default, we adopt 10 Myr as the lifetime of such birth clouds, and fix their gas-to-star ratio and dust-to-gas ratio to 10 and 0.01, respectively \citep{Trayford2020}.  To compensate for the dust allocated to birth clouds, the mass in the diffuse dust component is uniformly reduced to maintain the same total dust mass, $M_{\rm dust}$.  For particular combinations of input parameters (a low $M_{\rm dust}/M_{\rm star}$, high $f_{\rm cov}$ and SFH yielding abundant young stars) the above prescription can imply that insufficient dust is present to populate the birth clouds.  We therefore consider the parameter set as unphysical in such a situation and will not include the respective toy model in our library.  The final parameter defining the birth cloud population is the gas surface density of the clouds, $\Sigma_{\rm gas,BC}$, which we keep at a constant 35 $M_{\odot}/{\rm pc}^2$, informed by the constancy of Galactic Giant Molecular Cloud (GMC) surface densities \citep{Lada2020}.

\subsubsection{Observer's perspective}
\label{sec:observer}

For axisymmetric geometries, the viewing angle of the observer is fully captured by the inclination $\theta$.  The projected dust column and associated effective attenuation as well as reddening all increase from face-on (0$^{\circ}$) to edge-on (90$^{\circ}$) perspectives, although the imprint of orientation is reduced for clumpier dust distributions.  \citet{Zhang2023} document and interpret these trends for large galaxy samples spanning 2.5 dex in mass, and ranging from the nearby universe out to $z \sim 2.5$.

For a more generalized approach considering also galaxies of triaxial shape, the observer's perspective would need to be expanded to the full set of $(\theta, \phi)$ viewing angles on a sphere.  Indeed, deviations from axisymmetry, in the form of galaxies with prolate shapes, appear to be more common at high redshifts \citep[e.g.,][]{van2014b, Zhang2019, Zhang2023, Pandya2024}.  We note, however, that this signal of prolateness becomes most apparent at low masses, where dust levels are generally low and the kind of panchromatic resolved observations as ideally suited for {\tt SE3D} analysis remain less common.  We therefore concentrate in this initial exploration on the axisymmetric case.

\subsection{Producing spectral distributions}
\label{sec:SDs}

Having defined our toy models, we now describe the setup used to run radiative transfer on them (Section\ \ref{sec:SKIRT_RT}) and to subsequently extract their SDs (i.e., wavelength dependent properties; Section\ \ref{sec:cube2SDs}).

\subsubsection{Running SKIRT radiative transfer
}
\label{sec:SKIRT_RT}

We utilize the 3D dust radiative transfer code SKIRT by \citet{Camps2015a, Camps2020} to generate mock rest-frame UV-to-mm observations for any toy model galaxy.  Briefly, SKIRT adopts a Monte Carlo approach to trace the propagation of photon packages through a dusty medium.  It accounts for scattering, absorption and re-emission by dust, including the stochastic heating of dust grains giving rise to PAH emission \citep{Camps2015b}.  For a review on numerical implementations of these processes, we refer the reader to \citet{Steinacker2013}.

In practice, our toy model galaxies are set up with $10^5$ star particles, $10^6$ dust particles, and the Monte Carlo radiative transfer is computed using $10^7$ photon packages.  In detail, we apply a resampling step to stellar particles younger than 100 Myr, following \citet{Trayford2017} and documented further in Z25.  The effective number of star particles therefore exceeds $10^5$.  We tested that with these numbers the resulting SDs are converged.

Throughout this work, we adopt the THEMIS dust model (The Heterogeneous dust Evolution Model for Interstellar Solids, \citealt{Jones2017}), which specifies the dust composition and grain size distribution.

In constructing our SKIRT library, any run on a toy model galaxy has 15 cameras set up to record the emerging light, with inclinations drawn randomly from a flat distribution in $\cos(\theta)$.

As outlined in Section\ \ref{sec:toymodel}, the major components making up our toy model galaxies are the diffuse stellar and dust components, and compact birth clouds. Since the latter are significantly smaller than the whole galaxy, accurately calculating radiative transfer within the birth clouds is challenging to do in the same SKIRT run that captures the global galaxy profile. To combat this, we run SKIRT for individual birth clouds with a suitable zoom-in spatial resolution and construct a Milky-Way like birth cloud library, consisting of the following dimensions: 0.1 < age [Myr] < 10; 0.005 < $Z_{\rm{star}}$ [$Z_\odot$] < 2.5; 0.01 < $M_{\rm{dust}}/M_{\rm{star}}$ < 1; 3 < $\log(M_{\rm{star}})$ [$\log(M_\odot)$] < 6.

For each entry in the library, a high-resolution SKIRT run on a single birth cloud records the panchromatic SED, split into attenuated primary (stellar) emission, and secondary (dust) emission. The attenuation levels at 5500\r{A} are between 1.75 and 2.04. The SED is then divided by the birth cloud stellar mass with which it is generated, so that it can be entered as a stellar-mass-normalized SED in an input file (and SED family in SKIRT nomenclature) for later galaxy-scale radiative transfer runs. For a birth cloud particle with $({\rm{age}_i}, Z_i, M_{{\rm{dust}},i}/M_{{\rm{star}},i}, \log(M_{{\rm{star}},i}))$, the birth cloud SED can be found by interpolation from the birth cloud library. We use such birth cloud SED to account for the attenuation within birth clouds and calculate the subsequent radiative transfer with diffuse dust in the ISM. We ignore the small chance that light from birth clouds enters other birth clouds since they occupy a relatively small fraction of the volume in a galaxy. The radiative transfer on light emitted by diffuse stars that are not located in birth clouds does account for attenuation by both diffuse dust and any dust in birth clouds the photons encounter on the way out. The process is separated from that of birth cloud stars. We summarize the radiative transfer treatment of multiple components below:\\
1. Birth cloud attenuated primary SEDs + diffuse dust \\
2. Birth cloud secondary SEDs + diffuse dust \\
3. Diffuse star SEDs + birth cloud dust + diffuse dust \\

The SEDs and surface brightness maps from all three radiative transfer runs are then combined into the overall mock-observed SED and imaging. As a minor caveat, we point out that the above treatment of birth clouds, while common in the literature, is not fully self-consistent in terms of computing the temperature of the diffuse dust. However, any bias in dust temperature is expected to be minor as the escaping short-wavelength emission from birth clouds is much lower than the luminosity of the diffuse stellar component, and the diffuse dust is hence mainly heated by diffuse stars.\footnote{Alternative implementations for birth clouds, including photoionization modelling, exist in the literature \citep[e.g.,][]{Trayford2017,Kapoor2023}.  Albeit also not fully self-consistent, in that they do not account for the possibility of birth cloud dust obscuring background stars, their use may be worth exploring in future work, especially when expanding {\tt SE3D} to also incorporate line emission.}

The end product of the above procedure is a 3D data cube for every camera, sampling the spatial distribution of emerging light at 512 distinct wavelengths.

\subsubsection{From data cubes to spectral distributions}
\label{sec:cube2SDs}

We next distill the data cubes output by SKIRT into 4 spectral distributions: $F_{\lambda}(\lambda)$, $R_e(\lambda)$, $n(\lambda)$ and $q(\lambda)$. The PTS toolkit \citep{Camps2020} is used to extract $F_{\lambda}(\lambda)$. For $R_e(\lambda)$, $n(\lambda)$, and $q(\lambda)$, we process each data cube image slice from the UV to mm.

Firstly, we extract the half-light radii by growing an ellipse on the image whilst keeping the projected axis ratio $q$ constant. In the stellar (dust) regime, we use $q_{\rm{mass, star}}$ ($q_{\rm{mass, dust}}$), which is computed using the known stellar (dust) disk thickness $C_{\rm{star}}/R_{\rm star}$ ($C_{\rm dust}/R_{\rm dust}$),  and the viewing angle $\theta$. The semi-major axis of the ellipse which encapsulates 50\% of the light is defined as the half-light radius ($R_{50}$).

In a second step, we compute the S\'{e}rsic index, $n$, by integrating the surface brightness within elliptical apertures, yielding a cumulative distribution function (CDF, a.k.a. curve of growth). The best fit $n$ is computed by minimizing the $\chi^{2}$ between the fraction of light encompassed in seven percentile bins (10\%, .., 70\%), and various S\'{e}rsic profiles. This is done for every wavelength slice, effectively constructing $R_e(\lambda)$, $n(\lambda)$. 

For $q(\lambda)$, we use the following for each wavelength step:
\begin{equation}
    \begin{aligned}
    q_{\rm{light}} = q_{\rm{mass, star}}f_{\rm{star}} + q_{\rm{mass, dust}}(1-f_{\rm{star}}),
    \end{aligned}
    \label{axialratio_eq2}
\end{equation}

where $f_{\rm{star}}$ is the fraction of light directly contributed by stars (as opposed to dust re-emission) at a specific wavelength.

\subsection{Emulating spectral distributions}
\label{sec:emulator}

In recent years, there has been growing activity in machine learning (ML) applications to support extragalactic astrophysics research.  Of immediate relevance to this work are efforts to utilize ML techniques to predict/fit/interpret spectral energy distributions. Examples of such work include \citet{Alsing2020}, \citet{Dobbels2021}, \citet{Mathews2023}, \citet{Sethuram2023}, and \citet{Lovell2025}. For a more comprehensive overview, we refer the reader to \citet{Iyer2025}.

In our work, we employ ML for the primary purpose of computational speed up, relative to full-fledged physical calculations of RT using SKIRT (see Section\ \ref{sec:performance}).  A particular novelty we introduce is to emulate observations of global galaxy structure as well as SEDs.

In this Section, we first describe the SKIRT library generated for training, test and validation purposes (Section\ \ref{sec:library}).  We then cover the data pre-processing (Section\ \ref{sec:preprocessing}) before specifying the Bayesian neural network architecture and hyperparameter tuning (Section\ \ref{sec:BNN}).

\subsubsection{SKIRT library}
\label{sec:library}

To train our emulator, we first create a library of toy model galaxies treated with dust radiative transfer. We randomly sample the parameters describing the toy models from the distributions specified in Table \ref{tab:param_library}.  This yields a suite of 80,000 toy models, each with 15 random viewing angles ($\theta$). As alluded to in Section\ \ref{sec:toymodel}, our method of assigning dust to birth clouds implies that some sets of parameter values lead to situations where the total amount of dust in birth clouds exceeds the galaxy dust mass. In such cases, we flag the galaxy as unphysical and do not run SKIRT on them. We refer the reader to Appendix \ref{app:physical_unphysical} for more detail.

The remaining 61,261 physical toy models are split randomly into training (84\%), validation (8\%) and test (8\%) subsets, so as to ensure each subset contains representative samples.  With 15 cameras per toy model, this yields 768,915 sets of SDs to train on. We note that the toy models are split based on the SKIRT runs rather than at the level of individual viewing angles. Therefore, there are no cases where the same toy model galaxy exists in multiple subsets with different viewing angles. We further note that SKIRT has the functionality to return the SEDs of stellar and dust emission separately, which we both store.

\subsubsection{Data pre-processing}
\label{sec:preprocessing}

We consider 16 input parameters to our emulator.\footnote{Birth cloud parameters \#16-19 are kept constant, as indicated in Table\ \ref{tab:param_library}.}  For each of them, we compute the mean and standard deviation from the training set, and use this to normalize the inputs in training, test and validation sets.

While we collectively use the term {\it emulator} to refer to the algorithm mapping these inputs to the desired SDs, in practice the prediction of each of the following spectral distributions is done separately by its own Bayesian neural network: StellarSED, DustSED, SRD, SND, SQD.  This allows using architectures of lower complexity and separate, dedicated training and hyperparameter tuning for each of the SDs (Section\ \ref{sec:BNN}).  Moreover, instead of aiming to predict each of the 512 wavelength elements of the SDs individually, we apply Principal Component Analysis (PCA) as a means to reduce the dimensionality of the outputs \citep[see also, e.g.,][]{Alsing2020}.  To this end, we first normalize the SDs at a reference wavelength, where we adopt $\lambda_{\rm ref} = 1\ \mu$m, with the exception of the DustSED, for which we use $\lambda_{\rm ref} = 15\ \mu$m.  We then apply PCA to infer $n$ basis components (i.e., templates) which in superposition can reproduce the range of SD shapes, capturing 99\% of the variance for the StellarSED, DustSED, SRD, SND, and SQD. This results in 4, 5, 7, 12, and 2 PCA components for the StellarSED, DustSED, SRD, SND and SQD, respectively. Following this, we concatenate the zero point values which were previously used to normalize the SDs to the array of PCA basis coefficients.  We then normalize these outputs using the mean and standard deviation across the training sample. We train our emulator to predict the normalized PCA basis coefficients and the zero point for the set of 16 parameter inputs.  After denormalization, the respective spectral distribution can then be recovered from these outputs via
\begin{equation}
SD_{\lambda} = \sum _{i=1}^{n_{\rm PCA}} w_i\ {\rm PC}_i + ZP
\end{equation}

\subsubsection{BNN architecture, training and hyperparameter tuning}
\label{sec:BNN}

We make use of Bayesian Neural Networks (BNNs) in this work as they offer a systematic approach to quantify uncertainty in predictions by treating model weights as probability distributions rather than deterministic values as standard neural networks do. 
This allows BNNs to capture both aleatoric and epistemic uncertainty, leading to more robust predictions. We make use of \texttt{torchbnn}, a package designed by \citet{Lee2022}. 

A generous range of hyperparameters in our BNNs is tuned using \texttt{Optuna} \citep{Akiba2019}, allowing for flexible architecture designs. We determine the best hyperparameters over the course of 2500 trials, where each trial consists of training the model for 4000 epochs by minimizing the following loss function:
\begin{equation}
\begin{split}
{\rm Loss} = \frac{1}{n_{\rm obj}} \sum _{i=1}^{n_{\rm obj}} \left[ \frac{1}{n_{\rm PCA}} \sum _{j=1}^{n_{\rm PCA}} {\rm PC}_{j,{\rm var}}\ \left( w_{ij,{\rm pred}}\ - w_{ij,{\rm truth}} \right)^2 \right. \\
\left. +\ {\rm ZP}_{\rm weight}\ \left( ZP_{i,{\rm pred}} - ZP_{i,{\rm truth}} \right)^2 \vphantom{\sum _{j=1}^{n_{\rm PCA}}} \right] \\
+\ {\rm KL}_{\rm weight}\ KL_{\rm model} \vphantom{\sum ^{n_{\rm PCA}}}
\end{split}
\label{eq:lossfunc}
\end{equation}
where $n_{\rm obj}$ stands for the number of objects in the training batch, and ${\rm ZP_{weight}}$ and ${\rm KL_{weight}}$ are tuned hyperparameters describing the weighted importance of the difference between predicted and true zero points and of the deviation between the model posterior at that epoch and the prior (\textit{KL}$_{\rm model}$), respectively. PC$_{\rm var}$ describes what fraction of the variance in the data is captured by each principal component.  This information was directly computed and saved during the preprocessing step described in Section\ \ref{sec:preprocessing}.

Following the 4000 epochs of training for a given set of hyperparameters, we compute the normalized median absolute deviation (NMAD) of SD residuals as:
\begin{equation}
{\rm NMAD} = 1.4826 \times {\rm median}(|x_{\rm predict} - x_{\rm SKIRT}|),
\label{eq:NMAD}
\end{equation}
where $x$ represents any of the wavelength-dependent properties.  The median is taken over all (absolute) residuals across all training objects and wavelengths, and the factor 1.4826 ensures the metric matches the standard deviation in case of a Gaussian distribution of residuals.  The hyperparameters from the trial resulting in the lowest NMAD are adopted to select the most optimal model, and both the trained model and corresponding hyperparameters are saved for future use.\footnote{We do not minimize the NMAD directly during training, as it would require preloading the testing set and performing a costly computation at every epoch. This approach would significantly prolong the training process, particularly during hyperparameter tuning.}

For further detail regarding the extent of parameters we tune, we refer the reader to Appendix \ref{app:hyper}. 

\subsection{SE3D fitting}
\label{sec:fitting}

With an emulator which can efficiently predict rest-frame SDs in hand, we are now ready to fit observational data.  To this end, we employ the Bayesian affine invariant MCMC parameter space exploration as implemented in {\tt emcee} \citep{Foreman-Mackey2013}.  By default, we define the log-likelihood as the straight sum of log-likelihoods for each of the considered SDs (with the residuals computed in dex), although optionally weights can be applied to give some of the SDs more emphasis.  Priors on parameters can also be set, and do not need to match the distributions specified in Table\ \ref{tab:param_library}.

In every call to the model function, a few post-processing steps are applied  to bring the model to the observational domain.  First, the SDs are redshifted.\footnote{By default, we will assume the redshift to be accurately known from spectroscopy, and the Age (since onset of star formation) to match the age of the Universe at that redshift.  In the absence of a spectroscopic redshift, Age (parameter \#5 from Table\ \ref{tab:param_library}) could be left free, or foreseen of a prior informed by a photometric redshift code.}  Secondly, attenuation by the intergalactic medium (IGM) is applied following the prescription by \citet{Meiksin2006}.  Finally, the SDs are convolved with the relevant filter throughput curves to obtain model quantities that are directly comparable to the available measurements.  When applied to a catalogue of objects, parallelization is implemented in an ad hoc fashion, with individual objects being sent to their own thread for fitting.

\section{Results}
\label{sec:results}

\subsection{Performance of the emulator}
\label{sec:performance}

\subsubsection{Computational speed}
\label{sec:speed}

Making emulator predictions of all SDs for a single set of input parameters takes on average 0.018s on a single CPU.  In practice, it is more efficient to call the emulator for all MCMC walkers at once.  For example, when employing 500 walkers, making emulator predictions for each of them takes $\sim 0.03$s (an equivalent of 60$\mu$s per input set), with the post-processing to obtain model predictions in the relevant observed-frame wavebands adding an extra $\sim 0.05$s (19 filters considered for this case example).  For reference, the average time to run SKIRT on a toy model in our library was 4.5 minutes on 60 CPUs, with the preceding step of generating particle tables and the postprocessing into SDs each adding less than a minute.

\subsubsection{Case example SDs}
\label{sec:oneparam}

\begin{figure*}
    \centering
    \includegraphics[width=\linewidth]{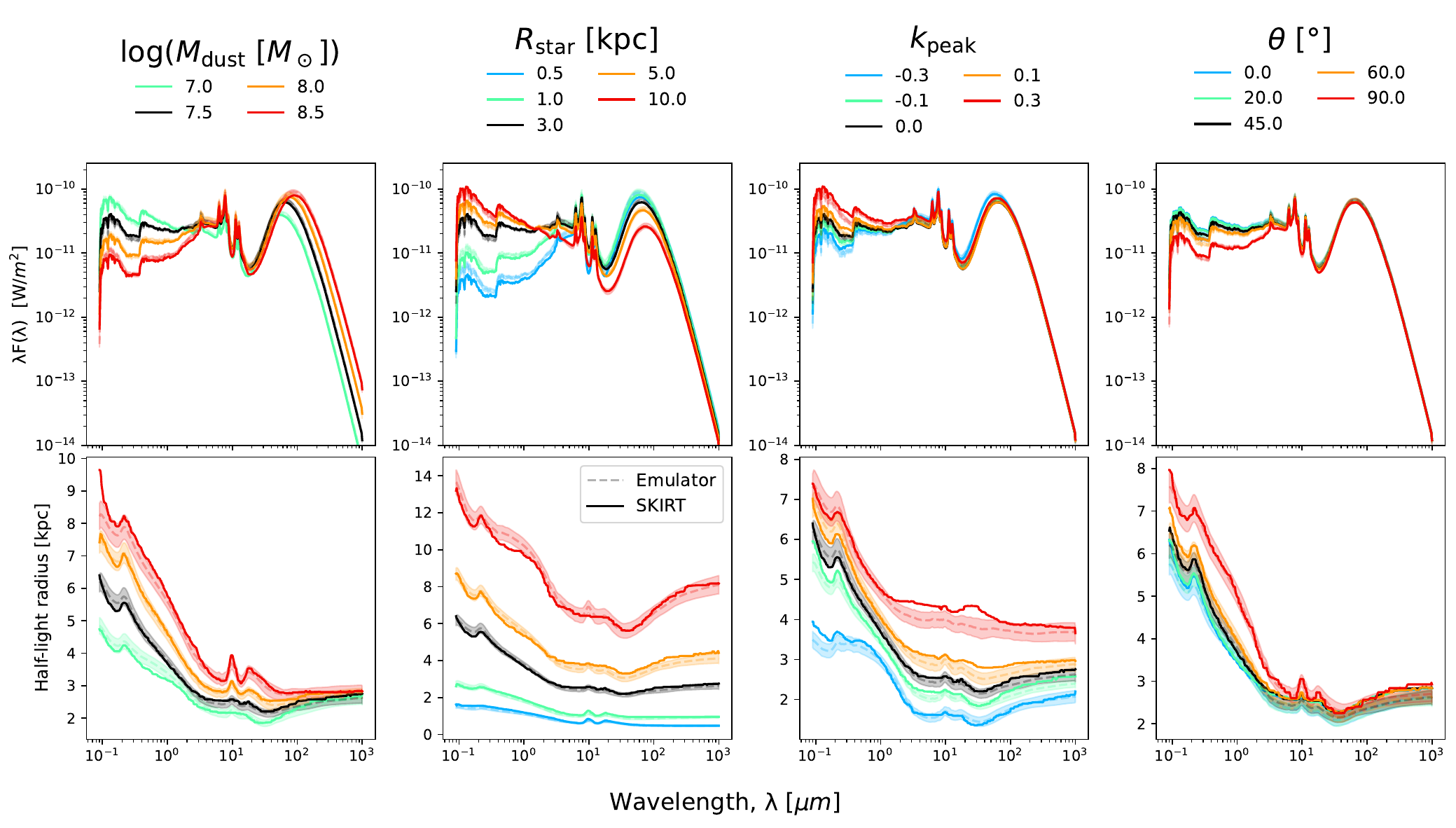}
    \caption{Effect of varying a small subset of model parameters on the predicted SEDs and SRDs. Each panel shows the result of modifying one parameter while keeping others fixed to the reference model (shown in black and described in Appendix\ \ref{app:oneparam}). From left to right, we vary the dust content, the scale of the galaxy, the radial gradient of the time of peak star formation, and the inclination. Note: because of how we defined parameters \#3, \#12, and \#13 in Table\ \ref{tab:param_library}, varying $R_{\rm star}$ effectively scales the entire toy model galaxy up/down, preserving geometric thickness and the relative distribution of dust and stars. Emulator predictions and uncertainties are plotted in dashed lines and shaded regions, whereas ground truth SKIRT outputs are plotted in solid lines.}
    \label{fig:oneparam}
\end{figure*}
By means of illustration, we show in Figure\ \ref{fig:oneparam} a set of case example SDs computed with SKIRT (solid curves) and predicted by our emulator (dashed curves).  The black curves in each panel correspond to the same reference toy model, whereas coloured curves display the impact of varying just one defining feature of the toy model at a time.  From left to right, these are, respectively, the total dust mass, the scale of the galaxy (for this, we scale all of $R_{\rm star}$, $R_{\rm dust}$, $C_{\rm star}$ and $C_{\rm dust}$ in lockstep), $k_{\rm peak}$ (controlling the age gradient), and the inclination.

A first takeaway is that the emulator predictions (dashed curves) match adequately the trends exhibited by the actual RT calculations (solid curves).  This includes the enhanced attenuation, boosted dust re-emission and shift to colder dust temperatures with increasing $M_{\rm dust}$.  In varying the distribution of stars and dust from extended to more compact, an increased attenuation can likewise be noted, but this time associated with a (modest) {\it increase} in dust temperature.  These patterns are the natural consequence of changes to the dust columns and radiation fields the dust grains are exposed to.  

In terms of structural observables, the general behaviour is for half-light radii to increase from rest-frame NIR to rest-frame UV wavelengths.  This is the case despite the identical spatial distribution of stellar and dust components adopted for the toy models shown in Figure\ \ref{fig:oneparam} (see Appendix\ \ref{app:oneparam} for examples where stellar or dust extent are varied separately, as well as a broader range of parameter variations).  As covered in more detail by \citet{Zhang2023}, the effect can be attributed to enhanced dust columns towards the galaxy centre.  The resulting elevated central attenuation then implies that the radii encompassing half the light grow larger than the corresponding half-mass radius.  We note that in detail, varying $M_{\rm dust}$ and varying the galaxy scale leave a different imprint on not only SEDs but also SRDs, despite both impacting the effective dust columns.

Finally, increasing $k_{\rm peak}$ from negative (stellar populations that form outside-in) to positive (inside-out) values has a modest impact on the galaxy-integrated SED, but leaves an appreciable imprint on the SRD shapes.

The impact of the observer's perspective is illustrated in the rightmost panels of Figure\ \ref{fig:oneparam}.  We note that how strongly and in which manner orientation effects are expressed can vary with the intrinsic make-up of the toy model.  For instance, if we had adopted a relatively low $M_{\rm dust}$ in combination with a SFH that attributes the bulk of available dust to clumps, little dependence on inclination would remain.  As another example, a geometry with diffuse dust that is radially extended with respect to the stars would yield fairly little obscuration under a face-on viewing angle (on account of reduced dust columns), but more efficient attenuation and reddening in the edge-on case, as more of the dust would act as a foreground screen.

\subsubsection{Overall performance}
\label{sec:overallperformance}

\begin{figure}
    \centering
    \includegraphics[width=\linewidth]{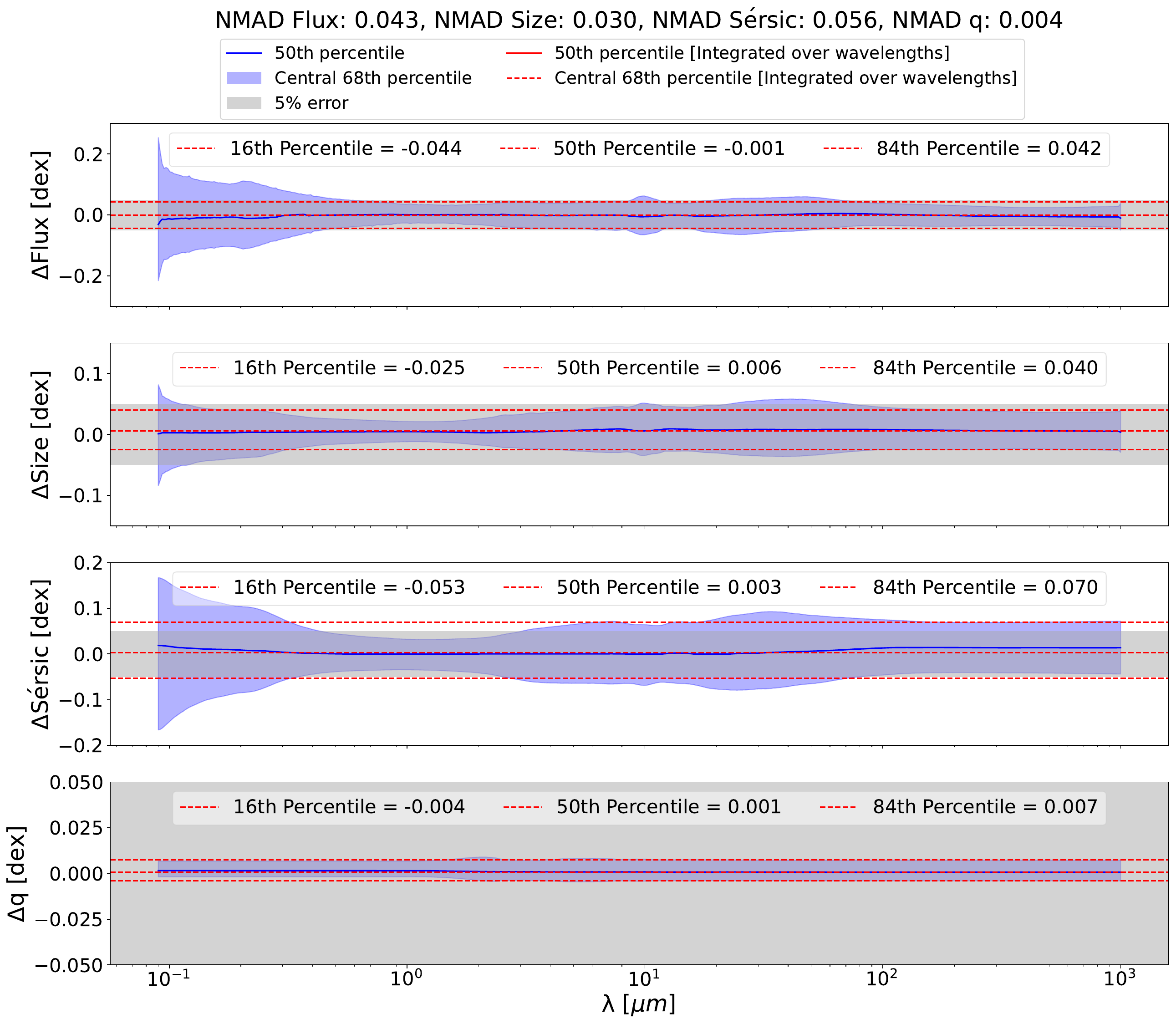}
    \caption{Validation of the emulator. Residuals, in dex, between 75,000 (unseen) SKIRT ground truth spectral distributions and the corresponding emulator predictions are displayed. The blue shaded regions showcase the central \nth{68} percentile with the running median shown as the blue solid line. Integrating over wavelengths, we further mark the \nth{16} (red dashed), \nth{50} (red solid) and \nth{84} (red dashed) percentiles for each spectral distribution. For reference, 5\% errors are marked in grey. The NMADs for Flux, Size, S\'{e}rsic index and projected axis ratio $q$ are 0.043, 0.030, 0.056, and 0.004 dex, respectively.}
    \label{fig:ML_Validation_Blue_Polygon}
\end{figure}

While the examples shown in Figure\ \ref{fig:oneparam} and discussed in the previous section convey emulator accuracy in a qualitative sense, they do so anecdotally and do not capture the full range of models the ML algorithm was trained to emulate.  Here, we validate our ML emulator by testing its ability to predict the SDs of 75,000 unseen toy models making up the testing set (see Section\ \ref{sec:library}).  The SDs computed with SKIRT serve as truth in this comparison, with Figure\ \ref{fig:ML_Validation_Blue_Polygon} displaying percentile ranges of the (emulator minus SKIRT) residuals as a function of wavelength.  We find the uncertainty in SED and SND predictions to be most wavelength dependent, with performance decreasing in the UV/optical regime.  This primarily arises due to the larger variations the emulator has to account for in this regime, due to a combination of dust and stellar population related factors.

The summary statistics listed at the top of Figure\ \ref{fig:ML_Validation_Blue_Polygon} capture the normalized median absolute deviation (NMAD) of residuals integrated across all wavelengths.  Computed as per Eq.\ \ref{eq:NMAD}, the metric captures the combination of random and systematic offsets between emulator prediction and truth.  However, Figure\ \ref{fig:ML_Validation_Blue_Polygon} illustrates that the NMAD are largely dominated by scatter as any systematic offsets are much smaller.  Aggregate NMAD values increase from 0.004 dex for the projected axis ratio to 0.030 dex and 0.043 dex for size and flux, to a slightly larger 0.056 dex for the S\'{e}rsic index.  Figure\ \ref{app:oneparam_App} in Appendix\ \ref{app:oneparam} illustrates that the relative performance is directly tied to the complexity of the SDs to be reproduced.  In the case of the SND, this is particularly true as there is no guarantee every image slice of a SKIRT output data cube is well described by the S\'{e}rsic functional form.  Some of the high frequency wiggles in SNDs may therefore be attributed to unstable S\'{e}rsic fitting of the curves of growth in the presence of such template mismatch.

\begin{figure}
    \centering
    \includegraphics[width=\linewidth]{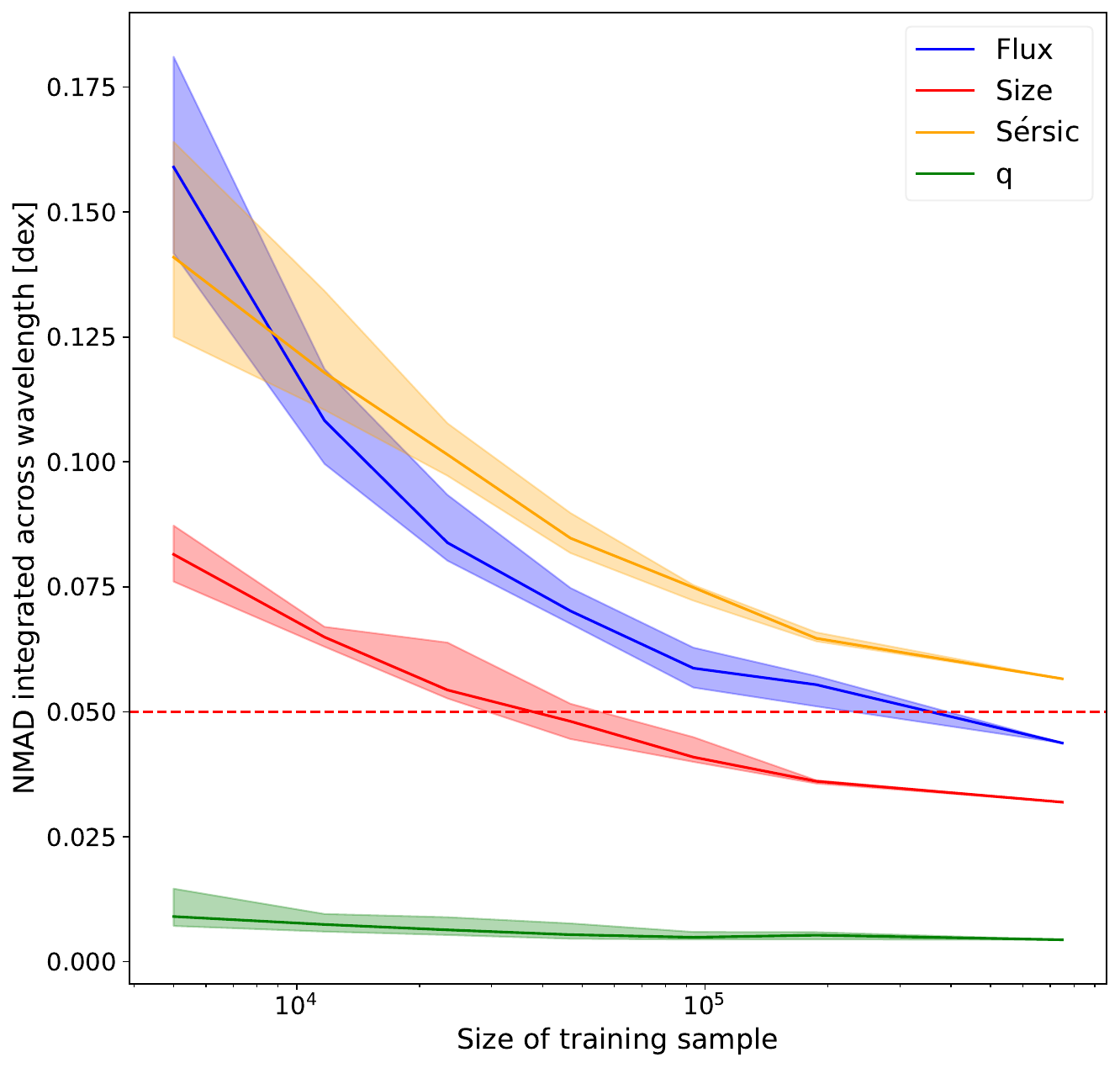}
    \caption{Emulator performance (NMAD) as evaluated on our testing set of 75,000 toy models, for varying training library sizes. The coloured polygons display the range of computed NMADs for a given library size, where for sizes smaller than the full library size, multiple iterations are computed by sampling disjoint subsets of the corresponding size from the remaining data.  The red dashed line indicates a 0.05 dex error for reference. Emulator accuracy improves with  increasing training sample size.}
    \label{fig:NMAD_Vs_LogNtrain}
\end{figure}

\begin{figure}
    \centering
    \includegraphics[width=\linewidth]{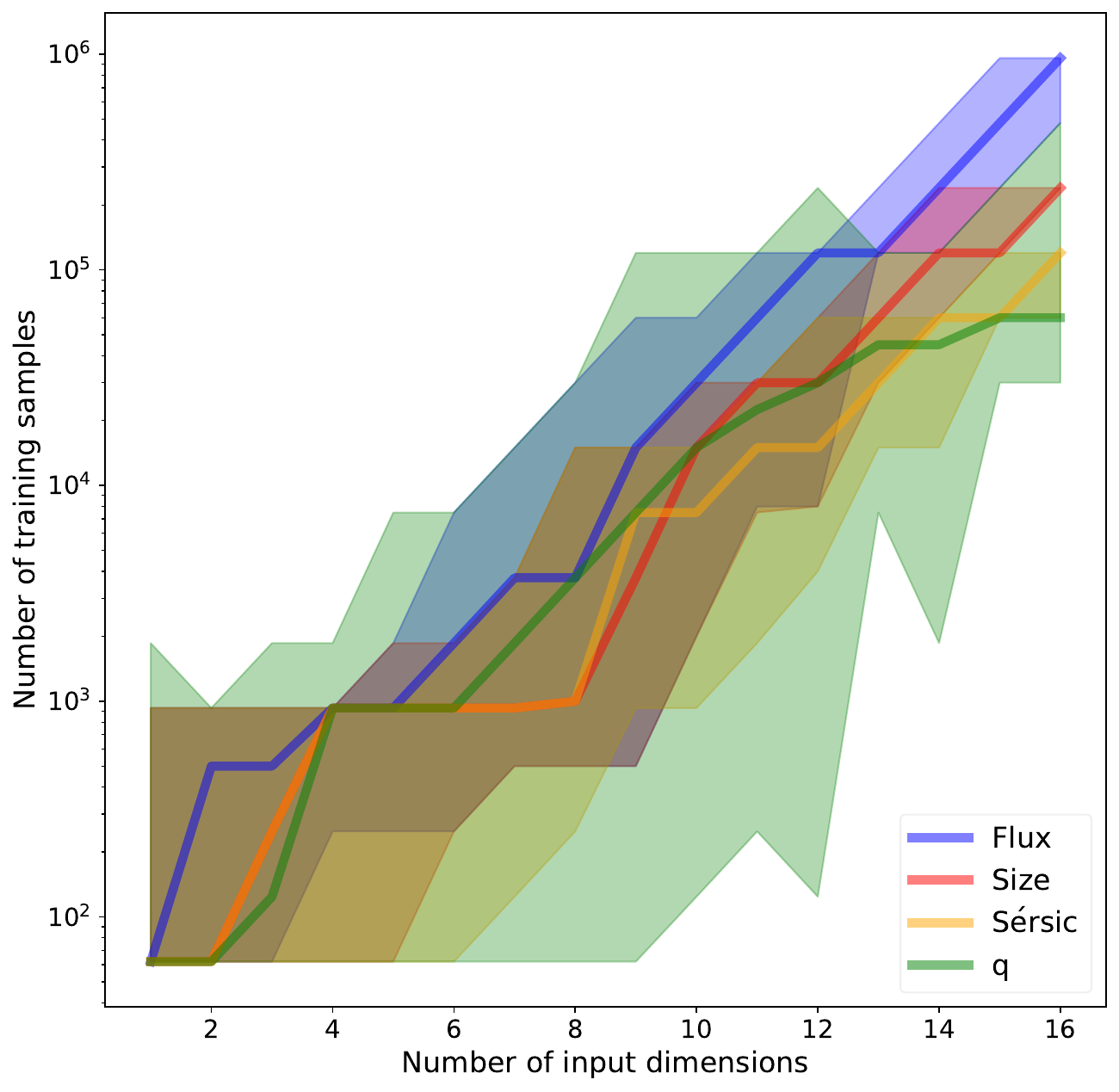}
    \caption{Number of training samples required to reach the NMAD performance of 0.043, 0.030, 0.056, and 0.004 dex for Flux, Size, S\'{e}rsic, and axial ratio ($q$), as a function of the number of input dimensions defining the toy model. The solid lines correspond to the running mean of the coloured polygons, which display the range of training samples required to reach the target NMAD, depending on the order in which the input dimensions are built up. Required training rows to reach our defined threshold accuracy, on average, double with each added input dimension.}
    \label{fig:toymodeldim}
\end{figure}

A common thread in supervised ML applications is the need for large and representative training samples.  We return to their representative nature in Section\ \ref{sec:UVJ}, but first consider the dependence on sample size.  Figure\ \ref{fig:NMAD_Vs_LogNtrain} shows how the wavelength-integrated NMAD values vary with size of the training sample.  The rightmost end points of each polygon correspond to the same NMAD values reported in Figure\ \ref{fig:ML_Validation_Blue_Polygon}, achieved by exploiting the full training library.  It is apparent that emulator performance deteriorates as smaller subsets of the library are considered for training, with the width of the polygons capturing the spread in performance metrics obtained for different independent subsets of the respective size.

Turned around, one can ask what training sample size would be required to achieve a similar emulator accuracy had we formulated a simpler toy model galaxy definition, of lower dimensionality (i.e., fewer than 16 parameters). The answer to this question is depicted in Figure\ \ref{fig:toymodeldim}, which shows the required training sample size as a function of toy model dimensionality.  As in detail the scaling with number of parameters describing the toy model depends on the order in which this complexity is built up (e.g., first introducing structural and dust parameters, and then SFH related parameters, or vice versa), we repeated the exercise multiple times.  The spread in results is captured by the width of the polygons in Figure\ \ref{fig:toymodeldim}. We conclude that the number of training samples scales exponentially with the toy model dimensionality, where an increase with one dimension requires, on average, doubling the training sample size.

\subsubsection{Dependence of performance on input parameters}

\begin{figure}
    \centering
    \includegraphics[width=\linewidth]{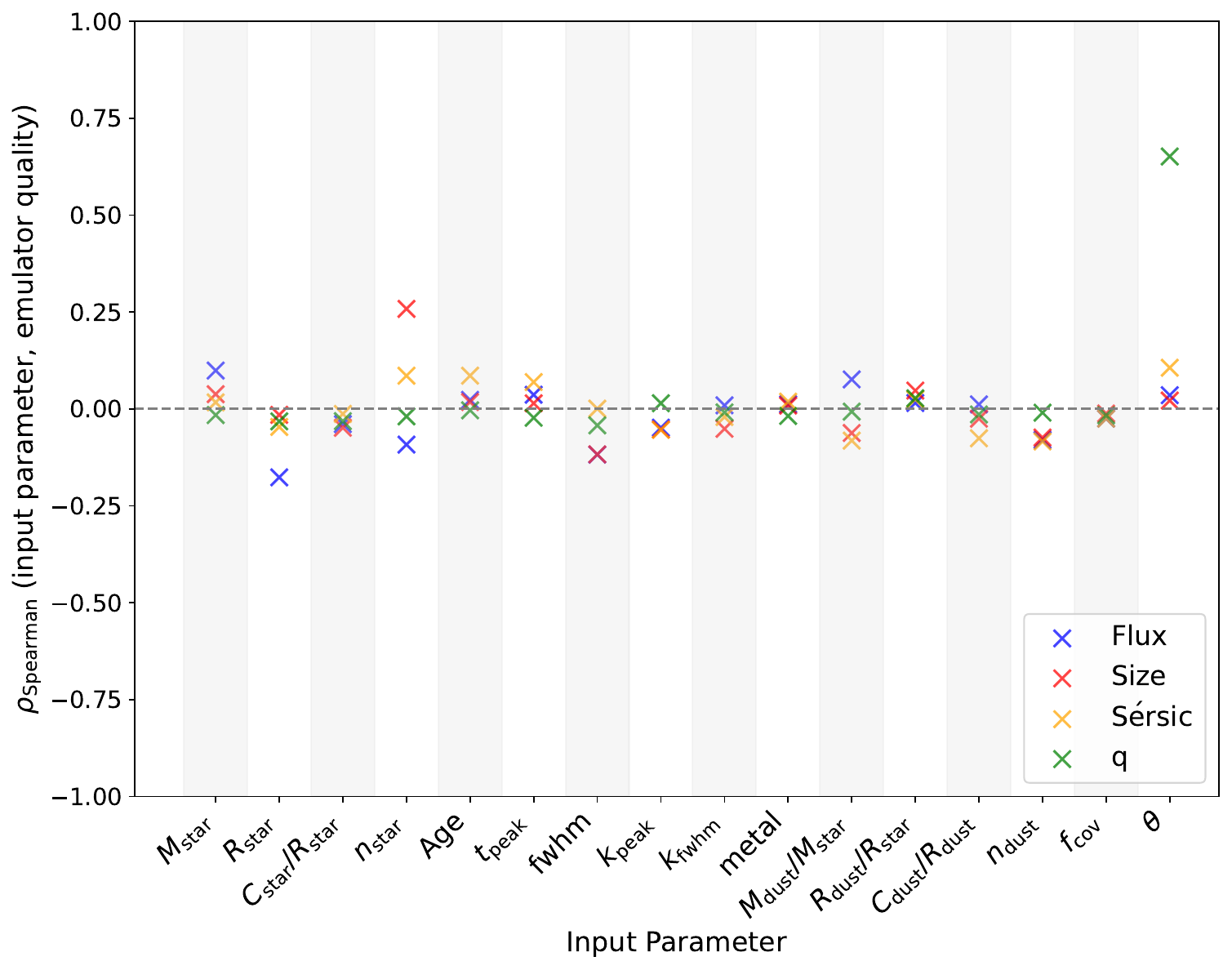}
    \caption{Spearman rank correlation coefficients between the input parameters in our library and the SD residuals (integrated across wavelength) for the respective toy model galaxies. The performance of the emulator remains approximately constant across the dynamic range sampled for each parameter.  Residuals in predicted $q$ formally increase with $\theta$, but in an absolute sense are negligible ($< 0.01$ dex). }
    \label{fig:SpearRankCorr_Inputs}
\end{figure}

Section\ \ref{sec:overallperformance} quantified the overall performance of the emulator.  We expand on this in Figure\ \ref{fig:SpearRankCorr_Inputs} by considering whether there are regimes of parameter space where the emulator performs more poorly.  For simplicity, we restrict ourselves to a consideration of each of the input parameters separately.  Specifically, we compute for each toy model galaxy the NMAD of SD residuals (i.e., emulator prediction minus SKIRT truth for all wavelength elements).  For each input parameter describing a physical feature of the toy models in our library, we then compute the Spearman rank correlation coefficient between the value of the respective input parameter and the NMAD for the corresponding toy model galaxies.  If, e.g., the emulator's SED predictions would strongly deteriorate with decreasing $R_{\rm star}$ this would then show as a prominently negative $\rho_{\rm Spearman}$ for Flux in the $R_{\rm star}$ column of Figure\ \ref{fig:SpearRankCorr_Inputs}.  No such high amplitude $\rho_{\rm Spearman}$ values are seen, implying that emulator performance is relatively stable across the dynamic range of input parameters considered.  The only exception is the projected axis ratio, for which the emulator accuracy reduces towards highly inclined orientations.  We do emphasize, however, that -while formally featuring a significant correlation- the residuals on $q$ are in an absolute sense small, remaining $<$ 0.01 dex even for edge-on cases.

We conclude that the emulator performs consistently well across the range of input parameters considered.  In the next section, we evaluate how this translates to emulator accuracy across the observational domain of galaxy colours.

\subsubsection{Performance across colour space}
\label{sec:UVJ}

\begin{figure}
    \centering
    \includegraphics[width=\linewidth]{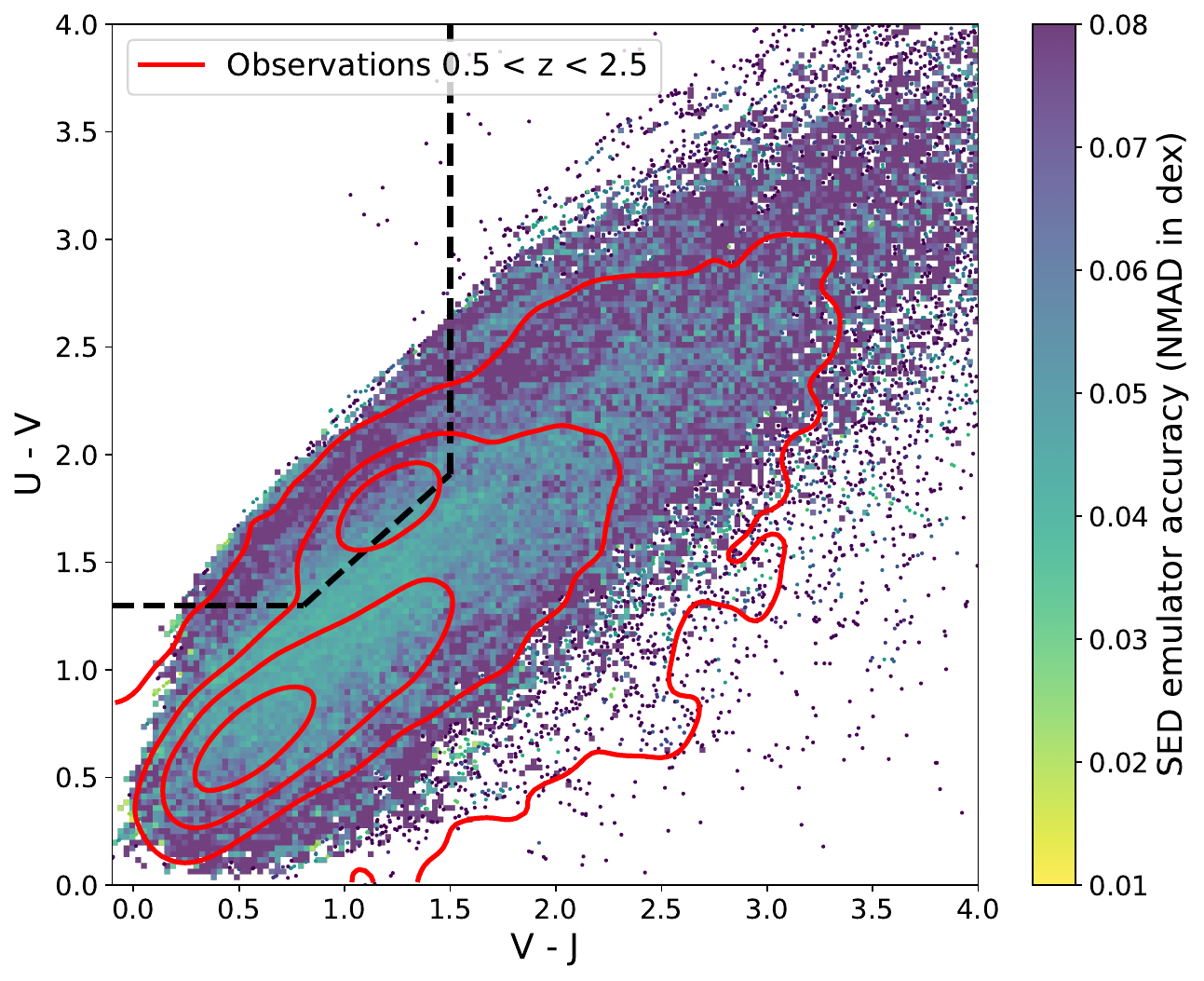} 
    \caption{$UVJ$ diagram illustrating the range in rest-frame colours spanned by our SKIRT training library, colour-coded by the median SED emulator performance in each bin. Red contours mark the observed DJA galaxy sample with $\log(M_{\rm star}) > 9.5$ and $0.5 < z < 2.5$.  The black dashed wedge denotes the quiescent/star-forming galaxy separation from \citet{Muzzin2013}. Our toy models span a larger extent of $UVJ$ space compared to observations and can capture the very dusty and attenuated regime of observed galaxies.}
    \label{fig:SKIRTLib_UV_Obs}
\end{figure}

Figure\ \ref{fig:SKIRTLib_UV_Obs} shows the rest-frame $UVJ$ colours of galaxies in our SKIRT training library, with the colour-coding of each cell denoting the median NMAD (computed by considering residuals between emulator predictions and SKIRT truth as per Eq.\ \ref{eq:NMAD}) of all toy models in the respective $UVJ$ bin.  For reference, we show the $UVJ$ colour distribution of a large sample of observed galaxies using red contours.  The sample of observed galaxies comprises of both quiescent and star-forming objects, of stellar mass $\log(M_{\rm star}~[M_\odot]) > 9.5$ and spanning a redshift range $0.5<z<2.5$.  It was extracted from the morpho-photometric catalogue of galaxies in the GOODS-N/S, CEERS, PRIMER-UDS and PRIMER-COSMOS fields released by the DAWN JWST Archive (DJA; \citealt{Genin2025}), with redshifts and rest-frame properties of the 17,207 objects computed using EAZY \citep{Brammer2008}.\footnote{\url{https://dawn-cph.github.io/dja/blog/2024/08/16/morphological-data/}}

The purpose of Figure\ \ref{fig:SKIRTLib_UV_Obs} is twofold.  First, it shows that the training library spans a broad swath of colour space, akin to and in some regions even exceeding that of the observed galaxy population.  This is reassuring as it implies the emulator should be able to account for the diversity of galaxy SED shapes without extrapolation.  Secondly, Figure\ \ref{fig:SKIRTLib_UV_Obs} does not exhibit strong variations in emulator performance across $UVJ$ space.  In particular, those regions where the emulator accuracy is slightly reduced (e.g., at the reddest $U-V$ colours in the quiescent wedge) are anyway scarcely populated by the observed population of galaxies.

In passing, we note that no attempt was made for the training library to match the distribution of real galaxies in detail (in colour space nor in intrinsic physical properties).  Real galaxies exhibit strong correlations between their stellar population properties on the one hand, and dust and structural properties on the other hand (e.g., star-forming galaxies being dustier and of lower S\'{e}rsic index than quiescent ones).  No such covariances were introduced in building the training library according to the distribution functions specified in Table\ \ref{tab:param_library}.  This was a conscious decision in order not to bias {\tt SE3D} fitting results on individual objects to baked-in, average population trends.

\subsection{Connection between physical parameters and observables}
\label{sec:connection}

\begin{figure}
    \centering
    \includegraphics[width=\linewidth]{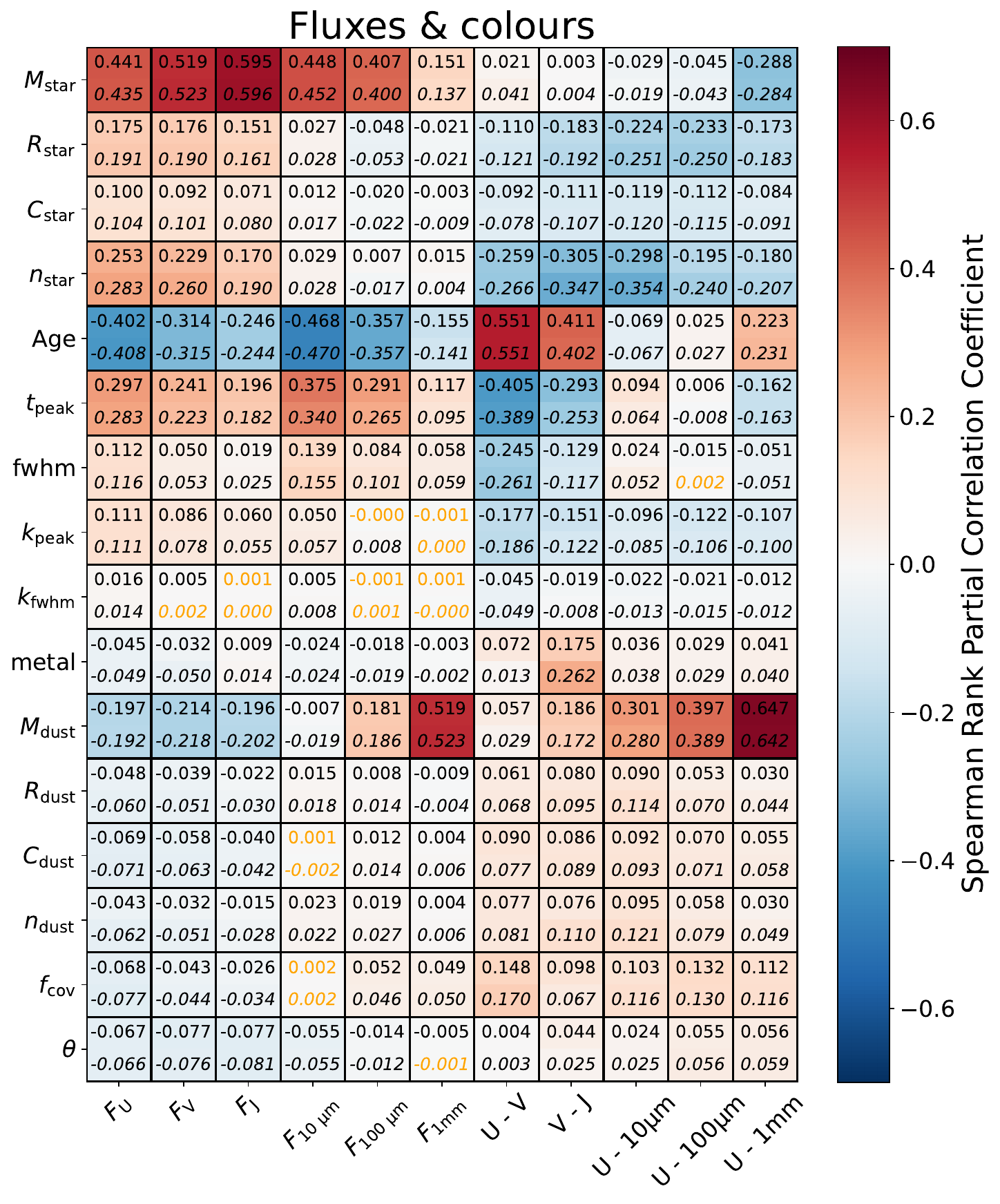}
    \caption{Spearman rank partial correlation coefficients between toy model galaxy parameters and rest-frame fluxes/colours. The top and bottom values in each cell correspond to the results obtained from the ML emulator and SKIRT ground truth, respectively. The consistency in colour within each cell confirms that the emulator has successfully learned the mapping between physical inputs to observables. Orange numbers mark cases where no significant correlation was found ($p > 0.05$).  As anticipated, galaxy colours are significantly influenced by SFH, and star-dust geometry.}
    \label{fig:PartialSpearRankCorr_Flux}
\end{figure}

\begin{figure}
    \centering
    \includegraphics[width=\linewidth]{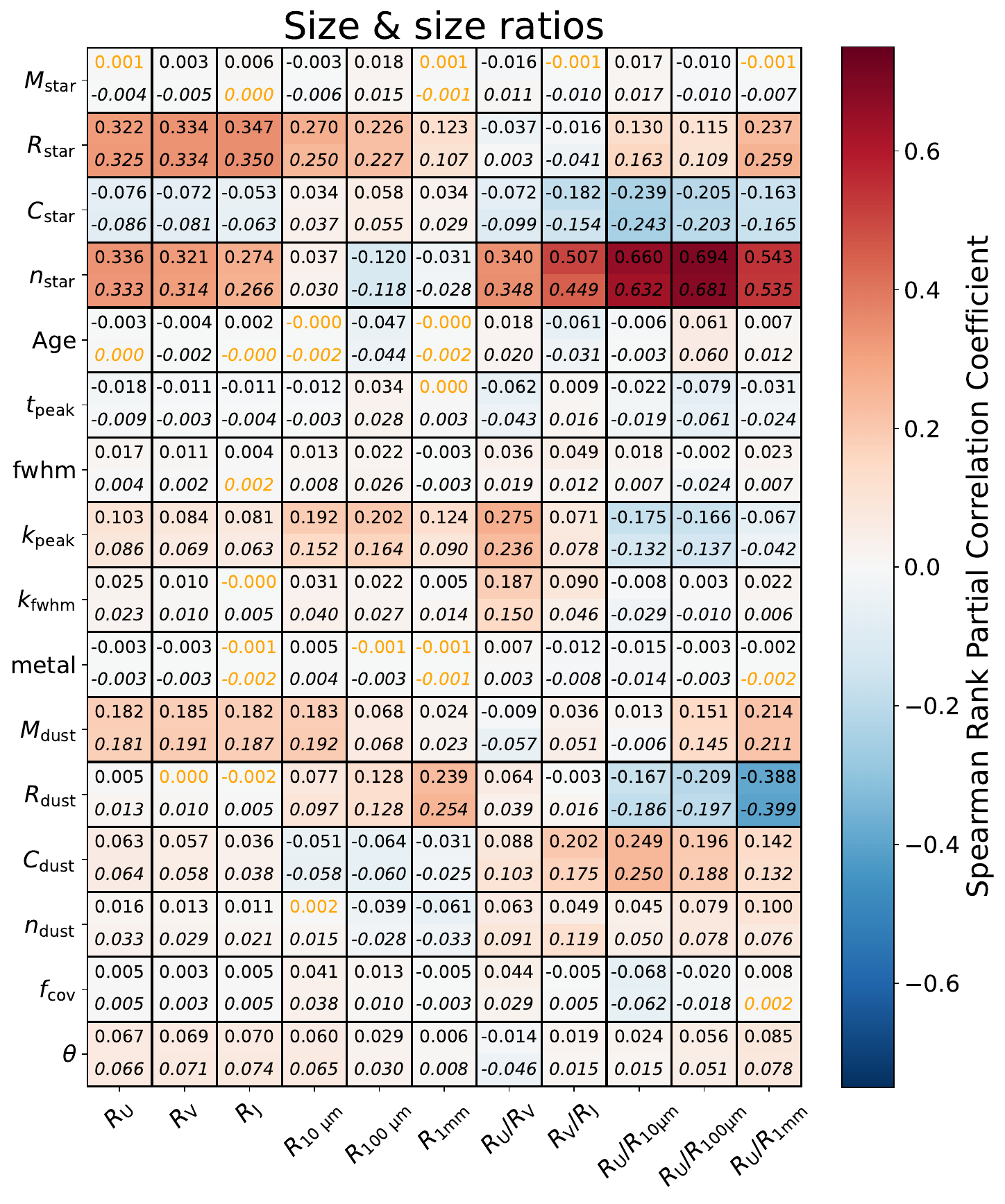}
    \caption{Same as Figure \ref{fig:PartialSpearRankCorr_Flux}, but considering sizes and size ratios.  These observables are most strongly influenced by star-dust geometry and the S\'{e}rsic index of the stellar distribution.}
    \label{fig:PartialSpearRankCorr_Size}
\end{figure}

\begin{figure}
    \centering
    \includegraphics[width=\linewidth]{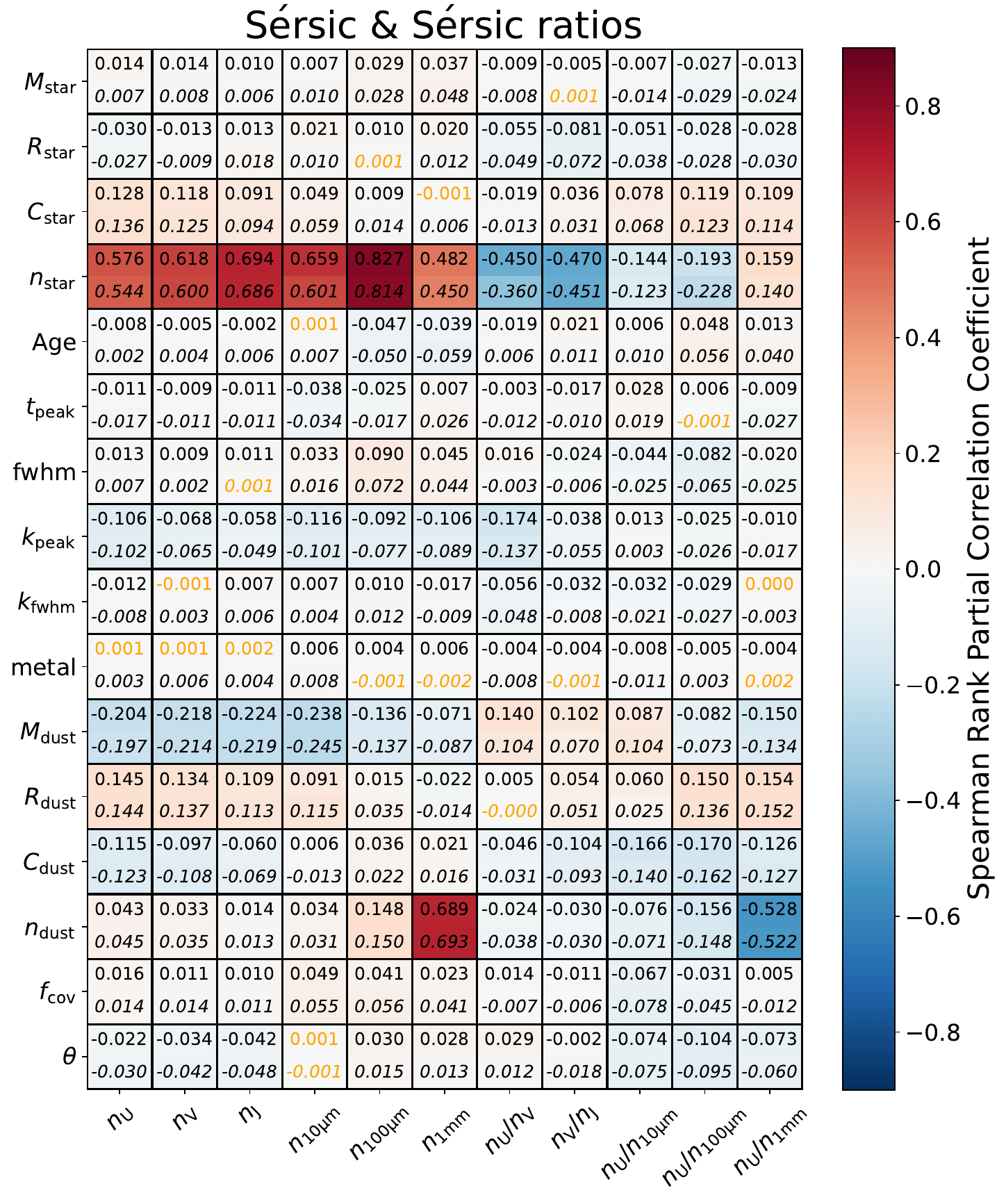}
    \caption{Same as Figure \ref{fig:PartialSpearRankCorr_Flux}, but for S\'{e}rsic and S\'{e}rsic ratios.  These observables are most strongly influenced by the S\'{e}rsic index for the projected stellar and dust distribution, and the star-dust geometry.}
    \label{fig:PartialSpearRankCorr_Sersic}
\end{figure}

The observables that can be extracted from panchromatic resolved observations of galaxies broadly divide into absolute quantities (fluxes, sizes, S\'{e}rsic indices) and specific quantities (colours, and ratios of size or S\'{e}rsic index as measured at two distinct wavelengths).  Each of these encode some information about the physical nature of the galaxy observed, although often a single observable will be impacted to a greater or lesser degree by multiple physical characteristics (consider, e.g., the well-known age-dust-metallicity degeneracy; \citealt{Bell2001, Walcher2011, DiazGarcia2015}).

In Figures\ \ref{fig:PartialSpearRankCorr_Flux} - \ref{fig:PartialSpearRankCorr_Sersic}, we visually represent a comprehensive overview of the connections between physical parameters (i.e., the inputs defining our toy models) and key observables that can be extracted from the observed SEDs, SRDs and SNDs, respectively.  Each square cell of these matrix diagrams zooms in on a specific input -- observable pair, with the colour coding depicting the Spearman rank partial correlation coefficient between them.  This statistic captures the degree of correlation once controlling for any correlations between the observable in question and any other input parameters.  In detail, we split each matrix cell into two, with the bottom number and associated colour being computed from the SKIRT ground truth, and the top   number and colour corresponding to the equivalent metric derived using the trained emulator.  Across all of Figures\ \ref{fig:PartialSpearRankCorr_Flux} - \ref{fig:PartialSpearRankCorr_Sersic}, the connections between input physical parameters and emerging observables as inferred by the emulator match closely the ground truth.  This serves as another way to demonstrate the emulator has correctly learned these complex mappings.

The matrix diagrams further serve as a sensitivity analysis, with darker shades implying that the respective observables are more informative about a particular physical feature.  In Figure\ \ref{fig:PartialSpearRankCorr_Flux}, we can for example read off the following well-known relations, to pick a few: (1) Increasing $M_{\rm star}$ boosts the SED normalization, and is most directly traced by the rest-NIR emission; (2) Dust mass is most sensitively probed via the Rayleigh-Jeans tail of the FIR SED (i.e., at 1 mm).  (3) Increasing the mean stellar age via an increased Age or decreased $t_{\rm peak}$ parameter results in redder $U-V$, and to a lesser extent redder $V-J$ colours.  (4) Increasing the inclination ($\theta$) has the anticipated effect of dimming the UV-to-NIR emission due to increasing projected dust columns.  As the dust-reprocessed radiation emerges largely isotropically, this leads to increased $U - IR$ colours.  However, one can also appreciate that such orientation effects are subtle compared to the stronger imprint left by intrinsic properties related to structure, SFH and dust.

Other patterns are perhaps less known or trivial: (1) Stellar metallicity is most strongly expressed via the rest-frame $V-J$ colour (see also \citealt{Nersesian2025}). (2) A more extended distribution of stars (i.e., increased $R_{\rm star}$) allows a larger fraction of the stellar emission to escape unhindered, associated with bluer integrated colours.  (3) The impact of $n_{\rm star}$ is qualitatively similar.  A higher S\'{e}rsic index does not only imply a more pronounced central cusp, but also a profile with more prominent wings.  The latter will suffer less attenuation and weigh more in the integrated colour, making the galaxy appear bluer.  (4) For the structural observables, it is again the connection with the S\'{e}rsic index of the stellar distribution that stands out (see the observed sizes and size ratios as displayed in Figure\ \ref{fig:PartialSpearRankCorr_Size}, as well as the S\'{e}rsic indices and their change with wavelength as shown in Figure\ \ref{fig:PartialSpearRankCorr_Sersic}).  Perhaps surprisingly so, as a rest-optical colour gradient (here captured via the $R_U/R_V$ size ratio) is often taken as a signature of either a stellar population gradient or a radially declining dust profile, not necessarily a probe of the shape of the intrinsic stellar distribution.  We note that there is an impact of stellar age gradients on the optical colour gradient too, expressed via a ($k_{\rm peak}$, $R_U/R_V$) and to a lesser extent ($k_{\rm fwhm}$, $R_U/R_V$) correlation, but it is comparatively of lower amplitude.  The sensitivity of $R_U/R_V$ to the amount of dust ($M_{\rm dust}$) or its spatial distribution ($R_{\rm dust}$, $C_{\rm dust}$, $n_{\rm dust}$, $f_{\rm cov}$) is yet lower.  When considering $R_V/R_J$, a clearer link to specifically $C_{\rm dust}$ is observed.  We attribute this to thicker dust distributions causing more of the dust to sit in the foreground, yielding more efficient reddening and radial variations thereof. Finally, in terms of size ratios between different wavelengths, the strongest correlation is found between $R_U/R_{\rm 100\mu m}$ and $n_{\rm star}$.  A qualitative description of this behaviour would be that for high-$n_{\rm star}$ models the bulk of stellar emission from the central cusp is reprocessed into compact FIR radiation, whereas the bulk of stellar emission from the profile wings escapes unhindered, yielding a large half-light radius in the $U$ band.

We conclude that the emulator has adequately learned the intrincate mappings between input parameters describing the physical make-up of galaxies on the one hand, and photometric plus structural observables quantified from the emerging light on the other hand.  Complementary to the sensitivity analysis presented here, the SDs for individually varying parameters in Appendix\ \ref{app:oneparam} provide an overview of how observables are linked to inputs.

\begin{figure}
\centering
\includegraphics[width=0.95\linewidth]{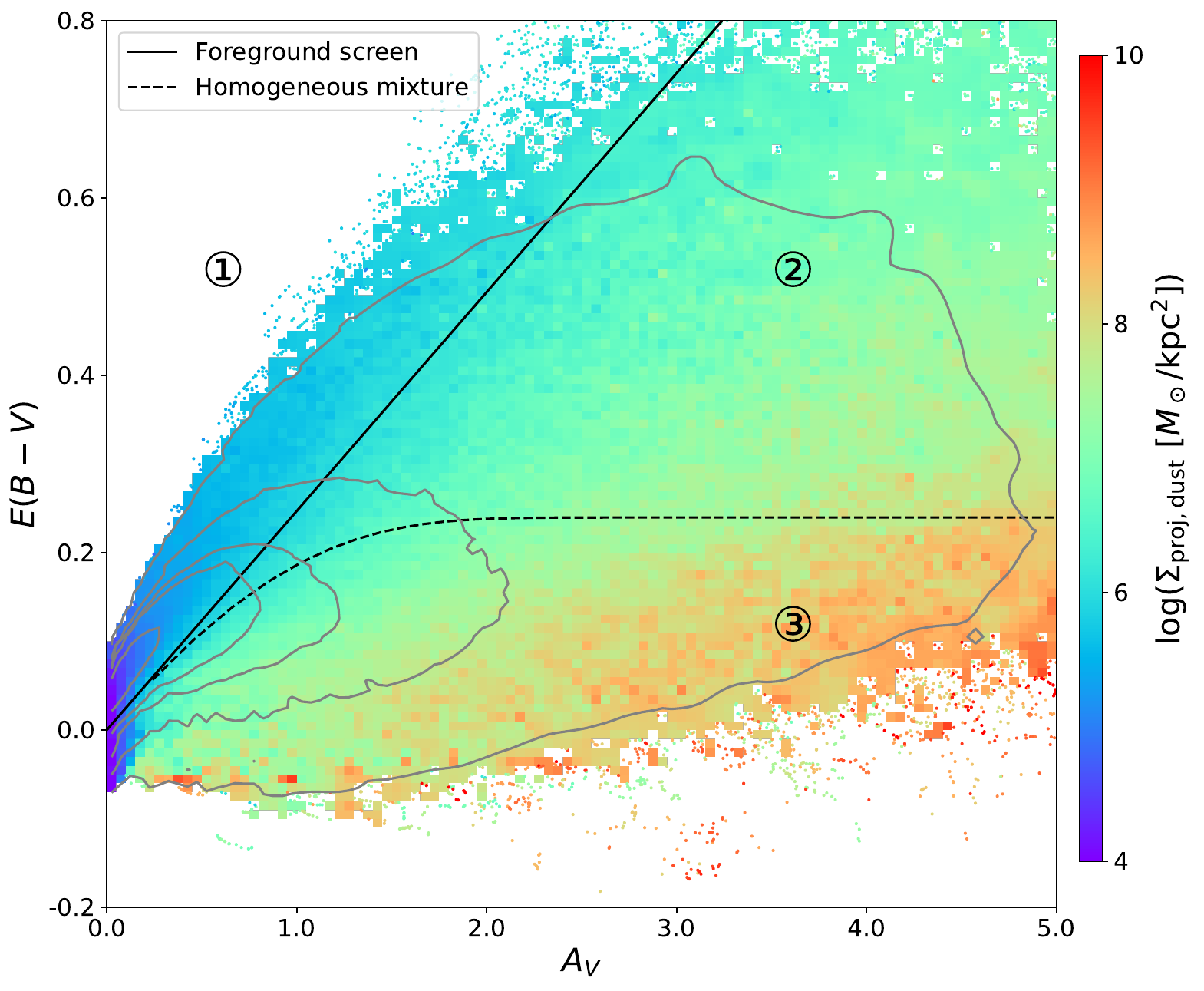}
\includegraphics[width=\linewidth]{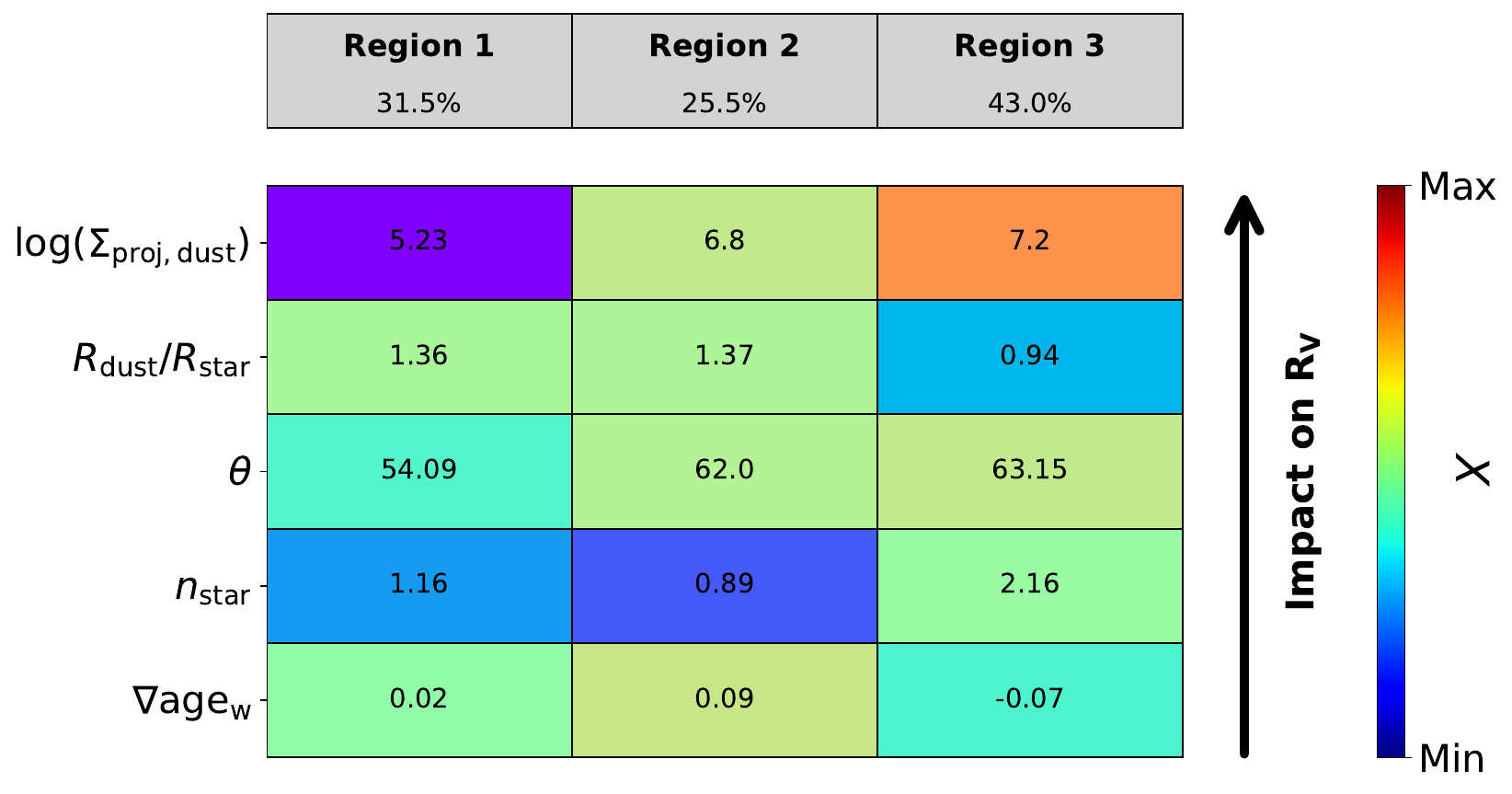}
\caption{{\it Top:} Distribution of galaxies in our toy model library across the $E(B-V)$ -- $A_V$ plane, colour coded by their projected dust surface mass density. {\it Bottom:} Median properties of toy models located in each region of the $E(B-V)$ -- $A_V$ diagram. Parameters are sorted based on their impact on the effective total-over-selective attenuation, $R_{\rm V}$ (see text for details). Regions 1, 2, and 3 contain 31.5\%, 25.5\%, and 43.0\% of the toy models within our SKIRT library, respectively.}
\label{fig:EBV_Av}
\end{figure}

\section{Discussion}
\label{sec:discussion}

In future work, we plan to exploit {\tt SE3D} to gain insight on the evolution of star-dust geometries and the development of stellar population gradients in galaxies across cosmic time.  A relevant question in this context remains when and in which galaxies signatures of inside-out growth \citep[e.g.,][]{Tacchella2018} versus an outside-in propagation of star formation \citep[e.g.,][]{Tadaki2017a} are picked up.

In the following, we first elaborate on the way dust reddening and attenuation are decoupled in our approach (Section\ \ref{sec:decouple}), followed by a discussion on some caveats of the {\tt SE3D} approach (Section\ \ref{sec:caveats}).  We then zoom out to take a broader look at potential alternative ML approaches to bridge the gap between direct observables and physical properties (Section\ \ref{sec:alternative}).

\subsection{Decoupling reddening and attenuation}
\label{sec:decouple}

One aspect that deserves emphasis is that in the {\tt SE3D} methodology the effects of reddening and attenuation by dust are decoupled, or at least not linearly tied as is the case when applying dust attenuation analytically via the assumption of a uniform foreground screen.  The top panel of Figure\ \ref{fig:EBV_Av} illustrates this by showing the effective reddening of the galaxy-integrated SED as a function of the total visual attenuation (observed minus dust-free $V$-band magnitude) for the toy model galaxies in our library.  For reference, and following Figure 10 in \citet{Wuyts2009a}, we indicate with solid and dashed lines the analytical forms for a uniform foreground screen and a homogeneous mixture of stars and dust, respectively.  For a \citet{Calzetti2000} reddening law, the former is given by $E(B-V) = A_V / 4.05$.  The latter is detailed by Equations 8 -- 12 in \citet{Wuyts2009a}.  Specifically, the analytical form for the homogeneous mixture relates to a scenario where a collection of equal-luminosity sources $i$ suffer a range of attenuations, with the $A_{V,i}$ distribution being flat up to some maximum $A_{V,{\rm max}}$.  We observe that the distribution of toy model galaxies in our library is wide-ranged, encompassing and even extending beyond the two analytical scenarios.  The fact that there is variation not only in the overall opacity but also in the ratio of effective attenuation to reddening, $R_V \equiv \frac{A_V}{E(B-V)}$, reflects the range in star-dust geometries and stellar population gradients explored.

To shed light on which conditions impact the resulting $R_V$ the most, we considered an array of physical properties. Among them were all those listed in Table\ \ref{tab:param_library}, as well as the following derived quantities which were regarded as potential candidates: the projected dust (and stellar) surface mass density, computed as $\Sigma_{\rm proj,dust} = \frac{0.5 M_{\rm dust}}{\pi R_{\rm dust}^2 q_{\rm dust}}$, the fraction of dust (stars) in birth clouds; the galaxy's specific star formation rate; the half-SFR radius; the mass-weighted stellar age, ${\rm age_w}$, and its radial gradient (defined as the difference between ${\rm age_w}$ outside minus inside $R_{\rm star}$); and finally the ratio of the length of the sightline through the dust distribution over the length of the sightline through the stellar distribution, $L_{\rm los, dust} / L_{\rm los, star}$, computed on the basis of the structural parameters $C_{\rm star}$, $R_{\rm star}$, $C_{\rm dust}$, $R_{\rm dust}$ and the viewing angle $\theta$.

Several of the parameters considered exhibit a statistically significant relation with $R_V$.  To rank them by importance, we first selected the parameter that had the highest (absolute) Spearman rank correlation coefficient with $R_V$, namely the projected dust surface mass density $\Sigma_{\rm proj,dust}$.  We then evaluated for all other parameters the partial Spearman rank correlation coefficient with $R_V$ while controlling for $\Sigma_{\rm proj,dust}$.  Following this procedure, $R_{\rm dust}/R_{\rm star}$ came out as contributing the most additional information on $R_V$.  As third ranked property, we subsequently selected the parameter which featured the largest (absolute) partial correlation coefficient with $R_V$ while controlling for the two most important ones, and so on.

In the bottom panel of Figure\ \ref{fig:EBV_Av}, we show a list of the five most important properties, ranked by their impact on $R_V$.  We detail their role below, quoting the partial Spearman rank correlation coefficient controlling for covariances with more important parameters as $\rho$:
\begin{itemize}
\item[(1)] $\Sigma_{\rm proj, dust}$ [$\rho = 0.73$]. As projected dust columns increase, so does $A_V$.  Initially, this is associated with increasing degrees of reddening also.  However, for the mixed star-dust geometries of our toy models, $E(B-V)$ saturates as a tracer of attenuation once projected dust columns become sufficiently high, leading to larger $R_V$ values. The fact that more opaque galaxies have shallower attenuation curves (i.e., higher $R_V$) is also seen in observations \citep{Salmon2016, Salim2018, Nersesian2025}. \\
\item[(2)] $R_{\rm dust}/R_{\rm star}$ [$\rho = 0.31$]. For a given dust distribution as seen by the observer, encapsulated by $\Sigma_{\rm proj, dust}$, increasing the size ratio of the dust versus stellar distribution has the net effect of placing the stars more centrally, hence sitting where the local dust columns are highest.  This amplifies the saturating effect described above.  At fixed $\Sigma_{\rm proj, dust}$, $R_V$ thus increases with increasing $R_{\rm dust}/R_{\rm star}$.  On the other hand, toy models of low $R_{\rm dust}/R_{\rm star}$ tend to feature on average higher dust columns in our library.  This explains why mock-observed toy models found in region 3 of the $E(B-V)$ -- $A_V$ plane have on average lower $R_{\rm dust}/R_{\rm star}$ (see Figure\ \ref{fig:EBV_Av}). \\
\item[(3)] $\theta$ [$\rho = 0.32$]. Edge-on viewing angles make for greyer attenuation laws (i.e., higher $R_V$, see also \citealt{Wild2011}; \citealt{Battisti2017}; \citealt{Trayford2020}; \citealt{Lu2022}; \citealt{Zhang2023}). \\
\item[(4)] $n_{\rm star}$ [$\rho = 0.18$]. As the stellar S\'{e}rsic index increases, both a central cusp and extended wings of the stellar distribution get more pronounced.  The cusp stars face the highest dust columns in our toy model description while stars in the outskirts are the least obscured.  The central, more dust-reddened stars weigh less in the overall SED than the outer, less dust-reddened stars.  This makes for a less efficient reddening in an integrated sense (i.e., higher $R_V$), especially when $n_{\rm star}$ is high. \\
\item[(5)] $\nabla {\rm age_w}$ [$\rho = -0.15$]. Negative values correspond to younger outskirts.  Paired with the fact that dust columns are lower at large radii, it causes the intrinsically bluest (youngest) stars to be the least attenuated, hence reducing the effective reddening of the integrated light and increasing $R_V$.  Indeed, in 2.4\% of the mock-observed toy models the integrated light is even bluer than in the absence of dust (i.e., the $E(B-V)$ is negative, albeit rarely going below -0.05).
\end{itemize}

We conclude that the projected dust column (colour coded in Figure\ \ref{fig:EBV_Av}) overwhelmingly dominates the variations in $R_V$ seen across the toy model library, but other factors make significant and explainable contributions too.

In analogy to Figure\ \ref{fig:SKIRTLib_UV_Obs} confirming reliable emulator performance across the $UVJ$ colour space, we verified that the emulator accuracy is encouragingly uniform across the $E(B-V)$ -- $A_V$ diagram.

\subsection{Caveats}
\label{sec:caveats}
A few assumptions and simplifications are introduced in the \texttt{SE3D} approach which are observationally and/or computationally motivated. In this section, we discuss caveats associated with our toy model galaxy description, provide reasoning behind our choices and briefly touch on their impact on observables.

\subsubsection{Galaxy structure}
\label{sec:GalaxyStructure}
The first simplification in our modelling concerns the treatment of galaxy structure, which inherently lacks complex structures seen in real galaxy observations, such as galactic bars and spiral arms. Our simplified description of galaxy structure is motivated by two main arguments.  Firstly, at high-z, a robust observational characterisation of subgalactic structures is challenging due to resolution and depth limitations.  As a result, global structural parameters (i.e., size, S\'{e}rsic index and axis ratio) are often the only observational constraints on galaxy structure available across a wide range of wavelengths.  At low-z, substructures are much better characterised thanks to the exquisite spatial resolution.  Indeed, dedicated high-resolution RT simulations including more complex structures are being developed to reproduce such rich, individual galaxy datasets \citep[e.g.][]{DeLooze2014, Nersesian2020b}. {\tt SE3D} aims to complement such efforts via a more population-wide exploration of galaxies in the distant Universe.  In order to do so efficiently, a second motivation for the simplified description of galaxy structure is to restrict the dimensionality of the toy models and the emulator.  This enables improved emulator accuracy and a more robust exploration of parameter space.

{\tt SE3D} can thus be regarded as a useful approach to fill the gap between dedicated individual-object modelling with high-resolution RT simulations on the one hand, and standard SED fitting algorithms on the other hand.  Compared to the latter, which effectively combine a zero-dimensional model of a galaxy with an effective attenuation law encoding the net impact of dust, {\tt SE3D}'s novel feature is that it factors in spatial information (i.e., measurements of global galactic structure) by employing 3D model geometries for stars and dust.  Whilst this is beneficial, the range of geometries necessarily falls short of the rich array of structures present in real galaxies.

Finally, we recognise that in strongly perturbed systems (e.g., galaxies undergoing interactions) the observed projected axis ratio may not necessarily serve as a reliable tracer of inclination.

\subsubsection{Star formation histories}
\label{sec:SFH}
The adoption of single-component star formation histories is wide-spread among galaxy-integrated SED fitting applications, albeit not without potential biases \citep[e.g.,][]{Leja2019}.  This includes the use of a log-normal SFHs as motivated by observations and simulations \citep[see, e.g.,][]{Gladders2013, Diemer2017}. Our SFH treatment goes beyond a single log-normal SFH for the entire galaxy, through the additional flexibility of a radial gradient in the peak time and/or fwhm (\#8 and \#9 in Table \ref{tab:param_library}), enabling a more varied family of SFHs. It allows for a rapid decline in star formation activity, but this then inherently requires first a rapid increase also, due to the nature of the log-normal functional form.  Furthermore, it does not capture burstiness and may therefore suffer from outshining effects as other applications adopting smooth SFHs do \citep{Papovich2001, Maraston2010, Pforr2012, Narayanan2024}. Categories of objects that may suffer most from potential SFH and especially burstiness related biases are very low-mass or very high-redshift galaxies, as their histories are thought to be most bursty \citep{Perry2025, Sun2023}. At the same time, such objects tend to be dust-poor \citep{RemyRuyer2015, Ferrara2016}, making them potentially less interesting to apply \texttt{SE3D} modelling to.

\subsubsection{Metallicity}
\label{sec:Metallicity}
We adopt a global stellar metallicity for an entire toy model galaxy in \texttt{SE3D}, a common assumption in SED fitting algorithms. In principle, improvements over this simplistic approach could come in the form of a gas regulator (a.k.a. bathtub) model, capable of self-consistently linking the star formation history (SFH) and chemical enrichment history (CEH) \citep{Weinberg2017, Belfiore2019}.  In practice, however, for such approaches to be realistic they require a non-trivial parametrization of depletion times and mass loading factors of galactic winds, which are challenging to constrain observationally, let alone from first principles.  A fruitful middle ground may therefore be to incorporate a mass-metallicity relation as prior while fitting, and this functionality is included as an option in the current fitting framework of \texttt{SE3D}.

\subsubsection{Dust models}
\label{sec:DustModels}
The treatment of dust in galaxy modelling comes in progressive levels of complexity: this begins with standard SED fitting which adopts a fixed attenuation law, and extends to modifications which allow for variations in attenuation law slope and/or strength of the 2175\AA\ bump feature \citep[e.g.,][]{Markov2025}. Beyond this, one can model explicitly the impact of varying star-dust geometries (as \texttt{SE3D} does), or the impact of distinct dust grain models (i.e., variations in grain composition and grain size distribution; see, e.g., \citealt{Zelko2020}). \citet{Matsumoto2025} analysed simulations that track the evolution of grain properties in a physically motivated way, and found the relative importance of star-dust geometry versus evolving grain size distributions to depend non-trivially on $A_V$ and inclination angle. Turning to observations, \citet{SachdevaNath2022} studied a large representative sample of local galaxies and concluded star-dust geometry effects dominate over dust grain evolution. At high redshift ($z > 5$), \citet{Markov2025} found typical attenuation curves to be flat and lacking a prominent UV bump, in line with expectations from dust primarily composed of large grains produced in Type II supernovae \citep{Makiya2022}. However, significant detections of a UV bump have also been reported for galaxies out to $z \sim 7$ \citep{Witstok2023, Ormerod2025}, indicative of an efficient production mechanism of small carbonaceous dust grains.

In future work, it would be useful to investigate the impact of the adopted dust grain model, if only as a robustness check on the results obtained, as degeneracies between dust grain properties and star-dust geometry may remain insurmountable for the typical data quality available \citep[see also][]{Lower2022}. Here, we briefly comment on the use of two different dust models readily implemented in SKIRT: THEMIS (our fiducial choice) and Zubko, an interstellar dust model derived by simultaneously fitting interstellar extinction, diffuse IR emission and elemental abundance constraints on dust \citep{Zubko2004}.  Our comparison for 100 randomly selected toy models from our library yields typical NMAD variations of 0.110, 0.023, 0.048 dex for the SED, SSD, and SND, where the strongest variations are primarily seen in the 3 - 30 $\mu$m (MIR) wavelength regime.

\subsubsection{Impact of model mismatch}
\label{sec:Z25}
For a quantitative analysis of the impact of various sources of model mismatch, we refer the reader to our companion paper (Z25) where we test {\tt SE3D} on mock observations of simulated TNG galaxies.  Cosmologically simulated galaxies inherently do feature a richer array of subgalactic structures in their stellar and ISM components (e.g., bars, spiral arms, clumps and dust holes), more complex SFHs (including bursts) and non-trivial metallicity distributions compared to our toy models.  They therefore serve as a useful benchmark.  Other than evaluating the combined impact on accuracy of recovered properties, we investigate the contribution of the aforementioned sources of model mismatch (structural, SFH and CEH) individually in Section 5.2 of Z25.

\subsection{Alternative ML approaches to connect physics and observables}
\label{sec:alternative}

The {\tt SE3D} approach presented in this paper is not the only conceivable application of ML techniques to connect physics to observables, or vice versa.  In this section, we touch briefly on potential alternatives, their merits and challenges.

\subsubsection{Forward vs backward modelling}
\label{sec:forward}

The {\tt SE3D} method can be considered as a form of forward modelling.  That is, the emulator is built to propagate forwards from physical inputs to observable outputs.  This has the advantage that continuous SDs for a given model can straightforwardly be displayed to gain intuition.  It further gives the user the freedom to specify their own priors when performing Bayesian inference, and evaluate the posterior distributions in a similar manner as when fitting simpler, analytical models.  As with any MCMC fitting, it does come at the cost of having to execute thousands of calls to a model function that generates trial SDs.

An alternative approach could therefore be to train a ML algorithm to map directly from observables in a set of wavebands to the desired physical properties.  A rich literature of data-driven photometric redshift estimation methods belong to this category \citep[see][and references therein]{Lin2022}.  Such an approach would undoubtedly be faster, but would complicate any treatment of priors.  It would further require a library of not just rest-frame SDs, but measurements within a large collection of filter throughput curves computed over a fine grid of redshifts.  As different filter set combinations would be available for objects observed in different deep fields, and not all of them would have resolved information, the classical ML problem of missing data would pose a possible hurdle.

\subsubsection{Sampling of training library}
\label{sec:sampling}

In constructing our training library, we drew values of input parameters randomly from pre-defined distribution functions (see Table\ \ref{tab:param_library}).  An alternative approach could have been to select equidistant points in parameter space via Latin Hypercube sampling (LHS, \citealt{McKay1979}).  We experimented on a training sample of limited size (15,000 sets of SDs), and found no evidence for an emulator trained on Latin-hypercube samples to outperform one that was trained on the same number of randomly selected samples.

Perhaps more fruitful could be the use of emulator accuracy metrics to determine in which regions of input parameter space additional training samples would be desirable.  One way to retrieve such metrics could be to assess residuals between emulator predictions and SKIRT ground truth, although by construction such diagnostics are only available where training samples already exist.  A second route could therefore be to rely on the confidence intervals predicted by the BNN itself.  The Bayesian nature of the neural network allows for efficient assessment of the spread in repeat predictions by the emulator for arbitrary positions in input parameter space.  We verified that the uncertainties predicted by the BNN correlate with those assessed empirically at high statistical significance ($p \ll 10^{-5}$), albeit with considerable scatter ($\rho_{\rm Spearman} = 0.4$).  We conclude that the BNN's uncertainty predictions can usefully serve as a guide for training sample augmentation.\footnote{In an absolute sense, we find the uncertainties predicted by the BNN to underestimate the empirical errors by a factor 1.25, indicating that the BNN is modestly overconfident in its predictions.} 

Finally, given that the size of the training sample has a critical impact on emulator performance (see Figure\ \ref{fig:NMAD_Vs_LogNtrain}), it is relevant to note potential avenues to more efficiently build large libraries.  First, RT calculations which only account for dust scattering and absorption and omit dust re-emission are faster to run, as they can skip the time-consuming calculation of equilibrium dust temperatures \citep[see the review][for details on this topic]{Steinacker2013}.  Of course, such shortcut is only of use to predict SDs in the stellar regime.  Secondly, \citet{RinoSilvestre2022} experimented with ML algorithms that learn a mapping from noisy (but fast) RT output computed with a low number of photon packages to high-fidelity (but slow) RT output based on large numbers of photon packages.  Their results are encouraging in terms of ability to de-noise the low photon count runs, albeit applied to a restricted number of test setups.

\subsubsection{Simulation based inference}
\label{sec:simbased}

Mock observations of galaxies simulated in a cosmological context could be used to train a ML application to identify physical properties that themselves are not directly observable.  Hydrodynamical simulations are for example commonly used to train algorithms for merger classification \citep[see, e.g.,][and references therein]{AvirettMackenzie2024}.  A common concern in this regard is how well the training on simulated examples translates to real galaxies (or galaxies simulated using other methods; see, e.g., \citealt{MargalefBentabol2024}).

In the context of analysing panchromatic resolved observations, such approach would in principle have the merit of using galaxy models with more physically motivated and complex characteristics (e.g., variations in SFH and structural make-up that aim to be more akin to those observed).  In practice, however, simulations of galaxy formation do not perfectly represent reality.  Those that simulate a large enough box to provide sufficient training samples tend to not explicitly incorporate dust (although see \citealt{Trayford2025}).  Moreover, when the effects of dust are introduced during post-processing with RT, they tend to struggle to span the full colour space occupied by observed galaxies \citep[][Z25]{Gebek2025}, a challenge that dates back to isolated galaxy simulations \citep{Wuyts2009b}.  Libraries of toy model galaxies comparatively suffer less from such a challenge, as conditions can be designed by hand to ensure representativeness.

\subsubsection{Image to image translation}
\label{sec:im2im}

While a reliable extraction of global structural parameters ($R_e$, $n$, $q$) is the best one can hope for in some bands, observations in other bands (in particular {\it JWST}/NIRCam) have a richer information content.  To make full use of this information, data-driven image to image (or cube to cube) translation techniques may be beneficial.  Among them, Generative Adversarial Networks (GANs, \citealt{Goodfellow2014}) and diffusion models \citep{Chen2024, Alsing2024} have gained popularity.  Such methods could conceivably ingest multi-wavelength images, alongside potential ancillary information such as associated PSFs, PSF FWHMs and/or target redshift, and output maps of physical quantities as well as their uncertainties.  The challenge here lies in the high dimensionality of the problem, and the creation of appropriate training sets.

\section{Conclusions}
\label{sec:summary}

In this paper we have introduced the {\tt SE3D} framework (Figure\ \ref{fig:schematic}), aimed at interpreting the physical make-up of galaxies via joint modelling of their available panchromatic and resolved information.  A machine learning emulator was developed to predict the wavelength-dependent properties of a diverse range of parametrized toy model galaxies.  The toy models varied in their stellar and dust content, their star formation history and stellar population gradients, as well as 3D spatial distributions of stars and dust.  The emulator was trained on an extensive library of toy model galaxies processed with 3D dust radiative transfer using the SKIRT code by \citet{Camps2020}.  The architecture makes use of Bayesian Neural Networks to predict the weights of Principal Component templates which in superposition reproduce the galaxy spectral energy distribution (SED), the galaxy's wavelength-dependent size (SRD), light profile (encoded as S\'{e}rsic index, SND) and projected axis ratio (SQD).  The emulator introduces a $>10^5\times$ speed up compared to actual radiative transfer calculations, enabling its use within Bayesian inference fitting of observational datasets in a computationally tractable manner.

We provide example spectral distributions predicted by the emulator for varying input parameters, and contrast them to SKIRT ground truth.  Across the UV-to-mm wavelength range, the SED, SRD, SND and SQD are predicted by the emulator at a typical accuracy of 0.043, 0.030, 0.056, and 0.004 dex. We evaluate the critical importance of training sample size and dimensionality of the toy model on emulator performance.  We further demonstrate how the emulator accuracy is stable across the dynamic range of input parameters it was trained on (see Table\ \ref{tab:param_library}) and across the rest-frame $UVJ$ colour space spanned by observed galaxies over a wide range in redshift.  Via the use of partial correlation matrix diagrams, we visualize the mapping between input physical quantities and output observables (fluxes, colours, sizes, size ratios, ...) at a range of wavelengths, and demonstrate how the emulator has successfully learned these interconnections.  In a companion paper (Z25), we further apply and test the method against mock-observed toy model and simulated galaxies.

A key goal of {\tt SE3D} is to uncover how star formation proceeded within galaxies, while also accounting for the range of (evolving) star-dust geometries in galaxies.  Specifically, we demonstrate how the range of geometries represented in our training library is expressed in a wide distribution across the $E(B-V)$ -- $A_V$ diagram, and we address which parameters most significantly impact the resulting total-over-selective attenuation $R_V$.  In order of importance, these are first and foremost the projected dust surface mass density ($\Sigma_{\rm proj, dust}$), and subsequently $R_{\rm dust}/R_{\rm star}$, $\theta$, and $n_{\rm star}$.

Besides scientific prospects, we also discuss in Section\ \ref{sec:discussion} the broader landscape of efforts to use ML techniques in mapping physical quantities to observables, or vice versa.  We consider pros and cons of alternative approaches and touch on potential future directions.

We conclude that ML emulators can be a fruitful tool in bridging the gap between large observational datasets and computationally demanding radiative transfer calculations.

\section*{Acknowledgements}

We thank the authors of SKIRT, Maarten Baes and Peter Camps, for making their radiative transfer code publicly available.  We also thank James Trayford, Andrea Gebek, Nick Andreadis, Shiyin Shen and XianZhong Zheng for valuable discussions on this work.  The authors gratefully acknowledge support from the Royal Society International Exchanges scheme (IES\textbackslash R2\textbackslash 242195).  SW acknowledges support from China's National Foreign Expert programme (H20240871).  The authors acknowledge the Tsinghua Astrophysics High-Performance Computing platform at Tsinghua University for providing computational and data storage resources that have contributed to the research results reported within this paper.

\section*{Data Availability}

A public release of the {\tt SE3D} ML emulator and fitting framework is envisioned as part of a forthcoming paper applying the code to an observational sample.  Derived data presented in this work will be shared upon reasonable request to the corresponding author.

\begin{table*}
\centering 
\resizebox{\linewidth}{!}{
\begin{tabular}{l|l|l|l} 
\hline \hline
\# & \textbf{Hyperparameter} & \textbf{Definition} & \textbf{Possible range/combinations} \\
\hline
\multicolumn{4}{|c|}{Architecture} \\
\hline
1. & $\rm{n}_{layers}$ & Number of layers & [1, 4] \\
2. & $\rm{n}_{nodes}$ & Number of nodes for each layer ($\rm{n}_{layers}$) & [64, 512] \\
3. & $\mu$ & Initialized mean for the BNN weight distribution & [-0.5, 0.5] \\
4. & $\sigma$ & Initialized standard deviation for the BNN weight distribution & [0.1, 0.5] \\
5. & $\alpha(w)$ & Activation Function for each layer ($\rm{n}_{layers}$)  & relu, tanh, sigmoid, leaky relu, gelu, gelu tanh, silu, celu \\
6. & Optimizer & Algorithms used to update neural networks weights & Adam, AdamW, NAdam, RMSProp \\
7. & Momentum & Applies smoothing to gradient descent & [0.85, 0.99] \\
8. & Weight decay & Regularization applied to Optimizer to prevent overfitting  &  [$1 \times 10^{-5}, 1 \times 10^{-2}$] \\
9. & $\beta_{1},\ \beta_{2}$ & Parameters controlling the decay rate of the first and second moments for gradients & [0.9, 0.999] \\ 
\hline
\multicolumn{4}{|c|}{Training} \\
\hline
10. & lr & Learning rate & [$1 \times 10^{-6}, 1 \times 10^{-2}$] \\
11. & Batch size & Number of data rows processed per training epoch & [32, 264] \\
12. & $\rm{grad_{clip}}$ & Value at which gradients are clipped every training epoch & [0.5, 10] \\
13. & $\rm{KL}_{weight}$ & Weight coefficient for the Kullback-Leibler divergence describing model uncertainty  & [0.0, 1.0] \\
14. & $\rm{ZP}_{weight}$ & Weight coefficient for the difference between predicted and true zero points & [0.0, 1.0]
\\
\hline 
\end{tabular}}
\caption{Parameters defining the BNN architecture and training, together with the options/ranges considered by {\tt Optuna} during hyperparameter tuning.}
\label{tab:hyper}
\end{table*}



\bibliographystyle{mnras}
\bibliography{bibliography}

@ARTICLE{Martorano2025,
       author = {{Martorano}, M. and {van der Wel}, A. and {Baes}, M. and {Bell}, E.~F. and {Brammer}, G. and {Franx}, M. and {Gebek}, A. and {Meidt}, S.~E. and {Miller}, T.~B. and {Nelson}, E. and {Nersesian}, A. and {Price}, S.~H. and {van Dokkum}, P. and {Whitaker}, K.~E. and {Wuyts}, S.},
        title = "{Evolution of the S{\'e}rsic index up to z = 2.5 from JWST and HST}",
      journal = {\aap},
         year = 2025,
        month = feb,
       volume = {694},
          eid = {A76},
        pages = {A76},
          doi = {10.1051/0004-6361/202452919},
archivePrefix = {arXiv},
       eprint = {2501.02956},
 primaryClass = {astro-ph.GA},
       adsurl = {https://ui.adsabs.harvard.edu/abs/2025A&A...694A..76M}
}

@ARTICLE{Martorano2026,
       author = {{Martorano}, M. and {van der Wel}, A. and {Gebek}, A. and {Baes}, M. and {Bell}, E.~F. and {Brammer}, G. and {Meidt}, S.~E. and {Nersesian}, A. and {Whitaker}, K. and {Wuyts}, S.},
        title = "{Evolution and mass dependence of UV-to-near-IR color gradients up to z = 2.5 from the Hubble Space Telescope and the James Webb Space Telescope}",
      journal = {\aap},
         year = 2026,
        month = jan,
       volume = {705},
          eid = {A236},
        pages = {A236},
          doi = {10.1051/0004-6361/202555974},
archivePrefix = {arXiv},
       eprint = {2512.02440},
 primaryClass = {astro-ph.GA},
       adsurl = {https://ui.adsabs.harvard.edu/abs/2026A&A...705A.236M}
}

@ARTICLE{Tinsley1968,
       author = {{Tinsley}, Beatrice M.},
        title = "{Evolution of the Stars and Gas in Galaxies}",
      journal = {\apj},
         year = 1968,
        month = feb,
       volume = {151},
        pages = {547},
          doi = {10.1086/149455},
       adsurl = {https://ui.adsabs.harvard.edu/abs/1968ApJ...151..547T}
}

@ARTICLE{Guo2011,
       author = {{Guo}, Yicheng and {Giavalisco}, Mauro and {Cassata}, Paolo and {Ferguson}, Henry C. and {Dickinson}, Mark and {Renzini}, Alvio and {Koekemoer}, Anton and {Grogin}, Norman A. and {Papovich}, Casey and {Tundo}, Elena and {Fontana}, Adriano and {Lotz}, Jennifer M. and {Salimbeni}, Sara},
        title = "{Color and Stellar Population Gradients in Passively Evolving Galaxies at z \raisebox{-0.5ex}\textasciitilde 2 from HST/WFC3 Deep Imaging in the Hubble Ultra Deep Field}",
      journal = {\apj},
         year = 2011,
        month = jul,
       volume = {735},
       number = {1},
          eid = {18},
        pages = {18},
          doi = {10.1088/0004-637X/735/1/18},
archivePrefix = {arXiv},
       eprint = {1101.0843},
 primaryClass = {astro-ph.CO},
       adsurl = {https://ui.adsabs.harvard.edu/abs/2011ApJ...735...18G}
}

@ARTICLE{Guo2012,
       author = {{Guo}, Yicheng and {Giavalisco}, Mauro and {Ferguson}, Henry C. and {Cassata}, Paolo and {Koekemoer}, Anton M.},
        title = "{Multi-wavelength View of Kiloparsec-scale Clumps in Star-forming Galaxies at z \raisebox{-0.5ex}\textasciitilde 2}",
      journal = {\apj},
         year = 2012,
        month = oct,
       volume = {757},
       number = {2},
          eid = {120},
        pages = {120},
          doi = {10.1088/0004-637X/757/2/120},
archivePrefix = {arXiv},
       eprint = {1110.3800},
 primaryClass = {astro-ph.CO},
       adsurl = {https://ui.adsabs.harvard.edu/abs/2012ApJ...757..120G}
}

@ARTICLE{vanderWel2024,
       author = {{van der Wel}, Arjen and {Martorano}, Marco and {H{\"a}u{\ss}ler}, Boris and {Nedkova}, Kalina V. and {Miller}, Tim B. and {Brammer}, Gabriel B. and {van de Ven}, Glenn and {Leja}, Joel and {Bezanson}, Rachel S. and {Muzzin}, Adam and {Marchesini}, Danilo and {de Graaff}, Anna and {Nelson}, Erica J. and {Kriek}, Mariska and {Bell}, Eric F. and {Franx}, Marijn},
        title = "{Stellar Half-mass Radii of 0.5 z < 2.3 Galaxies: Comparison with JWST/NIRCam Half-light Radii}",
      journal = {\apj},
         year = 2024,
        month = jan,
       volume = {960},
       number = {1},
          eid = {53},
        pages = {53},
          doi = {10.3847/1538-4357/ad02ee},
archivePrefix = {arXiv},
       eprint = {2307.03264},
 primaryClass = {astro-ph.GA},
       adsurl = {https://ui.adsabs.harvard.edu/abs/2024ApJ...960...53V}
}

@ARTICLE{Maraston2005,
       author = {{Maraston}, Claudia},
        title = "{Evolutionary population synthesis: models, analysis of the ingredients and application to high-z galaxies}",
      journal = {\mnras},
         year = 2005,
        month = sep,
       volume = {362},
       number = {3},
        pages = {799-825},
          doi = {10.1111/j.1365-2966.2005.09270.x},
archivePrefix = {arXiv},
       eprint = {astro-ph/0410207},
 primaryClass = {astro-ph},
       adsurl = {https://ui.adsabs.harvard.edu/abs/2005MNRAS.362..799M}
}

@ARTICLE{Conroy2009,
       author = {{Conroy}, Charlie and {Gunn}, James E. and {White}, Martin},
        title = "{The Propagation of Uncertainties in Stellar Population Synthesis Modeling. I. The Relevance of Uncertain Aspects of Stellar Evolution and the Initial Mass Function to the Derived Physical Properties of Galaxies}",
      journal = {\apj},
         year = 2009,
        month = jul,
       volume = {699},
       number = {1},
        pages = {486-506},
          doi = {10.1088/0004-637X/699/1/486},
archivePrefix = {arXiv},
       eprint = {0809.4261},
 primaryClass = {astro-ph},
       adsurl = {https://ui.adsabs.harvard.edu/abs/2009ApJ...699..486C}
}

@ARTICLE{Camps2015b,
       author = {{Camps}, Peter and {Misselt}, Karl and {Bianchi}, Simone and {Lunttila}, Tuomas and {Pinte}, Christophe and {Natale}, Giovanni and {Juvela}, Mika and {Fischera}, Joerg and {Fitzgerald}, Michael P. and {Gordon}, Karl and {Baes}, Maarten and {Steinacker}, J{\"u}rgen},
        title = "{Benchmarking the calculation of stochastic heating and emissivity of dust grains in the context of radiative transfer simulations}",
      journal = {\aap},
         year = 2015,
        month = aug,
       volume = {580},
          eid = {A87},
        pages = {A87},
          doi = {10.1051/0004-6361/201525998},
archivePrefix = {arXiv},
       eprint = {1506.05304},
 primaryClass = {astro-ph.IM},
       adsurl = {https://ui.adsabs.harvard.edu/abs/2015A&A...580A..87C}
}

@ARTICLE{Pandya2024,
       author = {{Pandya}, Viraj and {Zhang}, Haowen and {Huertas-Company}, Marc and {Iyer}, Kartheik G. and {McGrath}, Elizabeth and {Barro}, Guillermo and {Finkelstein}, Steven L. and {K{\"u}mmel}, Martin and {Hartley}, William G. and {Ferguson}, Henry C. and {Kartaltepe}, Jeyhan S. and {Primack}, Joel and {Dekel}, Avishai and {Faber}, Sandra M. and {Koo}, David C. and {Bryan}, Greg L. and {Somerville}, Rachel S. and {Amor{\'\i}n}, Ricardo O. and {Arrabal Haro}, Pablo and {Bagley}, Micaela B. and {Bell}, Eric F. and {Bertin}, Emmanuel and {Costantin}, Luca and {Dav{\'e}}, Romeel and {Dickinson}, Mark and {Feldmann}, Robert and {Fontana}, Adriano and {Gavazzi}, Raphael and {Giavalisco}, Mauro and {Grazian}, Andrea and {Grogin}, Norman A. and {Guo}, Yuchen and {Hahn}, ChangHoon and {Holwerda}, Benne W. and {Kewley}, Lisa J. and {Kirkpatrick}, Allison and {Kocevski}, Dale D. and {Koekemoer}, Anton M. and {Lotz}, Jennifer M. and {Lucas}, Ray A. and {Papovich}, Casey and {Pentericci}, Laura and {P{\'e}rez-Gonz{\'a}lez}, Pablo G. and {Pirzkal}, Nor and {Ravindranath}, Swara and {Rose}, Caitlin and {Schefer}, Marc and {Simons}, Raymond C. and {Straughn}, Amber N. and {Tacchella}, Sandro and {Trump}, Jonathan R. and {de la Vega}, Alexander and {Wilkins}, Stephen M. and {Wuyts}, Stijn and {Yang}, Guang and {Yung}, L.~Y. Aaron},
        title = "{Galaxies Going Bananas: Inferring the 3D Geometry of High-redshift Galaxies with JWST-CEERS}",
      journal = {\apj},
         year = 2024,
        month = mar,
       volume = {963},
       number = {1},
          eid = {54},
        pages = {54},
          doi = {10.3847/1538-4357/ad1a13},
archivePrefix = {arXiv},
       eprint = {2310.15232},
 primaryClass = {astro-ph.GA},
       adsurl = {https://ui.adsabs.harvard.edu/abs/2024ApJ...963...54P}
}

@ARTICLE{Carnall2019,
       author = {{Carnall}, Adam C. and {Leja}, Joel and {Johnson}, Benjamin D. and {McLure}, Ross J. and {Dunlop}, James S. and {Conroy}, Charlie},
        title = "{How to Measure Galaxy Star Formation Histories. I. Parametric Models}",
      journal = {\apj},
         year = 2019,
        month = mar,
       volume = {873},
       number = {1},
          eid = {44},
        pages = {44},
          doi = {10.3847/1538-4357/ab04a2},
archivePrefix = {arXiv},
       eprint = {1811.03635},
 primaryClass = {astro-ph.GA},
       adsurl = {https://ui.adsabs.harvard.edu/abs/2019ApJ...873...44C}
}

@ARTICLE{Zhang2023,
       author = {{Zhang}, Junkai and {Wuyts}, Stijn and {Cutler}, Sam E. and {Mowla}, Lamiya A. and {Brammer}, Gabriel B. and {Momcheva}, Ivelina G. and {Whitaker}, Katherine E. and {van Dokkum}, Pieter and {F{\"o}rster Schreiber}, Natascha M. and {Nelson}, Erica J. and {Schady}, Patricia and {Villforth}, Carolin and {Wake}, David and {van der Wel}, Arjen},
        title = "{Dust attenuation, dust content, and geometry of star-forming galaxies}",
      journal = {\mnras},
         year = 2023,
        month = sep,
       volume = {524},
       number = {3},
        pages = {4128-4147},
          doi = {10.1093/mnras/stad2066},
archivePrefix = {arXiv},
       eprint = {2307.02568},
 primaryClass = {astro-ph.GA},
       adsurl = {https://ui.adsabs.harvard.edu/abs/2023MNRAS.524.4128Z}
}

@ARTICLE{Zibetti2009,
       author = {{Zibetti}, Stefano and {Charlot}, St{\'e}phane and {Rix}, Hans-Walter},
        title = "{Resolved stellar mass maps of galaxies - I. Method and implications for global mass estimates}",
      journal = {\mnras},
         year = 2009,
        month = dec,
       volume = {400},
       number = {3},
        pages = {1181-1198},
          doi = {10.1111/j.1365-2966.2009.15528.x},
archivePrefix = {arXiv},
       eprint = {0904.4252},
 primaryClass = {astro-ph.CO},
       adsurl = {https://ui.adsabs.harvard.edu/abs/2009MNRAS.400.1181Z}
}

@ARTICLE{Abdurrouf2022,
       author = {{Abdurro'uf} and {Lin}, Yen-Ting and {Hirashita}, Hiroyuki and {Morishita}, Takahiro and {Tacchella}, Sandro and {Akiyama}, Masayuki and {Takeuchi}, Tsutomu T. and {Wu}, Po-Feng},
        title = "{Dissecting Nearby Galaxies with piXedfit. I. Spatially Resolved Properties of Stars, Dust, and Gas as Revealed by Panchromatic SED Fitting}",
      journal = {\apj},
         year = 2022,
        month = feb,
       volume = {926},
       number = {1},
          eid = {81},
        pages = {81},
          doi = {10.3847/1538-4357/ac439a},
archivePrefix = {arXiv},
       eprint = {2110.03158},
 primaryClass = {astro-ph.GA},
       adsurl = {https://ui.adsabs.harvard.edu/abs/2022ApJ...926...81A}
}

@ARTICLE{Verstocken2020,
       author = {{Verstocken}, Sam and {Nersesian}, Angelos and {Baes}, Maarten and {Viaene}, S{\'e}bastien and {Bianchi}, Simone and {Casasola}, Viviana and {Clark}, Christopher J.~R. and {Davies}, Jonathan I. and {De Looze}, Ilse and {De Vis}, Pieter and {Dobbels}, Wouter and {Galliano}, Fr{\'e}d{\'e}ric and {Jones}, Anthony P. and {Madden}, Suzanne C. and {Mosenkov}, Aleksandr V. and {Tr{\v{c}}ka}, Ana and {Xilouris}, Emmanuel M.},
        title = "{High-resolution, 3D radiative transfer modelling. II. The early-type spiral galaxy M 81}",
      journal = {\aap},
         year = 2020,
        month = may,
       volume = {637},
          eid = {A24},
        pages = {A24},
          doi = {10.1051/0004-6361/201935770},
archivePrefix = {arXiv},
       eprint = {2004.03615},
 primaryClass = {astro-ph.GA},
       adsurl = {https://ui.adsabs.harvard.edu/abs/2020A&A...637A..24V}
}

@ARTICLE{Nersesian2020a,
       author = {{Nersesian}, Angelos and {Verstocken}, Sam and {Viaene}, S{\'e}bastien and {Baes}, Maarten and {Xilouris}, Emmanuel M. and {Bianchi}, Simone and {Casasola}, Viviana and {Clark}, Christopher J.~R. and {Davies}, Jonathan I. and {De Looze}, Ilse and {De Vis}, Pieter and {Dobbels}, Wouter and {Fritz}, Jacopo and {Galametz}, Maud and {Galliano}, Fr{\'e}d{\'e}ric and {Jones}, Anthony P. and {Madden}, Suzanne C. and {Mosenkov}, Aleksandr V. and {Tr{\v{c}}ka}, Ana and {Ysard}, Nathalie},
        title = "{High-resolution, 3D radiative transfer modelling. III. The DustPedia barred galaxies}",
      journal = {\aap},
         year = 2020,
        month = may,
       volume = {637},
          eid = {A25},
        pages = {A25},
          doi = {10.1051/0004-6361/201936176},
archivePrefix = {arXiv},
       eprint = {2004.03616},
 primaryClass = {astro-ph.GA},
       adsurl = {https://ui.adsabs.harvard.edu/abs/2020A&A...637A..25N}
}

@ARTICLE{Nersesian2020b,
       author = {{Nersesian}, Angelos and {Viaene}, S{\'e}bastien and {De Looze}, Ilse and {Baes}, Maarten and {Xilouris}, Emmanuel M. and {Smith}, Matthew W.~L. and {Bianchi}, Simone and {Casasola}, Viviana and {Cassar{\`a}}, Letizia P. and {Clark}, Christopher J.~R. and {Dobbels}, Wouter and {Fritz}, Jacopo and {Galliano}, Fr{\'e}d{\'e}ric and {Madden}, Suzanne C. and {Mosenkov}, Aleksandr V. and {Tr{\v{c}}ka}, Ana},
        title = "{High-resolution, 3D radiative transfer modelling. V. A detailed model of the M 51 interacting pair}",
      journal = {\aap},
         year = 2020,
        month = nov,
       volume = {643},
          eid = {A90},
        pages = {A90},
          doi = {10.1051/0004-6361/202038939},
archivePrefix = {arXiv},
       eprint = {2009.07280},
 primaryClass = {astro-ph.GA},
       adsurl = {https://ui.adsabs.harvard.edu/abs/2020A&A...643A..90N}
}

@ARTICLE{Viaene2020,
       author = {{Viaene}, S. and {Nersesian}, A. and {Fritz}, J. and {Verstocken}, S. and {Baes}, M. and {Bianchi}, S. and {Casasola}, V. and {Cassar{\`a}}, L. and {Clark}, C. and {Davies}, J. and {De Looze}, I. and {De Vis}, P. and {Dobbels}, W. and {Galametz}, M. and {Galliano}, F. and {Jones}, A. and {Madden}, S. and {Mosenkov}, A. and {Trcka}, A. and {Xilouris}, E.~M. and {Ysard}, N.},
        title = "{High-resolution, 3D radiative transfer modelling. IV. AGN-powered dust heating in NGC 1068}",
      journal = {\aap},
         year = 2020,
        month = jun,
       volume = {638},
          eid = {A150},
        pages = {A150},
          doi = {10.1051/0004-6361/202037476},
archivePrefix = {arXiv},
       eprint = {2005.01720},
 primaryClass = {astro-ph.GA},
       adsurl = {https://ui.adsabs.harvard.edu/abs/2020A&A...638A.150V}
}

@ARTICLE{DeLooze2014,
       author = {{De Looze}, Ilse and {Fritz}, Jacopo and {Baes}, Maarten and {Bendo}, George J. and {Cortese}, Luca and {Boquien}, M{\'e}d{\'e}ric and {Boselli}, Alessandro and {Camps}, Peter and {Cooray}, Asantha and {Cormier}, Diane and {Davies}, Jon I. and {De Geyter}, Gert and {Hughes}, Thomas M. and {Jones}, Anthony P. and {Karczewski}, Oskar {\L}. and {Lebouteiller}, Vianney and {Lu}, Nanyao and {Madden}, Suzanne C. and {R{\'e}my-Ruyer}, Aur{\'e}lie and {Spinoglio}, Luigi and {Smith}, Matthew W.~L. and {Viaene}, Sebastien and {Wilson}, Christine D.},
        title = "{High-resolution, 3D radiative transfer modeling. I. The grand-design spiral galaxy M 51}",
      journal = {\aap},
         year = 2014,
        month = nov,
       volume = {571},
          eid = {A69},
        pages = {A69},
          doi = {10.1051/0004-6361/201424747},
archivePrefix = {arXiv},
       eprint = {1409.3857},
 primaryClass = {astro-ph.GA},
       adsurl = {https://ui.adsabs.harvard.edu/abs/2014A&A...571A..69D}
}

@ARTICLE{Lada2020,
       author = {{Lada}, Charles J. and {Dame}, T.~M.},
        title = "{The Mass-Size Relation and the Constancy of GMC Surface Densities in the Milky Way}",
      journal = {\apj},
         year = 2020,
        month = jul,
       volume = {898},
       number = {1},
          eid = {3},
        pages = {3},
          doi = {10.3847/1538-4357/ab9bfb},
archivePrefix = {arXiv},
       eprint = {2006.08632},
 primaryClass = {astro-ph.GA},
       adsurl = {https://ui.adsabs.harvard.edu/abs/2020ApJ...898....3L}
}

@ARTICLE{Trayford2017,
       author = {{Trayford}, James W. and {Camps}, Peter and {Theuns}, Tom and {Baes}, Maarten and {Bower}, Richard G. and {Crain}, Robert A. and {Gunawardhana}, Madusha L.~P. and {Schaller}, Matthieu and {Schaye}, Joop and {Frenk}, Carlos S.},
        title = "{Optical colours and spectral indices of z = 0.1 eagle galaxies with the 3D dust radiative transfer code skirt}",
      journal = {\mnras},
         year = 2017,
        month = sep,
       volume = {470},
       number = {1},
        pages = {771-799},
          doi = {10.1093/mnras/stx1051},
archivePrefix = {arXiv},
       eprint = {1705.02331},
 primaryClass = {astro-ph.GA},
       adsurl = {https://ui.adsabs.harvard.edu/abs/2017MNRAS.470..771T}
}

@ARTICLE{Conroy2013,
       author = {{Conroy}, Charlie},
        title = "{Modeling the Panchromatic Spectral Energy Distributions of Galaxies}",
      journal = {\araa},
         year = 2013, month = aug,
       volume = {51}, number = {1}, pages = {393-455},
       adsurl = {https://ui.adsabs.harvard.edu/abs/2013ARA&A..51..393C},
       doi = {10.1146/annurev-astro-082812-141017},
}

@ARTICLE{van2014b,
       author = {{van der Wel}, A. and {Chang}, Yu-Yen and {Bell}, E.~F. and {Holden}, B.~P. and {Ferguson}, H.~C. and {Giavalisco}, M. and {Rix}, H. -W. and {Skelton}, R. and {Whitaker}, K. and {Momcheva}, I. and {Brammer}, G. and {Kassin}, S.~A. and {Martig}, M. and {Dekel}, A. and {Ceverino}, D. and {Koo}, D.~C. and {Mozena}, M. and {van Dokkum}, P.~G. and {Franx}, M. and {Faber}, S.~M. and {Primack}, J.},
        title = "{Geometry of Star-forming Galaxies from SDSS, 3D-HST, and CANDELS}",
      journal = {\apjl},
         year = 2014,
        month = sep,
       volume = {792},
       number = {1},
          eid = {L6},
        pages = {L6},
          doi = {10.1088/2041-8205/792/1/L6},
archivePrefix = {arXiv},
       eprint = {1407.4233},
 primaryClass = {astro-ph.GA},
       adsurl = {https://ui.adsabs.harvard.edu/abs/2014ApJ...792L...6V}
}

@ARTICLE{Zhang2019,
       author = {{Zhang}, Haowen and {Primack}, Joel R. and {Faber}, S.~M. and {Koo}, David C. and {Dekel}, Avishai and {Chen}, Zhu and {Ceverino}, Daniel and {Chang}, Yu-Yen and {Fang}, Jerome J. and {Guo}, Yicheng and {Lin}, Lin and {Wel}, Arjen van der},
        title = "{The evolution of galaxy shapes in CANDELS: from prolate to discy}",
      journal = {\mnras},
         year = 2019,
        month = apr,
       volume = {484},
       number = {4},
        pages = {5170-5191},
          doi = {10.1093/mnras/stz339},
archivePrefix = {arXiv},
       eprint = {1805.12331},
 primaryClass = {astro-ph.GA},
       adsurl = {https://ui.adsabs.harvard.edu/abs/2019MNRAS.484.5170Z}
}

@ARTICLE{Camps2020,
       author = {{Camps}, P. and {Baes}, M.},
        title = "{SKIRT 9: Redesigning an advanced dust radiative transfer code to allow kinematics, line transfer and polarization by aligned dust grains}",
      journal = {Astronomy and Computing},
         year = 2020, month = apr,
       volume = {31}, eid = {100381}, pages = {100381},
       adsurl = {https://ui.adsabs.harvard.edu/abs/2020A&C....3100381C},
       doi = {10.1016/j.ascom.2020.100381},
}

@ARTICLE{Tadaki2020,
       author = {{Tadaki}, Ken-ichi and {Belli}, Sirio and {Burkert}, Andreas and {Dekel}, Avishai and {F{\"o}rster Schreiber}, Natascha M. and {Genzel}, Reinhard and {Hayashi}, Masao and {Herrera-Camus}, Rodrigo and {Kodama}, Tadayuki and {Kohno}, Kotaro and {Koyama}, Yusei and {Lee}, Minju M. and {Lutz}, Dieter and {Mowla}, Lamiya and {Nelson}, Erica J. and {Renzini}, Alvio and {Suzuki}, Tomoko L. and {Tacconi}, Linda J. and {{\"U}bler}, Hannah and {Wisnioski}, Emily and {Wuyts}, Stijn},
        title = "{Structural Evolution in Massive Galaxies at z {\ensuremath{\sim}} 2}",
      journal = {\apj},
         year = 2020,
        month = sep,
       volume = {901},
       number = {1},
          eid = {74},
        pages = {74},
          doi = {10.3847/1538-4357/abaf4a},
archivePrefix = {arXiv},
       eprint = {2009.01976},
 primaryClass = {astro-ph.GA},
       adsurl = {https://ui.adsabs.harvard.edu/abs/2020ApJ...901...74T}
}

@ARTICLE{Calzetti2000,
       author = {{Calzetti}, Daniela and {Armus}, Lee and {Bohlin}, Ralph C. and {Kinney}, Anne L. and {Koornneef}, Jan and {Storchi-Bergmann}, Thaisa},
        title = "{The Dust Content and Opacity of Actively Star-forming Galaxies}",
      journal = {\apj},
         year = 2000,
        month = apr,
       volume = {533},
       number = {2},
        pages = {682-695},
          doi = {10.1086/308692},
archivePrefix = {arXiv},
       eprint = {astro-ph/9911459},
 primaryClass = {astro-ph},
       adsurl = {https://ui.adsabs.harvard.edu/abs/2000ApJ...533..682C}
}

@ARTICLE{Jones2017,
       author = {{Jones}, A. and {Kauffmann}, G. and {D'Souza}, R. and {Bizyaev}, D. and {Law}, D. and {Haffner}, L. and {Bah{\'e}}, Y. and {Andrews}, B. and {Bershady}, M. and {Brownstein}, J. and {Bundy}, K. and {Cherinka}, B. and {Diamond-Stanic}, A. and {Drory}, N. and {Riffel}, R.~A. and {S{\'a}nchez}, S.~F. and {Thomas}, D. and {Wake}, D. and {Yan}, R. and {Zhang}, K.},
        title = "{SDSS IV MaNGA: Deep observations of extra-planar, diffuse ionized gas around late-type galaxies from stacked IFU spectra}",
      journal = {\aap},
         year = 2017,
        month = mar,
       volume = {599},
          eid = {A141},
        pages = {A141},
          doi = {10.1051/0004-6361/201629802},
archivePrefix = {arXiv},
       eprint = {1612.03920},
 primaryClass = {astro-ph.GA},
       adsurl = {https://ui.adsabs.harvard.edu/abs/2017A&A...599A.141J}
}

@ARTICLE{Reddy2018,
       author = {{Reddy}, Naveen A. and {Oesch}, Pascal A. and {Bouwens}, Rychard J. and {Montes}, Mireia and {Illingworth}, Garth D. and {Steidel}, Charles C. and {van Dokkum}, Pieter G. and {Atek}, Hakim and {Carollo}, Marcella C. and {Cibinel}, Anna and {Holden}, Brad and {Labb{\'e}}, Ivo and {Magee}, Dan and {Morselli}, Laura and {Nelson}, Erica J. and {Wilkins}, Steve},
        title = "{The HDUV Survey: A Revised Assessment of the Relationship between UV Slope and Dust Attenuation for High-redshift Galaxies}",
      journal = {\apj},
         year = 2018,
        month = jan,
       volume = {853},
       number = {1},
          eid = {56},
        pages = {56},
          doi = {10.3847/1538-4357/aaa3e7},
archivePrefix = {arXiv},
       eprint = {1705.09302},
 primaryClass = {astro-ph.GA},
       adsurl = {https://ui.adsabs.harvard.edu/abs/2018ApJ...853...56R}
}

@ARTICLE{Kriek2013,
       author = {{Kriek}, Mariska and {Conroy}, Charlie},
        title = "{The Dust Attenuation Law in Distant Galaxies: Evidence for Variation with Spectral Type}",
      journal = {\apjl},
         year = 2013,
        month = sep,
       volume = {775},
       number = {1},
          eid = {L16},
        pages = {L16},
          doi = {10.1088/2041-8205/775/1/L16},
archivePrefix = {arXiv},
       eprint = {1308.1099},
 primaryClass = {astro-ph.CO},
       adsurl = {https://ui.adsabs.harvard.edu/abs/2013ApJ...775L..16K}
}

@ARTICLE{Wild2011,
       author = {{Wild}, Vivienne and {Charlot}, St{\'e}phane and {Brinchmann}, Jarle and {Heckman}, Timothy and {Vince}, Oliver and {Pacifici}, Camilla and {Chevallard}, Jacopo},
        title = "{Empirical determination of the shape of dust attenuation curves in star-forming galaxies}",
      journal = {\mnras},
         year = 2011,
        month = nov,
       volume = {417},
       number = {3},
        pages = {1760-1786},
          doi = {10.1111/j.1365-2966.2011.19367.x},
archivePrefix = {arXiv},
       eprint = {1106.1646},
 primaryClass = {astro-ph.CO},
       adsurl = {https://ui.adsabs.harvard.edu/abs/2011MNRAS.417.1760W}
}

@ARTICLE{Salim2018,
       author = {{Salim}, Samir and {Boquien}, M{\'e}d{\'e}ric and {Lee}, Janice C.},
        title = "{Dust Attenuation Curves in the Local Universe: Demographics and New Laws for Star-forming Galaxies and High-redshift Analogs}",
      journal = {\apj},
         year = 2018,
        month = may,
       volume = {859},
       number = {1},
          eid = {11},
        pages = {11},
          doi = {10.3847/1538-4357/aabf3c},
archivePrefix = {arXiv},
       eprint = {1804.05850},
 primaryClass = {astro-ph.GA},
       adsurl = {https://ui.adsabs.harvard.edu/abs/2018ApJ...859...11S}
}

@ARTICLE{Barisic2020,
       author = {{Bari{\v{s}}i{\'c}}, Ivana and {Pacifici}, Camilla and {van der Wel}, Arjen and {Straatman}, Caroline and {Bell}, Eric F. and {Bezanson}, Rachel and {Brammer}, Gabriel and {D'Eugenio}, Francesco and {Franx}, Marijn and {van Houdt}, Josha and {Maseda}, Michael V. and {Muzzin}, Adam and {Sobral}, David and {Wu}, Po-Feng},
        title = "{Dust Attenuation Curves at z {\ensuremath{\sim}} 0.8 from LEGA-C: Precise Constraints on the Slope and 2175{\v{S}}Bump Strength}",
      journal = {\apj},
         year = 2020,
        month = nov,
       volume = {903},
       number = {2},
          eid = {146},
        pages = {146},
          doi = {10.3847/1538-4357/abba37},
archivePrefix = {arXiv},
       eprint = {2010.01147},
 primaryClass = {astro-ph.GA},
       adsurl = {https://ui.adsabs.harvard.edu/abs/2020ApJ...903..146B}
}

@ARTICLE{Reddy2023,
       author = {{Reddy}, Naveen A. and {Topping}, Michael W. and {Sanders}, Ryan L. and {Shapley}, Alice E. and {Brammer}, Gabriel},
        title = "{Paschen-line Constraints on Dust Attenuation and Star Formation at z   1-3 with JWST/NIRSpec}",
      journal = {\apj},
         year = 2023,
        month = may,
       volume = {948},
       number = {2},
          eid = {83},
        pages = {83},
          doi = {10.3847/1538-4357/acc869},
archivePrefix = {arXiv},
       eprint = {2301.07249},
 primaryClass = {astro-ph.GA},
       adsurl = {https://ui.adsabs.harvard.edu/abs/2023ApJ...948...83R}
}

@ARTICLE{Miller2023,
       author = {{Miller}, Tim B. and {van Dokkum}, Pieter and {Mowla}, Lamiya},
        title = "{Color Gradients and Half-mass Radii of Galaxies Out to z = 2 in the CANDELS/3D-HST Fields: Further Evidence for Important Differences in the Evolution of Mass-weighted and Light-weighted Sizes}",
      journal = {\apj},
         year = 2023, month = mar,
       volume = {945}, number = {2}, eid = {155}, pages = {155},
       adsurl = {https://ui.adsabs.harvard.edu/abs/2023ApJ...945..155M},
}

@ARTICLE{Miller2022,
       author = {{Miller}, Tim B. and {Whitaker}, Katherine E. and {Nelson}, Erica J. and {van Dokkum}, Pieter and {Bezanson}, Rachel and {Brammer}, Gabriel and {Heintz}, Kasper E. and {Leja}, Joel and {Suess}, Katherine A. and {Weaver}, John R.},
        title = "{Early JWST Imaging Reveals Strong Optical and NIR Color Gradients in Galaxies at z   2 Driven Mostly by Dust}",
      journal = {\apjl},
         year = 2022,
        month = dec,
       volume = {941},
       number = {2},
          eid = {L37},
        pages = {L37},
          doi = {10.3847/2041-8213/aca675},
archivePrefix = {arXiv},
       eprint = {2209.12954},
 primaryClass = {astro-ph.GA},
       adsurl = {https://ui.adsabs.harvard.edu/abs/2022ApJ...941L..37M}
}

@ARTICLE{Tacchella2018,
       author = {{Tacchella}, S. and {Carollo}, C.~M. and {F{\"o}rster Schreiber}, N.~M. and {Renzini}, A. and {Dekel}, A. and {Genzel}, R. and {Lang}, P. and {Lilly}, S.~J. and {Mancini}, C. and {Onodera}, M. and {Tacconi}, L.~J. and {Wuyts}, S. and {Zamorani}, G.},
        title = "{Dust Attenuation, Bulge Formation, and Inside-out Quenching of Star Formation in Star-forming Main Sequence Galaxies at z {\ensuremath{\sim}} 2}",
      journal = {\apj},
         year = 2018,
        month = may,
       volume = {859},
       number = {1},
          eid = {56},
        pages = {56},
          doi = {10.3847/1538-4357/aabf8b},
archivePrefix = {arXiv},
       eprint = {1704.00733},
 primaryClass = {astro-ph.GA},
       adsurl = {https://ui.adsabs.harvard.edu/abs/2018ApJ...859...56T}
}

@ARTICLE{Sethuram2023,
       author = {{Sethuram}, Snigdaa S. and {Cochrane}, Rachel K. and {Hayward}, Christopher C. and {Acquaviva}, Viviana and {Villaescusa-Navarro}, Francisco and {Popping}, Gerg{\"o} and {Wise}, John H.},
        title = "{Emulating radiative transfer with artificial neural networks}",
      journal = {\mnras},
         year = 2023,
        month = dec,
       volume = {526},
       number = {3},
        pages = {4520-4528},
          doi = {10.1093/mnras/stad2524},
archivePrefix = {arXiv},
       eprint = {2308.13648},
 primaryClass = {astro-ph.GA},
       adsurl = {https://ui.adsabs.harvard.edu/abs/2023MNRAS.526.4520S}
}

@ARTICLE{Alsing2024,
       author = {{Alsing}, Justin and {Thorp}, Stephen and {Deger}, Sinan and {Peiris}, Hiranya V. and {Leistedt}, Boris and {Mortlock}, Daniel and {Leja}, Joel},
        title = "{pop-cosmos: A Comprehensive Picture of the Galaxy Population from COSMOS Data}",
      journal = {\apjs},
         year = 2024,
        month = sep,
       volume = {274},
       number = {1},
          eid = {12},
        pages = {12},
          doi = {10.3847/1538-4365/ad5c69},
archivePrefix = {arXiv},
       eprint = {2402.00935},
 primaryClass = {astro-ph.GA},
       adsurl = {https://ui.adsabs.harvard.edu/abs/2024ApJS..274...12A}
}

@ARTICLE{Wuyts2009a,
       author = {{Wuyts}, Stijn and {Franx}, Marijn and {Cox}, Thomas J. and {Hernquist}, Lars and {Hopkins}, Philip F. and {Robertson}, Brant E. and {van Dokkum}, Pieter G.},
        title = "{Recovering Stellar Population Properties and Redshifts from Broadband Photometry of Simulated Galaxies: Lessons for SED Modeling}",
      journal = {\apj},
         year = 2009,
        month = may,
       volume = {696},
       number = {1},
        pages = {348-369},
          doi = {10.1088/0004-637X/696/1/348},
archivePrefix = {arXiv},
       eprint = {0901.4337},
 primaryClass = {astro-ph.CO},
       adsurl = {https://ui.adsabs.harvard.edu/abs/2009ApJ...696..348W}
}

@ARTICLE{Muzzin2013,
       author = {{Muzzin}, Adam and {Marchesini}, Danilo and {Stefanon}, Mauro and {Franx}, Marijn and {Milvang-Jensen}, Bo and {Dunlop}, James S. and {Fynbo}, J.~P.~U. and {Brammer}, Gabriel and {Labb{\'e}}, Ivo and {van Dokkum}, Pieter},
        title = "{A Public K$_{s}$ -selected Catalog in the COSMOS/ULTRAVISTA Field: Photometry, Photometric Redshifts, and Stellar Population Parameters}",
      journal = {\apjs},
         year = 2013,
        month = may,
       volume = {206},
       number = {1},
          eid = {8},
        pages = {8},
          doi = {10.1088/0067-0049/206/1/8},
archivePrefix = {arXiv},
       eprint = {1303.4410},
 primaryClass = {astro-ph.CO},
       adsurl = {https://ui.adsabs.harvard.edu/abs/2013ApJS..206....8M}
}

@ARTICLE{Diemer2017,
       author = {{Diemer}, Benedikt and {Sparre}, Martin and {Abramson}, Louis E. and {Torrey}, Paul},
        title = "{Log-normal Star Formation Histories in Simulated and Observed Galaxies}",
      journal = {\apj},
         year = 2017,
        month = apr,
       volume = {839},
       number = {1},
          eid = {26},
        pages = {26},
          doi = {10.3847/1538-4357/aa68e5},
archivePrefix = {arXiv},
       eprint = {1701.02308},
 primaryClass = {astro-ph.GA},
       adsurl = {https://ui.adsabs.harvard.edu/abs/2017ApJ...839...26D}
}

@ARTICLE{Brammer2008,
       author = {{Brammer}, Gabriel B. and {van Dokkum}, Pieter G. and {Coppi}, Paolo},
        title = "{EAZY: A Fast, Public Photometric Redshift Code}",
      journal = {\apj},
         year = 2008, month = oct,
       volume = {686}, number = {2}, pages = {1503-1513},
       adsurl = {https://ui.adsabs.harvard.edu/abs/2008ApJ...686.1503B},
       doi = {10.1086/591786},
}

@ARTICLE{Wuyts2012,
       author = {{Wuyts}, Stijn and {F{\"o}rster Schreiber}, Natascha M. and {Genzel}, Reinhard and {Guo}, Yicheng and {Barro}, Guillermo and {Bell}, Eric F. and {Dekel}, Avishai and {Faber}, Sandra M. and {Ferguson}, Henry C. and {Giavalisco}, Mauro and {Grogin}, Norman A. and {Hathi}, Nimish P. and {Huang}, Kuang-Han and {Kocevski}, Dale D. and {Koekemoer}, Anton M. and {Koo}, David C. and {Lotz}, Jennifer and {Lutz}, Dieter and {McGrath}, Elizabeth and {Newman}, Jeffrey A. and {Rosario}, David and {Saintonge}, Amelie and {Tacconi}, Linda J. and {Weiner}, Benjamin J. and {van der Wel}, Arjen},
        title = "{Smooth(er) Stellar Mass Maps in CANDELS: Constraints on the Longevity of Clumps in High-redshift Star-forming Galaxies}",
      journal = {\apj},
         year = 2012,
        month = jul,
       volume = {753},
       number = {2},
          eid = {114},
        pages = {114},
          doi = {10.1088/0004-637X/753/2/114},
archivePrefix = {arXiv},
       eprint = {1203.2611},
 primaryClass = {astro-ph.CO},
       adsurl = {https://ui.adsabs.harvard.edu/abs/2012ApJ...753..114W}
}

@ARTICLE{Bruzual2003,
       author = {{Bruzual}, G. and {Charlot}, S.},
        title = "{Stellar population synthesis at the resolution of 2003}",
      journal = {\mnras},
         year = 2003, month = oct,
       volume = {344}, number = {4}, pages = {1000-1028},
       adsurl = {https://ui.adsabs.harvard.edu/abs/2003MNRAS.344.1000B},
       doi = {10.1046/j.1365-8711.2003.06897.x},
}

@ARTICLE{Chabrier2003,
       author = {{Chabrier}, Gilles},
        title = "{Galactic Stellar and Substellar Initial Mass Function}",
      journal = {\pasp},
         year = 2003, month = jul,
       volume = {115}, number = {809}, pages = {763-795},
       adsurl = {https://ui.adsabs.harvard.edu/abs/2003PASP..115..763C},
       doi = {10.1086/376392},
}

@ARTICLE{Foreman-Mackey2013,
       author = {{Foreman-Mackey}, Daniel and {Hogg}, David W. and {Lang}, Dustin and {Goodman}, Jonathan},
        title = "{emcee: The MCMC Hammer}",
      journal = {\pasp},
         year = 2013,
        month = mar,
       volume = {125},
       number = {925},
        pages = {306},
          doi = {10.1086/670067},
archivePrefix = {arXiv},
       eprint = {1202.3665},
 primaryClass = {astro-ph.IM},
       adsurl = {https://ui.adsabs.harvard.edu/abs/2013PASP..125..306F}
}

@ARTICLE{Magnelli2023,
       author = {{Magnelli}, Benjamin and {G{\'o}mez-Guijarro}, Carlos and {Elbaz}, David and {Daddi}, Emanuele and {Papovich}, Casey and {Shen}, Lu and {Arrabal Haro}, Pablo and {Bagley}, Micaela B. and {Bell}, Eric F. and {Buat}, V{\'e}ronique and {Costantin}, Luca and {Dickinson}, Mark and {Finkelstein}, Steven L. and {Gardner}, Jonathan P. and {Jim{\'e}nez-Andrade}, Eric F. and {Kartaltepe}, Jeyhan S. and {Koekemoer}, Anton M. and {Lyu}, Yipeng and {P{\'e}rez-Gonz{\'a}lez}, Pablo G. and {Pirzkal}, Nor and {Tacchella}, Sandro and {de la Vega}, Alexander and {Wuyts}, Stijn and {Yang}, Guang and {Yung}, L.~Y. Aaron and {Zavala}, Jorge},
        title = "{CEERS: MIRI deciphers the spatial distribution of dust-obscured star formation in galaxies at 0.1 < z < 2.5}",
      journal = {\aap},
         year = 2023,
        month = oct,
       volume = {678},
          eid = {A83},
        pages = {A83},
          doi = {10.1051/0004-6361/202347052},
archivePrefix = {arXiv},
       eprint = {2305.19331},
 primaryClass = {astro-ph.GA},
       adsurl = {https://ui.adsabs.harvard.edu/abs/2023A&A...678A..83M}
}

@ARTICLE{Boquien2019,
       author = {{Boquien}, M. and {Burgarella}, D. and {Roehlly}, Y. and {Buat}, V. and {Ciesla}, L. and {Corre}, D. and {Inoue}, A.~K. and {Salas}, H.},
        title = "{CIGALE: a python Code Investigating GALaxy Emission}",
      journal = {\aap},
         year = 2019,
        month = feb,
       volume = {622},
          eid = {A103},
        pages = {A103},
          doi = {10.1051/0004-6361/201834156},
archivePrefix = {arXiv},
       eprint = {1811.03094},
 primaryClass = {astro-ph.GA},
       adsurl = {https://ui.adsabs.harvard.edu/abs/2019A&A...622A.103B}
}

@ARTICLE{daCunha2008,
       author = {{da Cunha}, Elisabete and {Charlot}, St{\'e}phane and {Elbaz}, David},
        title = "{A simple model to interpret the ultraviolet, optical and infrared emission from galaxies}",
      journal = {\mnras},
         year = 2008,
        month = aug,
       volume = {388},
       number = {4},
        pages = {1595-1617},
          doi = {10.1111/j.1365-2966.2008.13535.x},
archivePrefix = {arXiv},
       eprint = {0806.1020},
 primaryClass = {astro-ph},
       adsurl = {https://ui.adsabs.harvard.edu/abs/2008MNRAS.388.1595D}
}

@ARTICLE{Lovell2022,
       author = {{Lovell}, C.~C. and {Geach}, J.~E. and {Dav{\'e}}, R. and {Narayanan}, D. and {Coppin}, K.~E.~K. and {Li}, Q. and {Franco}, M. and {Privon}, G.~C.},
        title = "{An orientation bias in observations of submillimetre galaxies}",
      journal = {\mnras},
         year = 2022,
        month = sep,
       volume = {515},
       number = {3},
        pages = {3644-3655},
          doi = {10.1093/mnras/stac2008},
archivePrefix = {arXiv},
       eprint = {2106.11588},
 primaryClass = {astro-ph.GA},
       adsurl = {https://ui.adsabs.harvard.edu/abs/2022MNRAS.515.3644L}
}

@ARTICLE{Qin2022,
       author = {{Qin}, Jianbo and {Zheng}, Xian Zhong and {Fang}, Min and {Pan}, Zhizheng and {Wuyts}, Stijn and {Shi}, Yong and {Peng}, Yingjie and {Gonzalez}, Valentino and {Bian}, Fuyan and {Huang}, Jia-Sheng and {Gu}, Qiu-Sheng and {Liu}, Wenhao and {Tan}, Qinghua and {Shi}, Dong Dong and {Ren}, Jian and {Zhang}, Yuheng and {Qiao}, Man and {Wen}, Run and {Liu}, Shuang},
        title = "{Systematic biases in determining dust attenuation curves through galaxy SED fitting}",
      journal = {\mnras},
         year = 2022,
        month = mar,
       volume = {511},
       number = {1},
        pages = {765-783},
          doi = {10.1093/mnras/stac132},
archivePrefix = {arXiv},
       eprint = {2201.05467},
 primaryClass = {astro-ph.GA},
       adsurl = {https://ui.adsabs.harvard.edu/abs/2022MNRAS.511..765Q}
}

@ARTICLE{Salim2020,
       author = {{Salim}, Samir and {Narayanan}, Desika},
        title = "{The Dust Attenuation Law in Galaxies}",
      journal = {\araa},
         year = 2020,
        month = aug,
       volume = {58},
        pages = {529-575},
          doi = {10.1146/annurev-astro-032620-021933},
archivePrefix = {arXiv},
       eprint = {2001.03181},
 primaryClass = {astro-ph.GA},
       adsurl = {https://ui.adsabs.harvard.edu/abs/2020ARA&A..58..529S}
}

@ARTICLE{Steinacker2013,
       author = {{Steinacker}, J{\"u}rgen and {Baes}, Maarten and {Gordon}, Karl D.},
        title = "{Three-Dimensional Dust Radiative Transfer*}",
      journal = {\araa},
         year = 2013,
        month = aug,
       volume = {51},
       number = {1},
        pages = {63-104},
          doi = {10.1146/annurev-astro-082812-141042},
archivePrefix = {arXiv},
       eprint = {1303.4998},
 primaryClass = {astro-ph.IM},
       adsurl = {https://ui.adsabs.harvard.edu/abs/2013ARA&A..51...63S}
}

@ARTICLE{Camps2015a,
       author = {{Camps}, P. and {Baes}, M.},
        title = "{SKIRT: An advanced dust radiative transfer code with a user-friendly architecture}",
      journal = {Astronomy and Computing},
         year = 2015,
        month = mar,
       volume = {9},
        pages = {20-33},
          doi = {10.1016/j.ascom.2014.10.004},
archivePrefix = {arXiv},
       eprint = {1410.1629},
 primaryClass = {astro-ph.IM},
       adsurl = {https://ui.adsabs.harvard.edu/abs/2015A&C.....9...20C}
}

@ARTICLE{Liu2017,
       author = {{Liu}, F.~S. and {Jiang}, Dongfei and {Faber}, S.~M. and {Koo}, David C. and {Yesuf}, Hassen M. and {Tacchella}, Sandro and {Mao}, Shude and {Wang}, Weichen and {Guo}, Yicheng and {Fang}, Jerome J. and {Barro}, Guillermo and {Zheng}, Xianzhong and {Jia}, Meng and {Tong}, Wei and {Liu}, Lu and {Meng}, Xianmin},
        title = "{The Origins of UV-optical Color Gradients in Star-forming Galaxies at z {\ensuremath{\sim}} 2: Predominant Dust Gradients but Negligible sSFR Gradients}",
      journal = {\apjl},
         year = 2017,
        month = jul,
       volume = {844},
       number = {1},
          eid = {L2},
        pages = {L2},
          doi = {10.3847/2041-8213/aa7cf5},
archivePrefix = {arXiv},
       eprint = {1707.00226},
 primaryClass = {astro-ph.GA},
       adsurl = {https://ui.adsabs.harvard.edu/abs/2017ApJ...844L...2L}
}

@ARTICLE{Suess2019,
       author = {{Suess}, Katherine A. and {Kriek}, Mariska and {Price}, Sedona H. and {Barro}, Guillermo},
        title = "{Half-mass Radii for {\ensuremath{\sim}}7000 Galaxies at 1.0 {\ensuremath{\leq}} z {\ensuremath{\leq}} 2.5: Most of the Evolution in the Mass-Size Relation Is Due to Color Gradients}",
      journal = {\apj},
         year = 2019,
        month = jun,
       volume = {877},
       number = {2},
          eid = {103},
        pages = {103},
          doi = {10.3847/1538-4357/ab1bda},
archivePrefix = {arXiv},
       eprint = {1904.10992},
 primaryClass = {astro-ph.GA},
       adsurl = {https://ui.adsabs.harvard.edu/abs/2019ApJ...877..103S}
}

@ARTICLE{Liu2016,
       author = {{Liu}, F.~S. and {Jiang}, Dongfei and {Guo}, Yicheng and {Koo}, David C. and {Faber}, S.~M. and {Zheng}, Xianzhong and {Yesuf}, Hassen M. and {Barro}, Guillermo and {Li}, Yao and {Li}, Dingpeng and {Wang}, Weichen and {Mao}, Shude and {Fang}, Jerome J.},
        title = "{The UV-Optical Color Gradients in Star-forming Galaxies at 0.5 < z < 1.5: Origins and Link to Galaxy Assembly}",
      journal = {\apjl},
         year = 2016,
        month = may,
       volume = {822},
       number = {2},
          eid = {L25},
        pages = {L25},
          doi = {10.3847/2041-8205/822/2/L25},
archivePrefix = {arXiv},
       eprint = {1604.05780},
 primaryClass = {astro-ph.GA},
       adsurl = {https://ui.adsabs.harvard.edu/abs/2016ApJ...822L..25L}
}

@ARTICLE{Trayford2020,
       author = {{Trayford}, James W. and {Lagos}, Claudia del P. and {Robotham}, Aaron S.~G. and {Obreschkow}, Danail},
        title = "{Fade to grey: systematic variation of galaxy attenuation curves with galaxy properties in the EAGLE simulations}",
      journal = {\mnras},
         year = 2020,
        month = jan,
       volume = {491},
       number = {3},
        pages = {3937-3951},
          doi = {10.1093/mnras/stz3234},
archivePrefix = {arXiv},
       eprint = {1908.08956},
 primaryClass = {astro-ph.GA},
       adsurl = {https://ui.adsabs.harvard.edu/abs/2020MNRAS.491.3937T}
}

@ARTICLE{Markov2025,
       author = {{Markov}, V. and {Gallerani}, S. and {Pallottini}, A. and {Brada{\v{c}}}, M. and {Carniani}, S. and {Tripodi}, R. and {Noirot}, G. and {Di Mascia}, F. and {Parlanti}, E. and {Martis}, N.},
        title = "{Unveiling the trends between dust attenuation and galaxy properties at z {\ensuremath{\sim}} 2{\ensuremath{-}}12 with the James Webb Space Telescope}",
      journal = {\aap},
         year = 2025,
        month = oct,
       volume = {702},
          eid = {A33},
        pages = {A33},
          doi = {10.1051/0004-6361/202555182},
archivePrefix = {arXiv},
       eprint = {2504.12378},
 primaryClass = {astro-ph.GA},
       adsurl = {https://ui.adsabs.harvard.edu/abs/2025A&A...702A..33M}
}

@ARTICLE{Ormerod2025,
       author = {{Ormerod}, Katherine and {Witstok}, Joris and {Smit}, Renske and {de Graaff}, Anna and {Helton}, Jakob M. and {Maseda}, Michael V. and {Shivaei}, Irene and {Bunker}, Andrew J. and {Carniani}, Stefano and {D'Eugenio}, Francesco and {Bhatawdekar}, Rachana and {Chevallard}, Jacopo and {Franx}, Marijn and {Kumari}, Nimisha and {Maiolino}, Roberto and {Rinaldi}, Pierluigi and {Robertson}, Brant and {Tacchella}, Sandro},
        title = "{Detection of the 2175 {\r{A}} UV bump at z > 7: evidence for rapid dust evolution in a merging reionization-era galaxy}",
      journal = {\mnras},
         year = 2025,
        month = sep,
       volume = {542},
       number = {2},
        pages = {1136-1154},
          doi = {10.1093/mnras/staf1228},
archivePrefix = {arXiv},
       eprint = {2502.21119},
 primaryClass = {astro-ph.GA},
       adsurl = {https://ui.adsabs.harvard.edu/abs/2025MNRAS.542.1136O}
}

@ARTICLE{Belfiore2019,
       author = {{Belfiore}, F. and {Vincenzo}, F. and {Maiolino}, R. and {Matteucci}, F.},
        title = "{From `bathtub' galaxy evolution models to metallicity gradients}",
      journal = {\mnras},
         year = 2019,
        month = jul,
       volume = {487},
       number = {1},
        pages = {456-474},
          doi = {10.1093/mnras/stz1165},
archivePrefix = {arXiv},
       eprint = {1903.05105},
 primaryClass = {astro-ph.GA},
       adsurl = {https://ui.adsabs.harvard.edu/abs/2019MNRAS.487..456B}
}

@ARTICLE{Zelko2020,
       author = {{Zelko}, Ioana A. and {Finkbeiner}, Douglas P.},
        title = "{Implications of Grain Size Distribution and Composition for the Correlation between Dust Extinction and Emissivity}",
      journal = {\apj},
         year = 2020,
        month = nov,
       volume = {904},
       number = {1},
          eid = {38},
        pages = {38},
          doi = {10.3847/1538-4357/abbb8d},
archivePrefix = {arXiv},
       eprint = {2009.11869},
 primaryClass = {astro-ph.GA},
       adsurl = {https://ui.adsabs.harvard.edu/abs/2020ApJ...904...38Z}
}

@ARTICLE{Lower2022,
       author = {{Lower}, Sidney and {Narayanan}, Desika and {Leja}, Joel and {Johnson}, Benjamin D. and {Conroy}, Charlie and {Dav{\'e}}, Romeel},
        title = "{How Well Can We Measure Galaxy Dust Attenuation Curves? The Impact of the Assumed Star-dust Geometry Model in Spectral Energy Distribution Fitting}",
      journal = {\apj},
         year = 2022,
        month = may,
       volume = {931},
       number = {1},
          eid = {14},
        pages = {14},
          doi = {10.3847/1538-4357/ac6959},
archivePrefix = {arXiv},
       eprint = {2203.00074},
 primaryClass = {astro-ph.GA},
       adsurl = {https://ui.adsabs.harvard.edu/abs/2022ApJ...931...14L}
}

@ARTICLE{Makiya2022,
       author = {{Makiya}, Ryu and {Hirashita}, Hiroyuki},
        title = "{Cosmic evolution of grain size distribution in galaxies using the {\ensuremath{\nu}}$^{2}$GC semi-analytical model}",
      journal = {\mnras},
         year = 2022,
        month = dec,
       volume = {517},
       number = {2},
        pages = {2076-2087},
          doi = {10.1093/mnras/stac2762},
archivePrefix = {arXiv},
       eprint = {2210.06176},
 primaryClass = {astro-ph.GA},
       adsurl = {https://ui.adsabs.harvard.edu/abs/2022MNRAS.517.2076M}
}

@ARTICLE{Papovich2001,
       author = {{Papovich}, Casey and {Dickinson}, Mark and {Ferguson}, Henry C.},
        title = "{The Stellar Populations and Evolution of Lyman Break Galaxies}",
      journal = {\apj},
         year = 2001,
        month = oct,
       volume = {559},
       number = {2},
        pages = {620-653},
          doi = {10.1086/322412},
archivePrefix = {arXiv},
       eprint = {astro-ph/0105087},
 primaryClass = {astro-ph},
       adsurl = {https://ui.adsabs.harvard.edu/abs/2001ApJ...559..620P}
}

@ARTICLE{Maraston2010,
       author = {{Maraston}, Claudia and {Pforr}, Janine and {Renzini}, Alvio and {Daddi}, Emanuele and {Dickinson}, Mark and {Cimatti}, Andrea and {Tonini}, Chiara},
        title = "{Star formation rates and masses of z \raisebox{-0.5ex}\textasciitilde 2 galaxies from multicolour photometry}",
      journal = {\mnras},
         year = 2010,
        month = sep,
       volume = {407},
       number = {2},
        pages = {830-845},
          doi = {10.1111/j.1365-2966.2010.16973.x},
archivePrefix = {arXiv},
       eprint = {1004.4546},
 primaryClass = {astro-ph.CO},
       adsurl = {https://ui.adsabs.harvard.edu/abs/2010MNRAS.407..830M}
}

@ARTICLE{Narayanan2024,
       author = {{Narayanan}, Desika and {Lower}, Sidney and {Torrey}, Paul and {Brammer}, Gabriel and {Cui}, Weiguang and {Dav{\'e}}, Romeel and {Iyer}, Kartheik G. and {Li}, Qi and {Lovell}, Christopher C. and {Sales}, Laura V. and {Stark}, Daniel P. and {Marinacci}, Federico and {Vogelsberger}, Mark},
        title = "{Outshining by Recent Star Formation Prevents the Accurate Measurement of High-z Galaxy Stellar Masses}",
      journal = {\apj},
         year = 2024,
        month = jan,
       volume = {961},
       number = {1},
          eid = {73},
        pages = {73},
          doi = {10.3847/1538-4357/ad0966},
archivePrefix = {arXiv},
       eprint = {2306.10118},
 primaryClass = {astro-ph.GA},
       adsurl = {https://ui.adsabs.harvard.edu/abs/2024ApJ...961...73N}
}

@ARTICLE{Pforr2012,
       author = {{Pforr}, Janine and {Maraston}, Claudia and {Tonini}, Chiara},
        title = "{Recovering galaxy stellar population properties from broad-band spectral energy distribution fitting}",
      journal = {\mnras},
         year = 2012,
        month = jun,
       volume = {422},
       number = {4},
        pages = {3285-3326},
          doi = {10.1111/j.1365-2966.2012.20848.x},
archivePrefix = {arXiv},
       eprint = {1203.3548},
 primaryClass = {astro-ph.CO},
       adsurl = {https://ui.adsabs.harvard.edu/abs/2012MNRAS.422.3285P}
}

@ARTICLE{Leja2019,
       author = {{Leja}, Joel and {Carnall}, Adam C. and {Johnson}, Benjamin D. and {Conroy}, Charlie and {Speagle}, Joshua S.},
        title = "{How to Measure Galaxy Star Formation Histories. II. Nonparametric Models}",
      journal = {\apj},
         year = 2019,
        month = may,
       volume = {876},
       number = {1},
          eid = {3},
        pages = {3},
          doi = {10.3847/1538-4357/ab133c},
archivePrefix = {arXiv},
       eprint = {1811.03637},
 primaryClass = {astro-ph.GA},
       adsurl = {https://ui.adsabs.harvard.edu/abs/2019ApJ...876....3L}
}

@ARTICLE{Carnall2018,
       author = {{Carnall}, A.~C. and {McLure}, R.~J. and {Dunlop}, J.~S. and {Dav{\'e}}, R.},
        title = "{Inferring the star formation histories of massive quiescent galaxies with BAGPIPES: evidence for multiple quenching mechanisms}",
      journal = {\mnras},
         year = 2018,
        month = nov,
       volume = {480},
       number = {4},
        pages = {4379-4401},
          doi = {10.1093/mnras/sty2169},
archivePrefix = {arXiv},
       eprint = {1712.04452},
 primaryClass = {astro-ph.GA},
       adsurl = {https://ui.adsabs.harvard.edu/abs/2018MNRAS.480.4379C}
}

@ARTICLE{Tadaki2017a,
       author = {{Tadaki}, Ken-ichi and {Genzel}, Reinhard and {Kodama}, Tadayuki and {Wuyts}, Stijn and {Wisnioski}, Emily and {F{\"o}rster Schreiber}, Natascha M. and {Burkert}, Andreas and {Lang}, Philipp and {Tacconi}, Linda J. and {Lutz}, Dieter and {Belli}, Sirio and {Davies}, Richard I. and {Hatsukade}, Bunyo and {Hayashi}, Masao and {Herrera-Camus}, Rodrigo and {Ikarashi}, Soh and {Inoue}, Shigeki and {Kohno}, Kotaro and {Koyama}, Yusei and {Mendel}, J. Trevor and {Nakanishi}, Kouichiro and {Shimakawa}, Rhythm and {Suzuki}, Tomoko L. and {Tamura}, Yoichi and {Tanaka}, Ichi and {{\"U}bler}, Hannah and {Wilman}, Dave J.},
        title = "{Bulge-forming Galaxies with an Extended Rotating Disk at z \raisebox{-0.5ex}\textasciitilde 2}",
      journal = {\apj},
         year = 2017,
        month = jan,
       volume = {834},
       number = {2},
          eid = {135},
        pages = {135},
          doi = {10.3847/1538-4357/834/2/135},
archivePrefix = {arXiv},
       eprint = {1608.05412},
 primaryClass = {astro-ph.GA},
       adsurl = {https://ui.adsabs.harvard.edu/abs/2017ApJ...834..135T}
}

@ARTICLE{Gladders2013,
       author = {{Gladders}, Michael D. and {Oemler}, Augustus and {Dressler}, Alan and {Poggianti}, Bianca and {Vulcani}, Benedetta and {Abramson}, Louis},
        title = "{The IMACS Cluster Building Survey. IV. The Log-normal Star Formation History of Galaxies}",
      journal = {\apj},
         year = 2013,
        month = jun,
       volume = {770},
       number = {1},
          eid = {64},
        pages = {64},
          doi = {10.1088/0004-637X/770/1/64},
archivePrefix = {arXiv},
       eprint = {1303.3917},
 primaryClass = {astro-ph.CO},
       adsurl = {https://ui.adsabs.harvard.edu/abs/2013ApJ...770...64G}
}

@ARTICLE{Chen2024,
       author = {{Chen}, Minshuo and {Mei}, Song and {Fan}, Jianqing and {Wang}, Mengdi},
        title = "{An Overview of Diffusion Models: Applications, Guided Generation, Statistical Rates and Optimization}",
      journal = {arXiv e-prints},
         year = 2024,
        month = apr,
          eid = {arXiv:2404.07771},
        pages = {arXiv:2404.07771},
          doi = {10.48550/arXiv.2404.07771},
archivePrefix = {arXiv},
       eprint = {2404.07771},
 primaryClass = {cs.LG},
       adsurl = {https://ui.adsabs.harvard.edu/abs/2024arXiv240407771C}
}

@ARTICLE{RinoSilvestre2022,
author = {Rino-Silvestre, João and González-Gaitán, Santiago and Stalevski, Marko and Smole, Majda and Guilherme-Garcia, Pedro and Carvalho, Joao and Mourao, Ana},
year = {2022},
month = {12},
pages = {7719-7760},
title = {EmulART: Emulating radiative transfer—a pilot study on autoencoder-based dimensionality reduction for radiative transfer models},
volume = {35},
journal = {Neural Computing and Applications},
doi = {10.1007/s00521-022-08071-x}
}

@article{Goodfellow2014,
author = {Goodfellow, Ian and Pouget-Abadie, Jean and Mirza, Mehdi and Xu, Bing and Warde-Farley, David and Ozair, Sherjil and Courville, Aaron and Bengio, Yoshua},
title = {Generative adversarial networks},
year = {2020},
issue_date = {November 2020},
publisher = {Association for Computing Machinery},
address = {New York, NY, USA},
volume = {63},
number = {11},
issn = {0001-0782},
url = {https://doi.org/10.1145/3422622},
doi = {10.1145/3422622},
journal = {Commun. ACM},
month = oct,
pages = {139–144},
numpages = {6}
}

@ARTICLE{Wuyts2009b,
       author = {{Wuyts}, Stijn and {Franx}, Marijn and {Cox}, Thomas J. and {F{\"o}rster Schreiber}, Natascha M. and {Hayward}, Christopher C. and {Hernquist}, Lars and {Hopkins}, Philip F. and {Labb{\'e}}, Ivo and {Marchesini}, Danilo and {Robertson}, Brant E. and {Toft}, Sune and {van Dokkum}, Pieter G.},
        title = "{Color Distributions, Number, and Mass Densities of Massive Galaxies at 1.5 < z < 3: Comparing Observations with Merger Simulations}",
      journal = {\apj},
         year = 2009,
        month = jul,
       volume = {700},
       number = {1},
        pages = {799-819},
          doi = {10.1088/0004-637X/700/1/799},
archivePrefix = {arXiv},
       eprint = {0905.2411},
 primaryClass = {astro-ph.CO},
       adsurl = {https://ui.adsabs.harvard.edu/abs/2009ApJ...700..799W}
}

@ARTICLE{Gebek2025,
       author = {{Gebek}, Andrea and {Diemer}, Benedikt and {Martorano}, Marco and {van der Wel}, Arjen and {Pantoni}, Lara and {Baes}, Maarten and {Gabrielpillai}, Austen and {Utsav Kapoor}, Anand and {Osinga}, Calvin and {Nersesian}, Angelos and {Matsumoto}, Kosei and {Gordon}, Karl},
        title = "{The mass-dependent UVJ diagram at cosmic noon: A challenge for galaxy evolution models and dust radiative transfer}",
      journal = {\aap},
         year = 2025,
        month = mar,
       volume = {695},
          eid = {A90},
        pages = {A90},
          doi = {10.1051/0004-6361/202452768},
archivePrefix = {arXiv},
       eprint = {2501.12008},
 primaryClass = {astro-ph.GA},
       adsurl = {https://ui.adsabs.harvard.edu/abs/2025A&A...695A..90G}
}

@ARTICLE{MargalefBentabol2024,
       author = {{Margalef-Bentabol}, B. and {Wang}, L. and {La Marca}, A. and {Blanco-Prieto}, C. and {Chudy}, D. and {Dom{\'\i}nguez-S{\'a}nchez}, H. and {Goulding}, A.~D. and {Guzm{\'a}n-Ortega}, A. and {Huertas-Company}, M. and {Martin}, G. and {Pearson}, W.~J. and {Rodriguez-Gomez}, V. and {Walmsley}, M. and {Bickley}, R.~W. and {Bottrell}, C. and {Conselice}, C. and {O'Ryan}, D.},
        title = "{Galaxy merger challenge: A comparison study between machine learning-based detection methods}",
      journal = {\aap},
         year = 2024,
        month = jul,
       volume = {687},
          eid = {A24},
        pages = {A24},
          doi = {10.1051/0004-6361/202348239},
archivePrefix = {arXiv},
       eprint = {2403.15118},
 primaryClass = {astro-ph.GA},
       adsurl = {https://ui.adsabs.harvard.edu/abs/2024A&A...687A..24M}
}

@ARTICLE{AvirettMackenzie2024,
       author = {{Avirett-Mackenzie}, M.~S. and {Villforth}, C. and {Huertas-Company}, M. and {Wuyts}, S. and {Alexander}, D.~M. and {Bonoli}, S. and {Lapi}, A. and {Lopez}, I.~E. and {Ramos Almeida}, C. and {Shankar}, F.},
        title = "{A post-merger enhancement only in star-forming Type 2 Seyfert galaxies: the deep learning view}",
      journal = {\mnras},
         year = 2024,
        month = mar,
       volume = {528},
       number = {4},
        pages = {6915-6933},
          doi = {10.1093/mnras/stae183},
archivePrefix = {arXiv},
       eprint = {2401.09632},
 primaryClass = {astro-ph.GA},
       adsurl = {https://ui.adsabs.harvard.edu/abs/2024MNRAS.528.6915A}
}

@article{McKay1979,
 ISSN = {00401706},
 URL = {http://www.jstor.org/stable/1268522},
 author = {M. D. McKay and R. J. Beckman and W. J. Conover},
 journal = {Technometrics},
 number = {2},
 pages = {239--245},
 publisher = {[Taylor & Francis, Ltd., American Statistical Association, American Society for Quality]},
 title = {A Comparison of Three Methods for Selecting Values of Input Variables in the Analysis of Output from a Computer Code},
 urldate = {2025-06-24},
 volume = {21},
 year = {1979}
}

@ARTICLE{Lin2022,
       author = {{Lin}, Q. and {Fouchez}, D. and {Pasquet}, J. and {Treyer}, M. and {Ait Ouahmed}, R. and {Arnouts}, S. and {Ilbert}, O.},
        title = "{Photometric redshift estimation with convolutional neural networks and galaxy images: Case study of resolving biases in data-driven methods}",
      journal = {\aap},
         year = 2022,
        month = jun,
       volume = {662},
          eid = {A36},
        pages = {A36},
          doi = {10.1051/0004-6361/202142751},
archivePrefix = {arXiv},
       eprint = {2202.09964},
 primaryClass = {astro-ph.IM},
       adsurl = {https://ui.adsabs.harvard.edu/abs/2022A&A...662A..36L}
}

@ARTICLE{Pricopi2025,
       author = {{Pricopi}, D. and {Popescu}, C.~C. and {Rushton}, M.~T. and {Murphy}, D. and {Inman}, C.~J. and {Toma}, R.},
        title = "{Uncovering the truth about M101, NGC 3938, and their significant others through radiative transfer}",
      journal = {\mnras},
         year = 2025,
        month = feb,
       volume = {537},
       number = {1},
        pages = {56-83},
          doi = {10.1093/mnras/stae2809},
archivePrefix = {arXiv},
       eprint = {2412.18686},
 primaryClass = {astro-ph.GA},
       adsurl = {https://ui.adsabs.harvard.edu/abs/2025MNRAS.537...56P}
}

@article{Kasim2021,
  author = {Kasim, M F and Watson-Parris, D and Deaconu, L and Oliver, S and Hatfield, P and Froula, D H and Gregori, G and Jarvis, M and Khatiwala, S and Korenaga, J and Topp-Mugglestone, J and Viezzer, E and Vinko, S M},
  journal = {Machine Learning: Science and Technology},
  title = {Building high accuracy emulators for scientific simulations with deep neural architecture search},
  year = {2021},
  volume = {3},
  number = {1},
  pages = {015013},
  doi = {10.1088/2632-2153/ac3ffa},
  publisher = {IOP Publishing Ltd}
}

@ARTICLE{Zhang2025,
       author = {{Zhang}, Junkai and {Ramnichal}, Steven and {Wuyts}, Stijn and {Li}, Cheng},
        title = "{SE3D: Testing the recovery of stellar population, dust and structural properties on mock-observed toy model and simulated galaxies}",
      journal = {arXiv e-prints},
         year = 2025,
        month = nov,
          eid = {arXiv:2511.19614},
        pages = {arXiv:2511.19614},
          doi = {10.48550/arXiv.2511.19614},
archivePrefix = {arXiv},
       eprint = {2511.19614},
 primaryClass = {astro-ph.GA},
       adsurl = {https://ui.adsabs.harvard.edu/abs/2025arXiv251119614Z}
}

@ARTICLE{Genin2025,
       author = {{Genin}, Aur{\'e}lien and {Shuntov}, Marko and {Brammer}, Gabe and {Allen}, Natalie and {Ito}, Kei and {Magdis}, Georgios and {Matharu}, Jasleen and {Oesch}, Pascal A. and {Toft}, Sune and {Valentino}, Francesco},
        title = "{DAWN JWST Archive: Morphology from profile fitting of over 340 000 galaxies in major JWST fields: Morphology evolution with redshift and galaxy type}",
      journal = {\aap},
         year = 2025,
        month = jul,
       volume = {699},
          eid = {A343},
        pages = {A343},
          doi = {10.1051/0004-6361/202555504},
archivePrefix = {arXiv},
       eprint = {2505.21622},
 primaryClass = {astro-ph.GA},
       adsurl = {https://ui.adsabs.harvard.edu/abs/2025A&A...699A.343G}
}

@ARTICLE{Lovell2025,
       author = {{Lovell}, Christopher C. and {Roper}, William J. and {Vijayan}, Aswin P. and {Wilkins}, Stephen M. and {Newman}, Sophie and {Seeyave}, Louise},
        title = "{Synthesizer: a Software Package for Synthetic Astronomical Observables}",
      journal = {The Open Journal of Astrophysics},
         year = 2025,
        month = oct,
       volume = {8},
          eid = {152},
        pages = {152},
          doi = {10.33232/001c.145766},
archivePrefix = {arXiv},
       eprint = {2508.03888},
 primaryClass = {astro-ph.IM},
       adsurl = {https://ui.adsabs.harvard.edu/abs/2025OJAp....8E.152L}
}

@ARTICLE{Tan2024,
       author = {{Tan}, Qing-Hua and {Daddi}, Emanuele and {Magnelli}, Benjamin and {Correa}, Camila A. and {Bournaud}, Fr{\'e}d{\'e}ric and {Adscheid}, Sylvia and {Zhang}, Shao-Bo and {Elbaz}, David and {G{\'o}mez-Guijarro}, Carlos and {Kalita}, Boris S. and {Liu}, Daizhong and {Liu}, Zhaoxuan and {Pety}, J{\'e}r{\^o}me and {Puglisi}, Annagrazia and {Schinnerer}, Eva and {Silverman}, John D. and {Valentino}, Francesco},
        title = "{In situ spheroid formation in distant submillimetre-bright galaxies}",
      journal = {\nat},
         year = 2024,
        month = dec,
       volume = {636},
       number = {8041},
        pages = {69-74},
          doi = {10.1038/s41586-024-08201-6},
archivePrefix = {arXiv},
       eprint = {2407.16578},
 primaryClass = {astro-ph.GA},
       adsurl = {https://ui.adsabs.harvard.edu/abs/2024Natur.636...69T}
}

@ARTICLE{Meiksin2006,
       author = {{Meiksin}, Avery},
        title = "{Colour corrections for high-redshift objects due to intergalactic attenuation}",
      journal = {\mnras},
         year = 2006,
        month = jan,
       volume = {365},
       number = {3},
        pages = {807-812},
          doi = {10.1111/j.1365-2966.2005.09756.x},
archivePrefix = {arXiv},
       eprint = {astro-ph/0512435},
 primaryClass = {astro-ph},
       adsurl = {https://ui.adsabs.harvard.edu/abs/2006MNRAS.365..807M}
}

@ARTICLE{Battisti2017,
       author = {{Battisti}, A.~J. and {Calzetti}, D. and {Chary}, R. -R.},
        title = "{Characterizing Dust Attenuation in Local Star-forming Galaxies: Inclination Effects and the 2175 {\r{A}} Feature}",
      journal = {\apj},
         year = 2017,
        month = dec,
       volume = {851},
       number = {2},
          eid = {90},
        pages = {90},
          doi = {10.3847/1538-4357/aa9a43},
archivePrefix = {arXiv},
       eprint = {1711.04814},
 primaryClass = {astro-ph.GA},
       adsurl = {https://ui.adsabs.harvard.edu/abs/2017ApJ...851...90B}
}

@ARTICLE{Lu2022,
       author = {{Lu}, Jiafeng and {Shen}, Shiyin and {Yuan}, Fang-Ting and {Shao}, Zhengyi and {Hou}, Jinliang and {Zheng}, Xianzhong},
        title = "{The Chocolate Chip Cookie Model: Dust Geometry of Milky Way-like Disk Galaxies}",
      journal = {\apj},
         year = 2022,
        month = oct,
       volume = {938},
       number = {2},
          eid = {139},
        pages = {139},
          doi = {10.3847/1538-4357/ac92e9},
archivePrefix = {arXiv},
       eprint = {2209.08515},
 primaryClass = {astro-ph.GA},
       adsurl = {https://ui.adsabs.harvard.edu/abs/2022ApJ...938..139L}
}

@ARTICLE{Trayford2025,
       author = {{Trayford}, James W. and {Schaye}, Joop and {Correa}, Camila and {Ploeckinger}, Sylvia and {Richings}, Alexander J. and {Chaikin}, Evgenii and {Schaller}, Matthieu and {Benitez-Llambay}, Alejandro and {Frenk}, Carlos and {Husko}, Filip},
        title = "{Modelling the evolution and influence of dust in cosmological simulations that include the cold phase of the interstellar medium}",
      journal = {arXiv e-prints},
         year = 2025,
        month = may,
          eid = {arXiv:2505.13056},
        pages = {arXiv:2505.13056},
          doi = {10.48550/arXiv.2505.13056},
archivePrefix = {arXiv},
       eprint = {2505.13056},
 primaryClass = {astro-ph.GA},
       adsurl = {https://ui.adsabs.harvard.edu/abs/2025arXiv250513056T}
}

@ARTICLE{Mathews2023,
       author = {{Mathews}, Elijah P. and {Leja}, Joel and {Speagle}, Joshua S. and {Johnson}, Benjamin D. and {Gibson}, Justus and {Nelson}, Erica J. and {Suess}, Katherine A. and {Tacchella}, Sandro and {Whitaker}, Katherine E. and {Wang}, Bingjie},
        title = "{As Simple as Possible but No Simpler: Optimizing the Performance of Neural Net Emulators for Galaxy SED Fitting}",
      journal = {\apj},
         year = 2023,
        month = sep,
       volume = {954},
       number = {2},
          eid = {132},
        pages = {132},
          doi = {10.3847/1538-4357/ace720},
archivePrefix = {arXiv},
       eprint = {2306.16442},
 primaryClass = {astro-ph.IM},
       adsurl = {https://ui.adsabs.harvard.edu/abs/2023ApJ...954..132M}
}

@ARTICLE{Dobbels2021,
       author = {{Dobbels}, Wouter and {Baes}, Maarten},
        title = "{Predicting far-infrared maps of galaxies via machine learning techniques}",
      journal = {\aap},
         year = 2021,
        month = nov,
       volume = {655},
          eid = {A34},
        pages = {A34},
          doi = {10.1051/0004-6361/202142084},
archivePrefix = {arXiv},
       eprint = {2110.01704},
 primaryClass = {astro-ph.GA},
       adsurl = {https://ui.adsabs.harvard.edu/abs/2021A&A...655A..34D}
}

@ARTICLE{Alsing2020,
       author = {{Alsing}, Justin and {Peiris}, Hiranya and {Leja}, Joel and {Hahn}, ChangHoon and {Tojeiro}, Rita and {Mortlock}, Daniel and {Leistedt}, Boris and {Johnson}, Benjamin D. and {Conroy}, Charlie},
        title = "{SPECULATOR: Emulating Stellar Population Synthesis for Fast and Accurate Galaxy Spectra and Photometry}",
      journal = {\apjs},
         year = 2020,
        month = jul,
       volume = {249},
       number = {1},
          eid = {5},
        pages = {5},
          doi = {10.3847/1538-4365/ab917f},
archivePrefix = {arXiv},
       eprint = {1911.11778},
 primaryClass = {astro-ph.IM},
       adsurl = {https://ui.adsabs.harvard.edu/abs/2020ApJS..249....5A}
}

@ARTICLE{Iyer2025,
       author = {{Iyer}, Kartheik G. and {Pacifici}, Camilla and {Calistro-Rivera}, Gabriela and {Lovell}, Christopher C.},
        title = "{The Spectral Energy Distributions of Galaxies}",
      journal = {arXiv e-prints},
         year = 2025,
        month = feb,
          eid = {arXiv:2502.17680},
        pages = {arXiv:2502.17680},
          doi = {10.48550/arXiv.2502.17680},
archivePrefix = {arXiv},
       eprint = {2502.17680},
 primaryClass = {astro-ph.GA},
       adsurl = {https://ui.adsabs.harvard.edu/abs/2025arXiv250217680I}
}

@inproceedings{Akiba2019, 
    author = {Akiba, Takuya and Sano, Shotaro and Yanase, Toshihiko and Ohta, Takeru and Koyama, Masanori}, 
    title = {Optuna: A Next-generation Hyperparameter Optimization Framework}, 
    year = {2019}, isbn = {9781450362016}, publisher = {Association for Computing Machinery}, 
    address = {New York, NY, USA}, 
    booktitle = {Proceedings of the 25th ACM SIGKDD International Conference on Knowledge Discovery \& Data Mining}, 
    pages = {2623–2631}, 
    numpages = {9}, 
    location = {Anchorage, AK, USA}, 
    series = {KDD '19} }

@ARTICLE{Lee2022,
  author={Lee, Sungyoon and Kim, Hoki and Lee, Jaewook},
  journal={IEEE Transactions on Pattern Analysis and Machine Intelligence}, 
  title={GradDiv: Adversarial Robustness of Randomized Neural Networks via Gradient Diversity Regularization}, 
  year={2023},
  volume={45},
  number={2},
  pages={2645-2651},
  doi={10.1109/TPAMI.2022.3169217}}

@ARTICLE{Bell2001,
       author = {{Bell}, Eric F. and {de Jong}, Roelof S.},
        title = "{Stellar Mass-to-Light Ratios and the Tully-Fisher Relation}",
      journal = {\apj},
         year = 2001,
        month = mar,
       volume = {550},
       number = {1},
        pages = {212-229},
          doi = {10.1086/319728},
archivePrefix = {arXiv},
       eprint = {astro-ph/0011493},
 primaryClass = {astro-ph},
       adsurl = {https://ui.adsabs.harvard.edu/abs/2001ApJ...550..212B}
}

@ARTICLE{Walcher2011,
       author = {{Walcher}, Jakob and {Groves}, Brent and {Budav{\'a}ri}, Tam{\'a}s and {Dale}, Daniel},
        title = "{Fitting the integrated spectral energy distributions of galaxies}",
      journal = {\apss},
         year = 2011,
        month = jan,
       volume = {331},
       number = {1},
        pages = {1-51},
          doi = {10.1007/s10509-010-0458-z},
archivePrefix = {arXiv},
       eprint = {1008.0395},
 primaryClass = {astro-ph.CO},
       adsurl = {https://ui.adsabs.harvard.edu/abs/2011Ap&SS.331....1W}
}

@ARTICLE{DiazGarcia2015,
       author = {{D{\'\i}az-Garc{\'\i}a}, L.~A. and {Cenarro}, A.~J. and {L{\'o}pez-Sanjuan}, C. and {Peralta de Arriba}, L. and {Ferreras}, I. and {Cervi{\~n}o}, M. and {M{\'a}rquez}, I. and {Masegosa}, J. and {del Olmo}, A. and {Perea}, J.},
        title = "{Stellar populations of galaxies in the ALHAMBRA survey up to z {\ensuremath{\sim}} 1. IV. Properties of quiescent galaxies on the stellar mass-size plane}",
      journal = {\aap},
         year = 2019,
        month = nov,
       volume = {631},
          eid = {A158},
        pages = {A158},
          doi = {10.1051/0004-6361/201935257},
archivePrefix = {arXiv},
       eprint = {1901.05983},
 primaryClass = {astro-ph.GA},
       adsurl = {https://ui.adsabs.harvard.edu/abs/2019A&A...631A.158D}
}

@ARTICLE{Salmon2016,
       author = {{Salmon}, Brett and {Papovich}, Casey and {Long}, James and {Willner}, S.~P. and {Finkelstein}, Steven L. and {Ferguson}, Henry C. and {Dickinson}, Mark and {Duncan}, Kenneth and {Faber}, S.~M. and {Hathi}, Nimish and {Koekemoer}, Anton and {Kurczynski}, Peter and {Newman}, Jeffery and {Pacifici}, Camilla and {P{\'e}rez-Gonz{\'a}lez}, Pablo G. and {Pforr}, Janine},
        title = "{Breaking the Curve with CANDELS: A Bayesian Approach to Reveal the Non-Universality of the Dust-Attenuation Law at High Redshift}",
      journal = {\apj},
         year = 2016,
        month = aug,
       volume = {827},
       number = {1},
          eid = {20},
        pages = {20},
          doi = {10.3847/0004-637X/827/1/20},
archivePrefix = {arXiv},
       eprint = {1512.05396},
 primaryClass = {astro-ph.GA},
       adsurl = {https://ui.adsabs.harvard.edu/abs/2016ApJ...827...20S}
}

@ARTICLE{Nersesian2025,
       author = {{Nersesian}, Angelos and {van der Wel}, Arjen and {Gallazzi}, Anna R. and {Kaushal}, Yasha and {Bezanson}, Rachel and {Zibetti}, Stefano and {Bell}, Eric F. and {D'Eugenio}, Francesco and {Leja}, Joel and {Martorano}, Marco and {Wu}, Po-Feng},
        title = "{More is better: Strong constraints on the stellar properties of LEGA-C z {\ensuremath{\sim}} 1 galaxies with Prospector}",
      journal = {\aap},
         year = 2025,
        month = mar,
       volume = {695},
          eid = {A86},
        pages = {A86},
          doi = {10.1051/0004-6361/202452662},
archivePrefix = {arXiv},
       eprint = {2502.03021},
 primaryClass = {astro-ph.GA},
       adsurl = {https://ui.adsabs.harvard.edu/abs/2025A&A...695A..86N}
}

@ARTICLE{Kapoor2023,
       author = {{Kapoor}, Anand Utsav and {Baes}, Maarten and {van der Wel}, Arjen and {Gebek}, Andrea and {Camps}, Peter and {Nersesian}, Angelos and {Meidt}, Sharon E. and {Smith}, Aaron and {Vicens}, Sebastien and {D'Eugenio}, Francesco and {Martorano}, Marco and {Barrientos}, Daniela and {Sartorio}, Nina Sanches},
        title = "{TODDLERS: a new UV-mm emission library for star-forming regions - I. Integration with SKIRT and public release}",
      journal = {\mnras},
         year = 2023,
        month = dec,
       volume = {526},
       number = {3},
        pages = {3871-3901},
          doi = {10.1093/mnras/stad2977},
archivePrefix = {arXiv},
       eprint = {2310.00388},
 primaryClass = {astro-ph.GA},
       adsurl = {https://ui.adsabs.harvard.edu/abs/2023MNRAS.526.3871K}
}

@ARTICLE{Sun2023,
       author = {{Sun}, Guochao and {Faucher-Gigu{\`e}re}, Claude-Andr{\'e} and {Hayward}, Christopher C. and {Shen}, Xuejian},
        title = "{Seen and unseen: bursty star formation and its implications for observations of high-redshift galaxies with JWST}",
      journal = {\mnras},
         year = 2023,
        month = dec,
       volume = {526},
       number = {2},
        pages = {2665-2672},
          doi = {10.1093/mnras/stad2902},
archivePrefix = {arXiv},
       eprint = {2305.02713},
 primaryClass = {astro-ph.GA},
       adsurl = {https://ui.adsabs.harvard.edu/abs/2023MNRAS.526.2665S}
}

@ARTICLE{Ferrara2016,
       author = {{Ferrara}, A. and {Viti}, S. and {Ceccarelli}, C.},
        title = "{The problematic growth of dust in high-redshift galaxies}",
      journal = {\mnras},
         year = 2016,
        month = nov,
       volume = {463},
       number = {1},
        pages = {L112-L116},
          doi = {10.1093/mnrasl/slw165},
archivePrefix = {arXiv},
       eprint = {1606.07214},
 primaryClass = {astro-ph.GA},
       adsurl = {https://ui.adsabs.harvard.edu/abs/2016MNRAS.463L.112F}
}

@ARTICLE{RemyRuyer2015,
       author = {{R{\'e}my-Ruyer}, A. and {Madden}, S.~C. and {Galliano}, F. and {Lebouteiller}, V. and {Baes}, M. and {Bendo}, G.~J. and {Boselli}, A. and {Ciesla}, L. and {Cormier}, D. and {Cooray}, A. and {Cortese}, L. and {De Looze}, I. and {Doublier-Pritchard}, V. and {Galametz}, M. and {Jones}, A.~P. and {Karczewski}, O. {\L}. and {Lu}, N. and {Spinoglio}, L.},
        title = "{Linking dust emission to fundamental properties in galaxies: the low-metallicity picture}",
      journal = {\aap},
         year = 2015,
        month = oct,
       volume = {582},
          eid = {A121},
        pages = {A121},
          doi = {10.1051/0004-6361/201526067},
archivePrefix = {arXiv},
       eprint = {1507.05432},
 primaryClass = {astro-ph.GA},
       adsurl = {https://ui.adsabs.harvard.edu/abs/2015A&A...582A.121R}
}

@ARTICLE{Weinberg2017,
       author = {{Weinberg}, David H. and {Andrews}, Brett H. and {Freudenburg}, Jenna},
        title = "{Equilibrium and Sudden Events in Chemical Evolution}",
      journal = {\apj},
         year = 2017,
        month = mar,
       volume = {837},
       number = {2},
          eid = {183},
        pages = {183},
          doi = {10.3847/1538-4357/837/2/183},
archivePrefix = {arXiv},
       eprint = {1604.07435},
 primaryClass = {astro-ph.GA},
       adsurl = {https://ui.adsabs.harvard.edu/abs/2017ApJ...837..183W}
}

@ARTICLE{Matsumoto2025,
       author = {{Matsumoto}, Kosei and {Sommovigo}, Laura and {Gebek}, Andrea and {Nagamine}, Kentaro and {Nersesian}, Angelos and {Baes}, Maarten and {De Looze}, Ilse and {van der Wel}, Arjen and {Somerville}, Rachel and {Romano}, Leonard E.~C. and {Cochrane}, Rachel K.},
        title = "{Evolution of galaxy attenuation curves driven by evolving dust mass and grain size distributions}",
      journal = {\aap},
         year = 2026,
        month = jan,
       volume = {705},
          eid = {A75},
        pages = {A75},
          doi = {10.1051/0004-6361/202555658},
archivePrefix = {arXiv},
       eprint = {2508.21157},
 primaryClass = {astro-ph.GA},
       adsurl = {https://ui.adsabs.harvard.edu/abs/2026A&A...705A..75M}
}

@ARTICLE{Witstok2023,
       author = {{Witstok}, Joris and {Shivaei}, Irene and {Smit}, Renske and {Maiolino}, Roberto and {Carniani}, Stefano and {Curtis-Lake}, Emma and {Ferruit}, Pierre and {Arribas}, Santiago and {Bunker}, Andrew J. and {Cameron}, Alex J. and {Charlot}, Stephane and {Chevallard}, Jacopo and {Curti}, Mirko and {de Graaff}, Anna and {D'Eugenio}, Francesco and {Giardino}, Giovanna and {Looser}, Tobias J. and {Rawle}, Tim and {Rodr{\'\i}guez del Pino}, Bruno and {Willott}, Chris and {Alberts}, Stacey and {Baker}, William M. and {Boyett}, Kristan and {Egami}, Eiichi and {Eisenstein}, Daniel J. and {Endsley}, Ryan and {Hainline}, Kevin N. and {Ji}, Zhiyuan and {Johnson}, Benjamin D. and {Kumari}, Nimisha and {Lyu}, Jianwei and {Nelson}, Erica and {Perna}, Michele and {Rieke}, Marcia and {Robertson}, Brant E. and {Sandles}, Lester and {Saxena}, Aayush and {Scholtz}, Jan and {Sun}, Fengwu and {Tacchella}, Sandro and {Williams}, Christina C. and {Willmer}, Christopher N.~A.},
        title = "{Carbonaceous dust grains seen in the first billion years of cosmic time}",
      journal = {\nat},
         year = 2023,
        month = sep,
       volume = {621},
       number = {7978},
        pages = {267-270},
          doi = {10.1038/s41586-023-06413-w},
archivePrefix = {arXiv},
       eprint = {2302.05468},
 primaryClass = {astro-ph.GA},
       adsurl = {https://ui.adsabs.harvard.edu/abs/2023Natur.621..267W}
}

@ARTICLE{SachdevaNath2022,
       author = {{Sachdeva}, Sonali and {Nath}, Biman B.},
        title = "{Star-dust geometry main determinant of dust attenuation in galaxies}",
      journal = {\mnras},
         year = 2022,
        month = jun,
       volume = {513},
       number = {1},
        pages = {L63-L67},
          doi = {10.1093/mnrasl/slac037},
archivePrefix = {arXiv},
       eprint = {2204.03478},
 primaryClass = {astro-ph.GA},
       adsurl = {https://ui.adsabs.harvard.edu/abs/2022MNRAS.513L..63S}
}

@ARTICLE{Perry2025,
       author = {{Perry}, Marissa N. and {Taylor}, Anthony J. and {Ch{\'a}vez Ortiz}, {\'O}scar A. and {Finkelstein}, Steven L. and {C.~K. Leung}, Gene and {Bagley}, Micaela B. and {Fern{\'a}ndez}, Vital and {Arrabal Haro}, Pablo and {Chworowsky}, Katherine and {Cleri}, Nikko J. and {Dickinson}, Mark and {Ellis}, Richard S. and {Kartaltepe}, Jeyhan S. and {Koekemoer}, Anton M. and {Pacucci}, Fabio and {Papovich}, Casey and {Pirzkal}, Nor and {Tacchella}, Sandro},
        title = "{The Prevalence of Bursty Star Formation in Low-mass Galaxies at z = 1─7 from H{\ensuremath{\alpha}}-to-UV Diagnostics}",
      journal = {\apj},
         year = 2025,
        month = nov,
       volume = {994},
       number = {1},
          eid = {14},
        pages = {14},
          doi = {10.3847/1538-4357/ae102f},
archivePrefix = {arXiv},
       eprint = {2510.05388},
 primaryClass = {astro-ph.GA},
       adsurl = {https://ui.adsabs.harvard.edu/abs/2025ApJ...994...14P}
}

@ARTICLE{Zubko2004,
       author = {{Zubko}, Viktor and {Dwek}, Eli and {Arendt}, Richard G.},
        title = "{Interstellar Dust Models Consistent with Extinction, Emission, and Abundance Constraints}",
      journal = {\apjs},
         year = 2004,
        month = jun,
       volume = {152},
       number = {2},
        pages = {211-249},
          doi = {10.1086/382351},
archivePrefix = {arXiv},
       eprint = {astro-ph/0312641},
 primaryClass = {astro-ph},
       adsurl = {https://ui.adsabs.harvard.edu/abs/2004ApJS..152..211Z}
}




\appendix

\section{Predicting toy models' physical status}
\label{app:physical_unphysical}

\begin{figure}
    \centering
    \includegraphics[width=\linewidth]{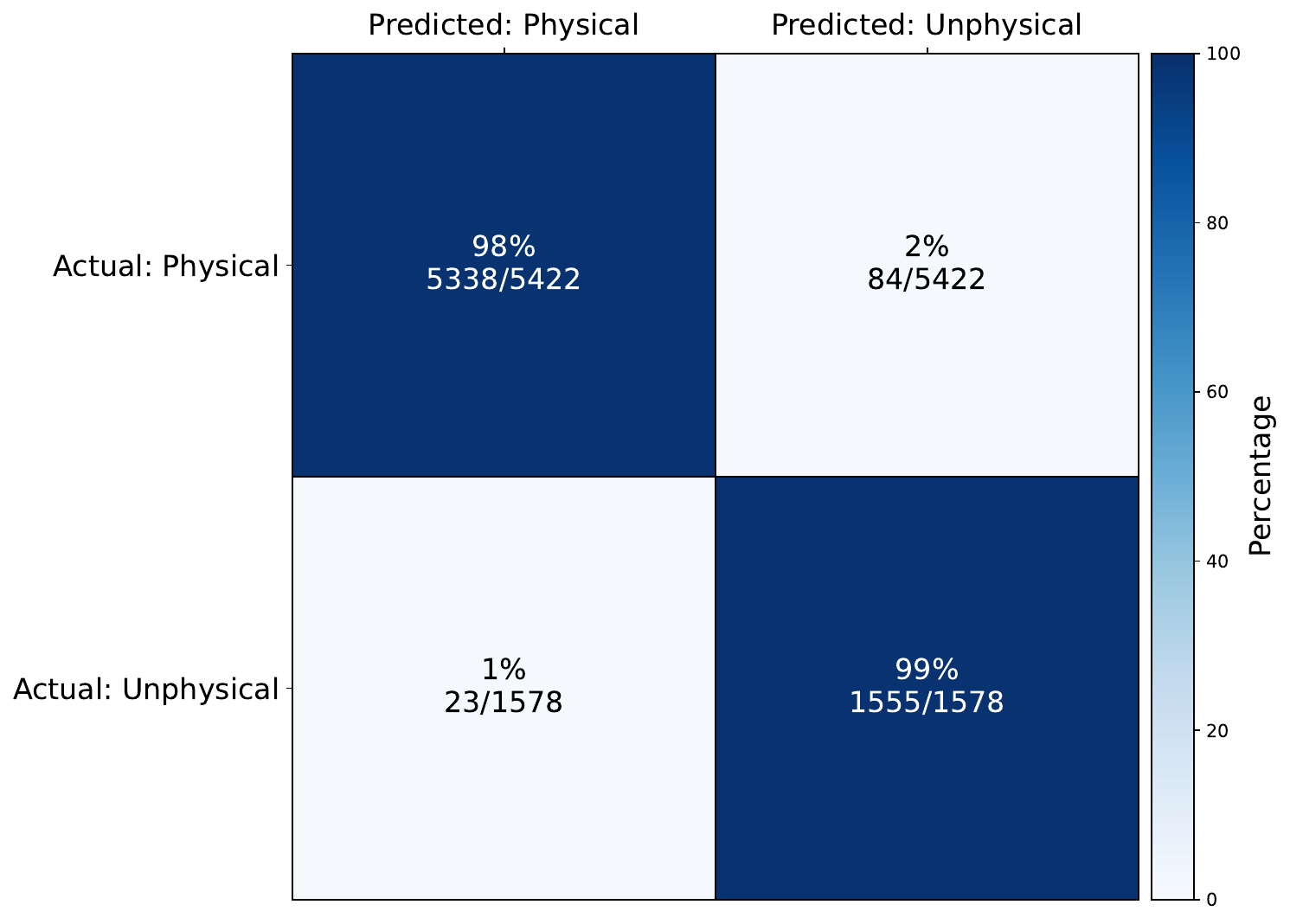}
    \caption{Validation of model predicting whether a toy model galaxy is physical or unphysical, as evaluated on an unseen testing set. 
    The fraction of correctly classified galaxies and false-positive predictions are displayed, demonstrating that predictions are made at high confidence.}
    \label{fig:Physical_Unphysical_ML}
\end{figure}

The parameter distributions specified in Table\ \ref{tab:param_library} by themselves do not guarantee that the mass of dust in birth clouds remains below the total dust mass specified for the galaxy.  As such a situation would be unphysical, the respective set of parameter values would be discarded from the training library.  Likewise, when doing {\tt SE3D} fitting, we want to prevent such unphysical combinations of parameters.  In order to efficiently identify whether a set of input parameters leads to a physical/unphysical status, we train a simple NN on the binary physical/unphysical flags recorded for the entries in our training library.  As shown in Figure\ \ref{fig:Physical_Unphysical_ML}, this tool performs at high ($> 98\%$) fidelity, and its virtually instantaneous output enables efficient implementation within the MCMC fitting.  In practice, for any unphysical set of parameters a prior of $-\infty$ is added to the computed log-likelihood, preventing walkers to explore this region of parameter space.

\section{Hyperparameter tuning}
\label{app:hyper}

In Table\ \ref{tab:hyper}, we specify the hyperparameters of our ML architecture and training process that are being tuned using {\tt Optuna} \citep{Akiba2019}. Each parameter is either uniformly sampled within its defined range or randomly selected from a discrete set of available options, depending on its type. This ensures a broad and unbiased exploration of hyperparameter space during the tuning process.  The optimization is done for each BNN separately, each responsible for predicting a particular SD type.

\section{Varying one parameter at a time}
\label{app:oneparam}

\begin{figure*}
    \centering
    \begin{subfigure}{\linewidth}
        \includegraphics[width=\linewidth]{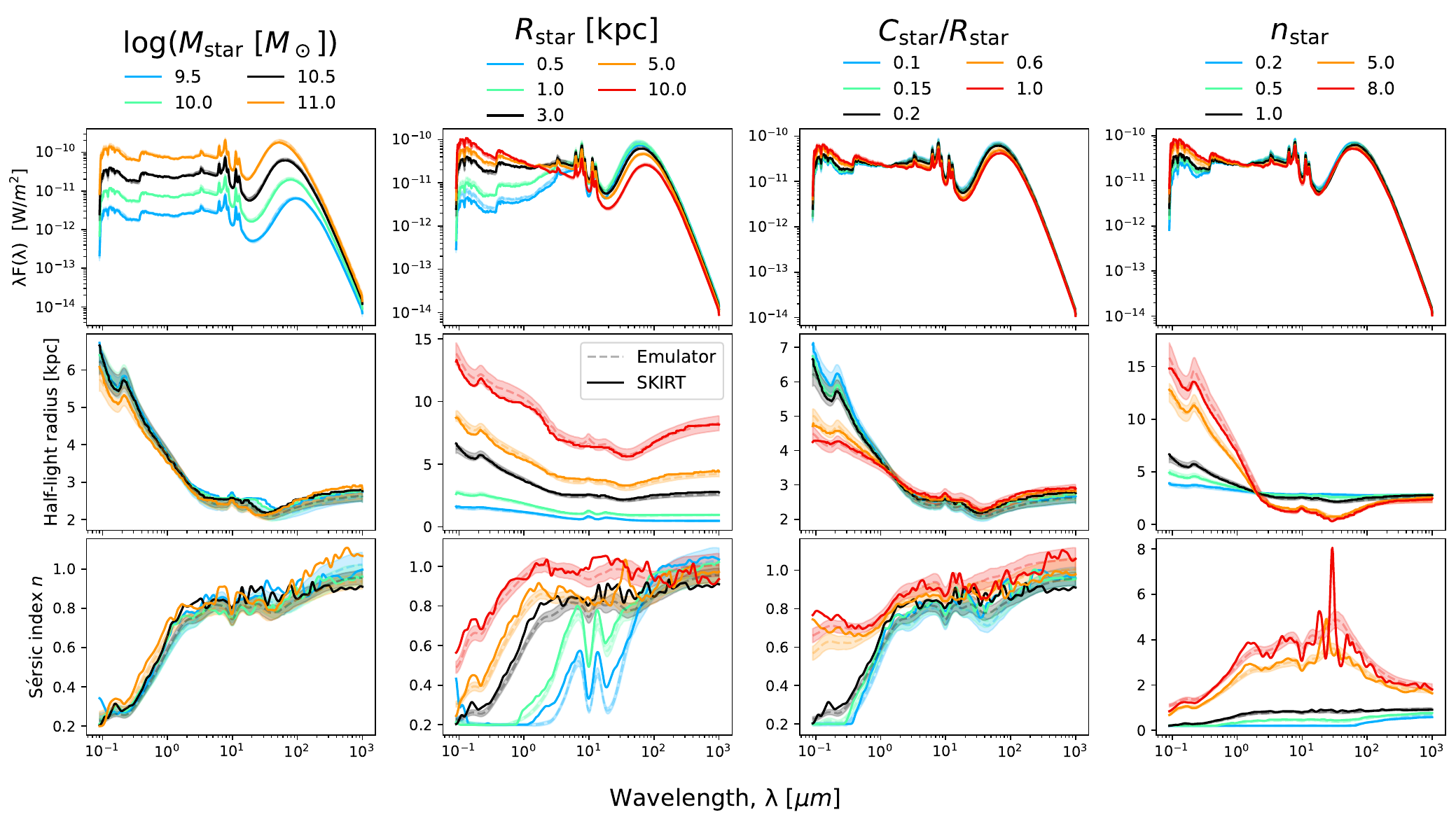}
    \end{subfigure}
    \\
    \begin{subfigure}{\linewidth}
        \includegraphics[width=\linewidth]{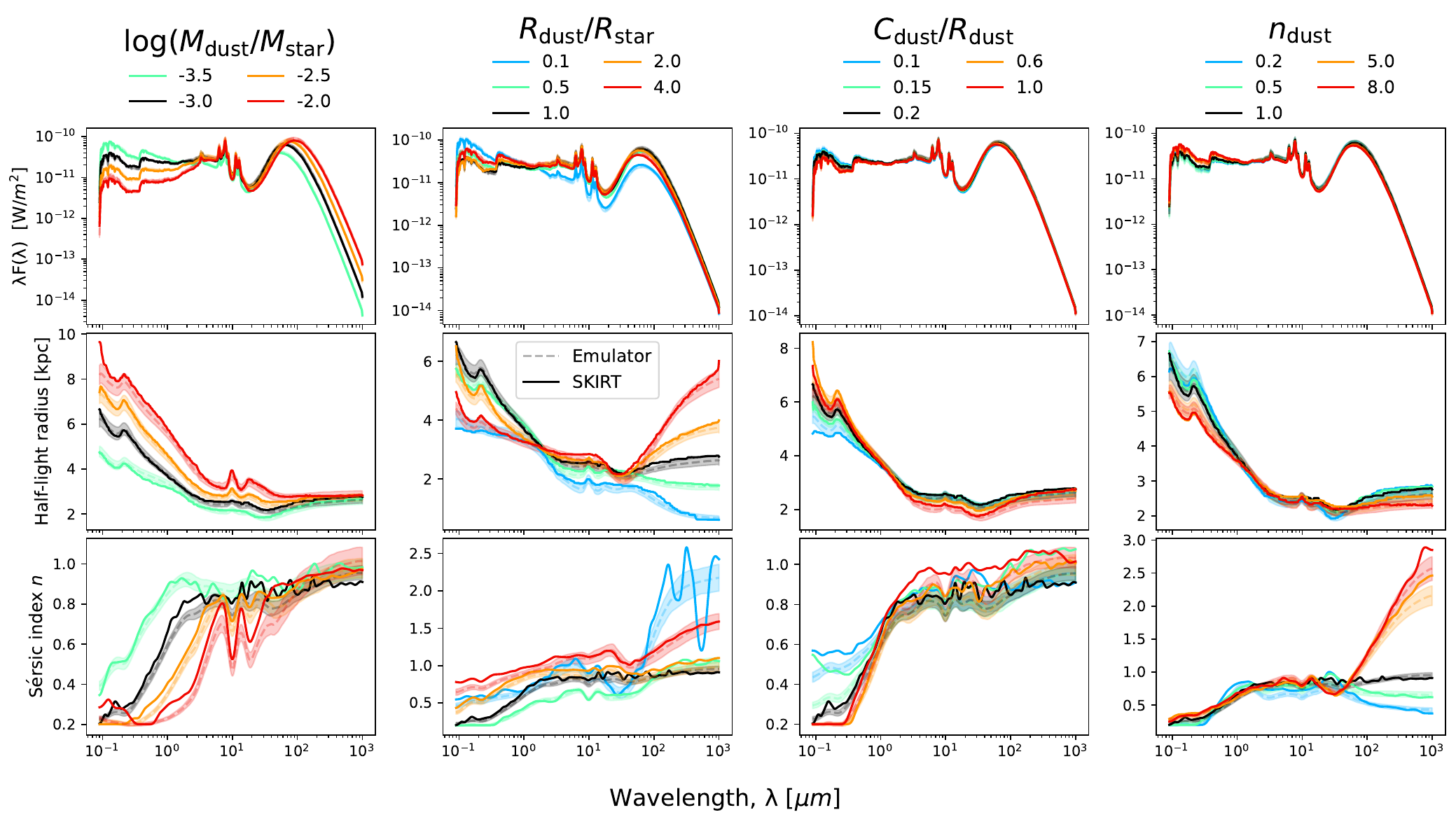}
    \end{subfigure}
    \caption{Effect of varying individual parameters on predicted SEDs, SRDs, and SNDs. Each panel displays the result of modifying one parameter while keeping others fixed to the reference model (shown in black). Dashed lines and shaded regions represent emulator predictions and corresponding uncertainties whereas solid lines display SKIRT ground truth. \textit{Top (left to right):} Varying log($M_{\rm star}$), $R_{\rm star}$, $C_{\rm star}$/$R_{\rm star}$, and $n_{\rm star}$. \textit{Bottom (left to right):} Varying log($M_{\rm dust}$/$M_{\rm star}$), $R_{\rm dust}$/$R_{\rm star}$, $C_{\rm dust}$/$R_{\rm dust}$, and $n_{\rm dust}$.}
    \label{app:oneparam_App}
\end{figure*}

\begin{figure*}
    \ContinuedFloat
    \centering
    \begin{subfigure}{\linewidth}
        \includegraphics[width=\linewidth]{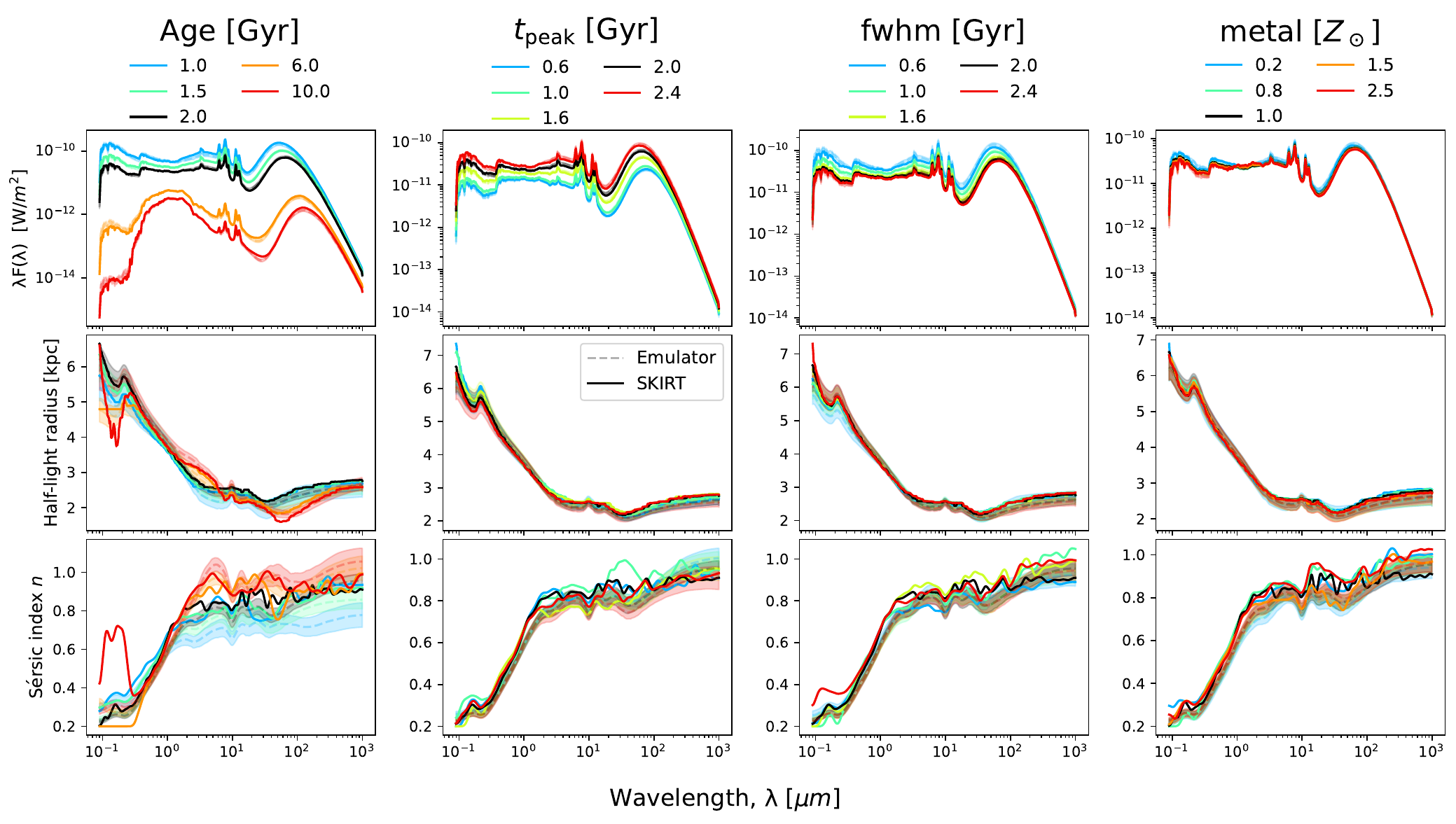}
    \end{subfigure}
    \\
    \begin{subfigure}{\linewidth}
        \includegraphics[width=\linewidth]{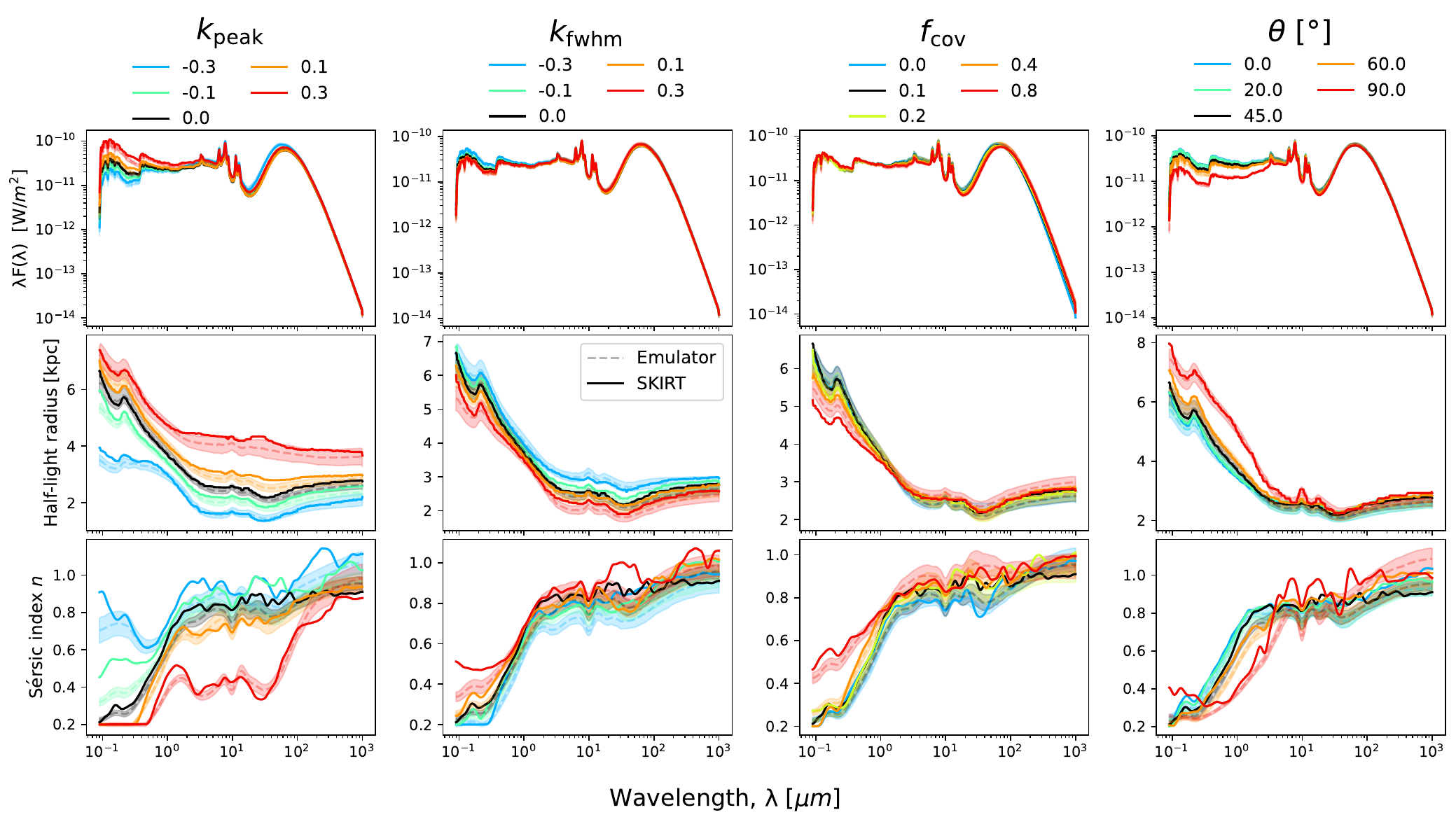}
    \end{subfigure}
    \caption*{Figure \ref{app:oneparam_App} continued. \textit{Top (left to right):} Varying Age, $t_{\rm peak}$, fwhm, metal. \textit{Bottom (left to right):} Varying $k_{\rm peak}$, $k_{\rm fwhm}$, $f_{\rm cov}$, $\theta$.}
\end{figure*}

Figure\ \ref{app:oneparam_App} displays the resulting SEDs, SRDs, and SNDs when varying one parameter at a time from the parameter distributions described in Table\ \ref{tab:param_library}. The same reference toy model, shown in black, is displayed in all panels of Figure\ \ref{app:oneparam_App} with the following parametrization: log($M_{\rm star}$) = 10.5, $R_{\rm star} = 3\ {\rm kpc}$, $C_{\rm star}$/$R_{\rm star}$ = 0.2, $n_{\rm star}$ = 1, log($M_{\rm dust}$) = 7.5,  Age = 2 Gyr, $t_{\rm peak}$ = 2 Gyr, fwhm = 2 Gyr, $k_{\rm peak} = 0$, $k_{\rm fwhm} = 0$, metal = $Z_{\odot}$, $R_{\rm dust}$/$R_{\rm star}$ = 1, $C_{\rm dust}$/$R_{\rm dust}$ = 0.2, $n_{\rm dust}$ = 1, $f_{\rm cov}$ = 0.1, and $\theta = 45^{\circ}$.  The SEDs shown correspond to observing the galaxies at a distance of 10 Mpc (and zero redshift).


\bsp	
\label{lastpage}
\end{document}